\documentclass[%
 aip,
 amsmath,amssymb,
preprint,%
]{revtex4-1}
\usepackage{amsmath}
\usepackage{amssymb}
\usepackage{physics}
\usepackage{graphicx}
\graphicspath{{./images/}}

\usepackage{array}
\newcolumntype{L}[1]{>{\raggedright\let\newline\\\arraybackslash\hspace{0pt}}m{#1}}
\newcolumntype{C}[1]{>{\centering\let\newline\\\arraybackslash\hspace{0pt}}m{#1}}
\newcolumntype{R}[1]{>{\raggedleft\let\newline\\\arraybackslash\hspace{0pt}}m{#1}}

\usepackage{adjustbox}

\usepackage{siunitx} %for easy & consistent units
	\DeclareSIUnit\molecule{molecule}
	\DeclareSIUnit\debye{D}
	\DeclareSIUnit\au{a.u.}
	\DeclareSIUnit\Buckingham{B}
\usepackage{color}
\usepackage{chngcntr}
\usepackage{hyperref}

\newcommand{\etal}{\textit{et al.}}

\newcommand{\ai}{\textit{ab initio}}
\newcommand{\Ai}{\textit{Ab initio}}

\newcommand{\Duo}{{\sc Duo}}
\newcommand{\Exocross}{{\sc Exocross}}

\newcommand{\tnu}{\tilde{\nu}}

\begin{document}

The following article has been accepted by Journal of Chemical Physics. After it is published, it will be found at \url{https://doi.org/10.1063/5.0063256}

\title{Calculation of electric quadrupole linestrengths for diatomic molecules: Application to the %$X$~$^1\Sigma^+$ states of 
H\textsubscript{2}, CO,  HF and O\textsubscript{2} molecules}

\author{W. Somogyi}
\email{wilfrid.somogyi.15@ucl.ac.uk}
\affiliation{Department of Physics and Astronomy, University College London, Gower Street, WC1E 6BT London, United Kingdom}

\author{S.N. Yurchenko}
\email{s.yurchenko@ucl.ac.uk}
\affiliation{Department of Physics and Astronomy, University College London, Gower Street, WC1E 6BT London, United Kingdom}%Lines break automatically or can be forced with \\

\author{A. Yachmenev}
\email{andrey.yachmenev@cfel.de}
\affiliation{Center for Free-Electron Laser Science CFEL, Deutsches Elektronen-Synchrotron DESY, 
Notkestraße 85, 22607 Hamburg, Germany; Center for Ultrafast Imaging, Universität Hamburg, Luruper 
Chaussee 
149, 22761 Hamburg, Germany}

\date{\today}

\begin{abstract}

We present a unified variational treatment of the electric quadrupole (E2) matrix elements, Einstein coefficients, and line strengths for general open-shell diatomic molecules in the general purpose  diatomic code \Duo. Transformation relations between the Cartesian representation (typically used in electronic structure calculations) to the tensorial representation (required for spectroscopic applications) of the electric quadrupole moment components are derived. The implementation has been validated against accurate theoretical calculations and experimental measurements of quadrupole intensities of \textsuperscript{1}H\textsubscript{2} available in the literature.   We also present accurate electronic structure calculations of the electric quadrupole moment functions for the $X^1\Sigma^+$ electronic states of CO and HF at the CCSD(T) and MRCI levels of theory, respectively, as well for the $a^1\Delta_g$ -- $b^1\Sigma_g^+$ quadrupole transition moment of O\textsubscript{2} with MRCI level of theory. Accurate infrared E2 line lists for \textsuperscript{12}C\textsuperscript{16}O and \textsuperscript{1}H\textsuperscript{19}F are provided. A demonstration of spectroscopic applications is presented  by simulating  E2 spectra for \textsuperscript{12}C\textsuperscript{16}O,  H\textsuperscript{19}F and \textsuperscript{16}O\textsubscript{2} (Noxon $a^1\Delta_g$ -- $b^1\Sigma_g^+$ band).
 
\end{abstract}

\maketitle

\vspace{3em}

\section{Introduction}

The electric dipole approximation is often used to treat the spectra of diatomic, or small polyatomic, molecules. For most systems, this is a valid approximation that produces good results. For homonuclear diatomic molecules however, electric dipole (E1) selection rules forbid pure rotational and vibrational transitions,  as well as parallel  electronic transitions
%between states of the same symmetry ($u \leftrightarrow u, g \leftrightarrow g$) \citep{68BuDiDu.CO,59Buckin.quadrupole} 
and electric quadrupole (E2) transitions and magnetic dipole (M1) become important.\citep{44ReSiRo.O2, 80Brault.O2, 81GoReRo.O2, 81RoGoxx.O2, 10GoKaCa.O2,11LeKaGo.O2, 86ReJeBr.N2, 17CeVaMo.N2,49Herzberg.H2, 98WoSiDa.H2, 12HuPaCh.H2, 19RoAbCz.H2, 03SeKaFi.S2, 14AuKrSp.He2+}   

This has implications for the spectra of several important molecules. The most famous example is the hydrogen molecule, which despite being the most abundant molecule in the Universe has no infrared electric dipole spectrum. The three lowest lying electronic states of another important molecule, O\textsubscript{2}, all have the same ({\it gerade}) symmetry and transitions between them are therefore electric dipole forbidden.\citep{14LiShSu.O2,12YuMiDr.O2,10GoKaCa.O2} Oxygen's significant absorption in the visible region comes from the electric quadrupole and magnetic dipole moments.

Even when electric dipole transitions are weakly allowed through interactions with other electronic states, E2 and M1 transitions may still be detectable, and their consideration is necessary for an accurate description of the molecule's spectrum,\citep{66TiSixx.CO,72TiSixx.CO,20CaKaYa.H2O,20CaSoSo.H2O} such as for the Cameron bands ($a^3\Pi$ -- $X^1\Sigma$) and fourth positive system ($A^1\Pi$ -- $X^1\Sigma$) of CO.\citep{66Krupenie.CO,72TiSixx.CO,96GlNuRe.CO}

E2 and M1 transitions prove difficult to measure experimentally, owing to their weak intensity and the long path lengths required for appreciable absorption. Electric quadrupole transition intensities are on the order of $10^{6}-10^{9}$ times smaller than electric dipole transition intensities.\cite{20CaKaYa.H2O,20CaSoSo.H2O,06BuJexx.method} Nevertheless, they are often present in atmospheric spectra, where sufficiently long path lengths are regularly achievable, and play an important role in geophysical and astrophysical applications.\citep{81RoGoxx.O2,81GoReRo.O2,95GoRiCa.O2,96GlNuRe.CO,10GoKaCa.O2,16DoWoMa.O2}

In spectroscopic applications, as used in e.g. the HITRAN database,\citep{HITRAN2016} the E2 intensities are usually represented by expressions in terms of effective electric quadrupole moment constants with the rotational line intensities modelled via H\"{o}nl-London factors.\citep{11MiBaSh.O2, 10GoKaCa.O2} Examples of variational methodology used for electric quadrupole intensities of open-shell diatomics molecules include earlier works by  \citet{65Chiu.quadrupole,90BaDCNa.O2,94BaNaxx.O2}. 

Exoplanetary atmospheric retrievals require high resolution molecular opacities across a wide spectral range for a variety of temperatures. This has been the ongoing focus of the ExoMol database, and to date molecular line lists  have been produced for more than 80 molecules and 190 isotopologues.\citep{20TeYuAl} However, several important homonuclear molecules, including N\textsubscript{2},\citep{19LaSpGr.N2} S\textsubscript{2} and the crucial biosignature molecule O\textsubscript{2}\citep{18MeReAr.O2, 19DoYoKl.O2, 16ScWoBe.O2, 21KiScxx.O2, 20LiDeUn.O2} have evaded rigorous treatment, due to the dipole-forbidden nature of their spectra. As a result, these molecules are currently missing from analyses of atmospheric  spectra of hot exoplanets, representing a significant obstacle to the characterisation of exoplanet atmospheres or indeed any high temperature environments. 

Here we present a  formulation of the electric quadrupole line intensities for a general (open-shell) diatomic molecule and an implementation of these E2 matrix element and linestrength expressions in the \Duo\ program \citep{Duo} - a powerful rovibronic variational program developed as part of the ExoMol project to solve the time-independent Schr\"{o}dinger equations and compute rovibronic spectra of diatomics. To the best of our knowledge, this work represents the first general computational methodology for generating  quadrupole  spectra of arbitrary  diatomic systems from first-principles,  which lays the foundations for future work to produce a complete molecular line list for O\textsubscript{2} and other homonuclear diatomics. 

%This work will facilitate possible detection of molecular oxygen in exoplanetary atmospheres, and enable us to better understand the atmospheric chemistry, geochemistry and habitability of Earth-like planets.

The structure of the paper is as follows.
Section \ref{sec:theoretical_background} introduces the rovibronic basis used by the \Duo\ program before presenting expressions for the electric quadrupole matrix elements, linestrengths and Einstein coefficients, for a general case of an arbitrary  diatomic molecule. We also show how the matrix element components in the Cartesian representation, commonly employed in electronic structure calculations, are related to the tensorial representation used by \Duo, and outline the approach taken to reconstruct the transformation between the two.
In section \ref{sec:demonstrations} we provide demonstrations for the \Duo\ implementation of electric quadrupole linestrength calculations, including a validation against accurate theoretical and experimental linestrengths for H\textsubscript{2}. We also present accurate quantum chemistry calculations of the electric quadrupole moment functions for CO and HF molecules, as well as infrared transition linestrengths for CO and HF molecules calculated using \Duo. These line lists are included into the ExoMol data base \url{www.exomol.com}, which aims to provide molecular spectroscopic data for studies of exoplanetary and other atmospheres. More challenging nuclear motion applications of electronic E2 spectra of open-shell diatomic molecules are underway. As an illustration of an open-shell application, an E2 spectrum for the electronic system  $a^1\Delta_g$ -- $b^1\Sigma_g^+$ (Noxon band) of O\textsubscript{2} is presented and compared to a  experimental spectrum from the literature. The spectroscopic model for each molecule, including \ai\ electric quadrupole moment functions $\Theta(r)$ is made available in the supplementary material  via \Duo\ input files. We also provide a list of calculated state energies and quantum numbers, as well cross-sections and line positions in the form of ExoMol line lists.\cite{20TeYuAl}

\section{Theoretical Background}
\label{sec:theoretical_background}

\subsection{Matrix Elements and Linestrengths}

\subsubsection{Rovibronic wavefunctions}

We  consider a calculation of electric quadrupole spectra for an arbitrary diatomic molecule between some generic rovibronic states. Our aim is to implement an E2 spectra module as part of the general diatomic code \Duo.\citep{Duo} The original \Duo\ program and its methodology is detailed extensively by \citet{Duo}. For the purpose of defining the matrix elements here, it suffices to simply introduce the definition of the quadrupole moment, the basis functions and the final eigenstates used by the \Duo\ program. \Duo\ uses the Hund's case (a) basis set in the following form: 
\begin{equation}
    \ket{\varphi_i} = 
    \ket{\xi \Lambda}\ket{S \Sigma}\ket{\xi v}\ket{J \Omega M}
\label{eq:duo_basis}
\end{equation}
where $J$ is the total angular momentum, $M$ is a projection of $J$ on the laboratory $Z$-axis in units of $\hbar$, $S$ is the total electronic spin angular momentum, $\Sigma$ is the projection of the spin of electrons on molecular $z$-axis, $\xi$ are indexes of the $\xi$-th electronic state, $\Lambda$ is the projection of the electronic angular momentum on molecular $z$-axis, $\Omega = \Lambda + \Sigma$ (projection of the total angular momentum on molecule $z$-axis) and $v$ is the vibrational quantum number.

The eigenfunctions corresponding to the final rovibronic eigenvalues are  expressed as linear combinations of the basis functions in Eq.~\eqref{eq:duo_basis}:
\begin{equation}
    \ket{\psi_{J M \tau}} = \sum_{\xi \Lambda S \Sigma v \Omega} C_{J\tau} (\xi \Lambda S \Sigma v \Omega) \ket{\xi,\Lambda} \ket{S \Sigma} \ket{\xi, v} \ket{J \Omega M}
	\label{eq:duo_eigenfunctions}
\end{equation}
where $C_{J \tau} (\xi \Lambda S \Sigma v \Omega) = C_{J \tau}(\varphi)$ are expansion coefficients obtained by solving a system of coupled rovibronic Schr\"odinger equations variationally, and $\tau$ is the symmetry of a rovibronic eigenstate. In case of a heteronuclear diatomic, $\tau$ is a parity $\tau = -$ (odd) or $+$ (even),\citep{93Kato.methods} which reflects how $\ket{\psi_{JM\tau}}$ transforms upon inversion or, equivalently, reflection through the molecule-fixed $xz$ plane. For a homonuclear molecule, the symmetry $\tau$ includes the parity with respect to the permutation of the nuclei and is traditionally represented by the combinations $+/-$ ($xz$-reflection) as well as  the $g/u$ parities   (molecular-fixed inversion), where $g$ and $u$ stand for `gerade' and `ungerade'. Generally the good quantum numbers are the total angular momentum $J$, the symmetry $\tau$ and the $g$ and $u$ parities (homonuclear molecules). It is also common to assign other quantum numbers according to the largest coefficient $C_{J\tau} (\varphi)$ in the basis set expansion.\citep{Duo}

\subsubsection{Electric Quadrupole Matrix Elements}

The Einstein $A$ coefficient for an E2 transition between a  lower state $i$ and an upper state $f$ is given in SI units, by:
%\citep{50Herzbe.quadrupole,82Hilbor.quadrupole,65Chiuxxx.quadrupole}
\begin{equation}
    A_{fi} = \frac{8\pi^5\nu_{fi}^5}{5\varepsilon_0 hc^5} \frac{1}{(2J_f + 1)} S_{fi}
    \label{eq:einstein_coefficient}
\end{equation}
where $\nu_{fi}$ [\si{\second^{-1}}] is the transition frequency, $\varepsilon_0$ [\si{\farad\metre^{-1}}] is the permittivity of free space, $h$ [\si{\joule\second}] is Planck's constant, $c$ [\si{\metre\per\second}] is the speed of light in a vaccum, and
\begin{equation}
S_{fi} = \left| M_{fi}^{(E2)} \right|^2 = \sum_{\alpha,\beta=x,y,z} \left| \mel{\psi_f}{Q_{\alpha\beta}}{\psi_i} \right|^2
\label{eq:general_quadrupole_linestrength}
\end{equation}
is the transition linestrength with unit $[\text{C}^2\cdot\text{m}^4]$ and the matrix elements are those of the quadrupole operator $Q_{\alpha\beta}$ ($\alpha,\beta=x, y$ or $z$) defined relative to the nuclear centre of mass by %Eq.~\eqref{eq:natural_quadrupole_moment}, and the traceless electric quadrupole (E2) moment operator has Cartesian components $Q_{ij}$ defined by
\begin{equation}
    Q_{\alpha\beta} = -\frac{3}{2} \sum_i e_i \left(r_{i,\alpha} r_{i,\beta} - \delta_{\alpha\beta} \frac{1}{3} r_{i}^2\right),
    \label{eq:natural_quadrupole_moment}
\end{equation}
where the sum runs over the nuclei and electrons with $e_i$ being the charge of the particle and $r_{i}$ its position vector in the molecule-fixed frame. We use the common convention of \citet{59Buckin.quadrupole}, used by many quantum chemistry programs such as \citet{MOLPRO}. Different sources employ definitions of the quadrupole moment with varying constant pre-factors, such as \citet{72Truhlar.CO}.

The \Duo\ rovibronic wavefunctions $\ket{\psi_{JM\tau}}$, and the transition linestrength in Eq.~\eqref{eq:general_quadrupole_linestrength} are defined in the laboratory-fixed frame. Meanwhile, the electric quadrupole moments in Eq.~\eqref{eq:natural_quadrupole_moment} are defined in the molecule-fixed frame. For the convenience of calculating matrix elements, the relationship between the molecule-fixed and laboratory-fixed components of tensor operators is best established using the algebra of irreducible tensors. Traceless symmetric quadrupole tensor of rank 2 can be expressed in terms of three irreducible tensors $Q^{(0)}$, $Q^{(1)}$ and $Q^{(2)}$ with ranks zero, one and two respectively. The components $Q^{(k)}_m$ with $-k \leq m \leq k$, are expressed in terms of the Cartesian $Q_{ij}$ via the following standard relations:\cite{06BuJexx.method, 02Longxx.quadrupole}
\begin{gather}
    Q_0^{(0)} = -\frac{1}{\sqrt{3}} \left( Q_{xx} + Q_{yy} + Q_{zz} \right) \label{QM00}\\[1em]
    Q_{0}^{(1)} = \frac{i}{\sqrt{2}} \left(Q_{xy} - Q_{yx}\right) \label{QM10}\\
    Q_{\pm 1}^{(1)} = -\frac{1}{2} \left[ Q_{xz} - Q_{zx} \pm i\left(Q_{zy} - Q_{yz} \right) \right] \label{QM11}\\[1em]
    Q_0^{(2)} = \frac{1}{\sqrt{6}} \left(2Q_{zz} - Q_{xx} - Q_{yy}\right) \label{QM20}\\
    Q_{\pm 1}^{(2)} = \frac{1}{2} \left[ \mp \left(Q_{xz} + Q_{zx}\right) - i \left(Q_{yz} + Q_{zy}\right) \right] \label{QM21}\\
    Q_{\pm 2}^{(2)} = \frac{1}{2} \left[ \left(Q_{xx} - Q_{yy}\right) \pm i \left(Q_{xy} + Q_{yx} \right) \right] \label{QM22}
\end{gather}
and transform  under rotation between the two frames as follows:\citep{93Kato.methods}
\begin{equation}
    Q_m^{(k)} = \sum_{m'} (-1)^{m-m'} Q_{m'}^{(k)} D_{-m,-m'}^{(k)},
    \label{eq:quadrupol_wigner_rotation}
\end{equation}
where $D_{-m,-m'}^{(k)}$ are the Wigner $D$-matrices. 
The traceless definition of the components $Q_{\alpha\beta}$ (Eq.~\eqref{eq:natural_quadrupole_moment}) and the property of being symmetric under interchange of the indices $\alpha,\beta$ implies that $Q^{(0)}_0 = Q^{(1)}_m = 0$, such that only the second rank components of the quadrupole moment are non-zero. This allows one to write the transition linestrength using the \Duo\ eigenfunctions (Eq.~\eqref{eq:duo_eigenfunctions}) as
\begin{equation}
    S_{fi} = g_{\rm ns} \sum_{M_i,M_f} \sum_{m=-2}^2 \left| \bra{\psi_{J_f M_f \tau_f}} Q_{m}^{(2)}  \ket{\psi_{J_i M_i \tau_i}}\right|^2,
    \label{eq:quadrupole_linestrength}
\end{equation}
where $g_\mathrm{ns}$ is a nuclear statistical weight that accounts for the degenerate nuclear spin components of the total nuclear-rovibronic wavefunction, see e.g. \citet{98BuJe.method}.

\citet{02Longxx.quadrupole} provides expressions that allows one to construct laboratory frame matrix element expressions for the electric polarisability tensor - also of rank two. Adapting the treatment, one can write the transition quadrupole moment matrix elements as:
\begin{equation}
\begin{split}
	S_{fi} = g_\mathrm{ns} (2J_i +1) (2J_f + 1)
	& \left| \sum_{\varphi_f} C_{J_i \tau_i}^* (\varphi_f) \sum_{\varphi_i} C_{J_f \tau_f} (\varphi_i) \sum_{m'} \delta_{S_f S_i} \delta_{\Sigma_f \Sigma_i} \times \vphantom{\begin{pmatrix} J \\ J \end{pmatrix}} \right. \\
			&\left. (-1)^{m'+\Omega_i} \mel{v_f}{\mel{\xi_f \Lambda_f}{Q^{(2)}_{m'}}{\xi_i \Lambda_i}}{v_i} 
			\begin{pmatrix}
				J_i & J_f & 2 \\
				-\Omega_i & \Omega_f & -m'
			\end{pmatrix}
			\right|^2,
	\label{eq:duo_quadrupole_linestrength}
\end{split}
\end{equation}
where Eq.~\eqref{eq:quadrupol_wigner_rotation} was used to transform from the laboratory frame to the molecular frame. Here, $m$ and $m'$ index components of the irreducible representation in the laboratory and molecular reference frames, respectively, and the following properties of the Wigner $D$-matrices, $D^{(k)}_{-m, -m'}$, have been used to express rotational matrix element in terms of the 3-$j$ symbols \citep{93Kato.methods}
\begin{align}
	\ket{JM\Omega} &= (-1)^{M-\Omega} \left( \frac{2J+1}{8\pi^2} \right)^\frac{1}{2} D^{(J)}_{-M, -\Omega}, \\
	\bra{JM\Omega} &= \left( \frac{2J+1}{8\pi^2} \right)^\frac{1}{2} D^{(J)}_{M,\Omega},
\end{align}
\begin{equation}
	\int D^C_{cc'} D^A_{aa'} D^B_{bb'} \sin \beta \dd{\beta} \dd{\alpha} \dd{\gamma}
	= 8\pi^2
	\begin{pmatrix}
		A & B & C \\
		a & b & c
	\end{pmatrix}
	\begin{pmatrix}
		A & B & C \\
		a' & b' & c'
	\end{pmatrix}
	\label{eq:3j_integral}
\end{equation}
with $\alpha$, $\beta$, and $\gamma$ the Euler angles. Additionally, the following standard property of the 3-$j$ symbols implies the 3-$j$ symbols containing $M_\textsc{i}$, $M_\textsc{f}$ and $m$, which arise as a result of Eq.~\eqref{eq:3j_integral}, can be summed over $M_\textrm{f}$, $M_\textrm{i}$ and $m$ and eliminated from Eq.~\eqref{eq:quadrupole_linestrength}
\begin{equation}
\sum_{m=-k}^k \sum_{M'=-J'}^{J'} \sum_{M''=-J''}^{J''}
    \begin{pmatrix}
		J'' & k & J' \\
		M'' & m & -M'
	\end{pmatrix}^2 = 1.
\end{equation}
If required, e.g for use with molecular dynamics programs such as \textsc{RichMol},\citep{RichMol2} Duo can explicitly calculate the laboratory frame components of the matrix elements. Note also that the 3-$j$ symbols are invariant under cyclic permutations of their columns and have the properties $|A - B| \leq C \leq |A + B|$, and $a + b + c = 0$. Together with the Kronecker deltas in Eq.~\eqref{eq:duo_quadrupole_linestrength} this implies following selection rules for E2 transitions:
\begin{equation}
    \Delta J = J_f - J_i = 0, \pm 1, \pm 2 
\end{equation}
and $\Delta S = \Delta \Sigma = 0$, such that
\begin{equation}
    \Delta \Lambda = \Lambda_f - \Lambda_i = -m = 0, \pm1, \pm2
    \label{eq:lambda_selection_rule}
\end{equation}
for all $\mel{\xi_f \Lambda_f}{Q^{(2)}_{m}}{\xi_i \Lambda_i}$ with $-2 \leq m \leq 2$ in Eq.~\eqref{eq:duo_quadrupole_linestrength}. These quantum number selection rules should be supplemented by the symmetry selection rules:
\begin{align}
    + \leftrightarrow +, & \quad  - \leftrightarrow - \\
    g \leftrightarrow g, & \quad  u \leftrightarrow u,
\end{align}
which arise as a result of the symmetric property of the quadrupole moment under coordinate inversion (Eq.~\eqref{eq:natural_quadrupole_moment}), and the requirement that the total matrix element is also symmetric under coordinate inversion, such that the integral over spatial coordinates is non-zero.

\subsubsection{Representation of \Ai{} Coupling Curves}

In this section we outline the procedure used by the \Duo\ program to transform coupling curves, specifically including the independent components of the quadrupole moment tensor, from the Cartesian representation commonly obtained from electronic structure calculations, to the tensorial, $\Lambda$-representation required by \Duo. The (transition) quadrupole moments in Eq.~\eqref{eq:duo_quadrupole_linestrength} are  $r$-dependent curves ($r$ is the vibrational coordinate) averaged over electronic coordinates:
\begin{equation}
    Q^{(2)}_{m}(r;\xi_f,\xi_i) = \mel{\xi_f \Lambda_f}{Q^{(2)}_{m}(r)}{\xi_i \Lambda_i}, 
    \label{eq:electronic_quadrupole_moment}
\end{equation}
where $\ket{\xi_i \Lambda_i}$ and $\ket{\xi_f \Lambda_f}$ are the corresponding  electronic wavefunctions. These curves are often obtained empirically by fitting analytical functions to experimental measurements of energies and linestrengths, or computed \ai\, using electronic structure programs such as those used in the present work (\textsc{MOLPRO} \cite{MOLPRO, MOLPRO2020} or the open-access software \textsc{CFOUR} \cite{CFOUR}). In electronic structure calculations the representations of the infinite symmetry groups for diatomic molecules $C_{\infty v}$ and $D_{\infty h}$ are commonly represented in terms of their Abelian subgroups $C_{2v}$ and $D_{2h}$ in order to facilitate the computation of physically-realised energy levels. For the practical purpose of transforming the electronic properties from the output of quantum chemistry programs to the representation required for the \Duo\ input, we also employ the representation of $C_{\infty v}$ and $D_{\infty h}$ in terms of the Abelian subgroups in the following derivation.

The irreducible Abelian representation of a matrix element of a given operator coupling electronic states $i$ and $f$, each with irreducible Abelian representations $G_i$ and $G_f$ respectively, must be contained within the Abelian group given by the direct product $G_i \times G_f$.\cite{06BuJexx.method} Moreover, it can be shown that there exists only one independent Cartesian quadrupole component that couples a given pair of irreducible representations within an Abelian symmetry group. Tables~\ref{tab:C2v_product_table} and \ref{tab:D2h_product_table} establish correlations between the products of  Cartesian vectors $r_x, r_y, r_z$, corresponding to components of the quadrupole moment operator in Eq.~\eqref{eq:natural_quadrupole_moment}, and the products of different irreducible representations, for $C_{2v}$ and $D_{2h}$ point groups, respectively. 

\begin{table}
    \caption{Irreducible representations for homonuclear symmetry groups, and corresponding components of electronic states. Appendix \ref{ap:molpro_enumeration} gives the same table with addition of the \textsc{Molpro} enumerations.}
    \label{tab:D2h_irreps}
    \begin{tabular}{c|r}
        Symmetry & Components \\
        \hline
        $A_g$    & $\Sigma_g^+$, $(\Delta_g)_{xx}$ \\
        $B_{1g}$ & $\Sigma_g^-$, $(\Delta_g)_{xy}$ \\
        $B_{2g}$ & $(\Pi_g)_x$ \\
        $B_{3g}$ & $(\Pi_g)_y$ \\
        $A_u$    & $\Sigma_u^-$, $(\Delta_u)_{xy}$ \\
        $B_{1u}$ & $\Sigma_u^+$, $(\Delta_u)_{xx}$ \\
        $B_{2u}$ & $(\Pi_u)_y$ \\
        $B_{3u}$ & $(\Pi_u)_x$
    \end{tabular}
\end{table}

\begin{table}
    \caption{Irreducible representations for heteronuclear symmetry groups, and corresponding components of electronic states. Appendix \ref{ap:molpro_enumeration} gives the same table with addition of the \textsc{Molpro} enumerations.}
    \label{tab:C2v_irreps}
    \begin{tabular}{c|r}
        Symmetry & Components \\
        \hline
        $A_1$ & $\Sigma^+$, $\Delta_{xx}$ \\
        $A_2$ & $\Sigma^-$, $\Delta_{xy}$ \\
        $B_1$ & $\Pi_x$ \\
        $B_2$ & $\Pi_y$ \\
    \end{tabular}
\end{table}

Eq. \eqref{eq:duo_quadrupole_linestrength} uses a tensorial representation of all electronic properties, including the electric quadrupole moments $Q^{(2)}_{m}(r)$. It is also convenient to represent the electronic basis functions  $\ket{\xi \Lambda}$ corresponding to doubly degenerate $\Lambda > 0$ states in the tensorial representation  with $\pm|\Lambda|$ as a good quantum number. These are related to the Cartesian components $\ket{\alpha}$ and $\ket{\beta}$ by \citep{Duo}
\begin{equation}
%	\ket{\pm \Lambda} = C_1^{(\pm|\Lambda|)} \ket{\alpha} + C_2^{(\pm|\Lambda|)} \ket{\beta}
	\ket{\xi, \pm |\Lambda|} = \frac{1}{\sqrt{2}} \big[ \ket{\alpha} \pm i \ket{\beta} \big],
\label{eq:lambda_from_cartesian}
\end{equation}
where $\ket{\alpha}$ and $\ket{\beta}$ are, for example, $\ket{\Pi_x}$ and $\ket{\Pi_y}$  ($|\Lambda|=1$), $\ket{\Delta_{xx}}$ and $\ket{\Delta_{xy}}$ ($|\Lambda|=2$) etc. as typically produced by electronic structure methods.

We now consider the unitary transformation from the Cartesian (`electronic structure') representation of the matrix elements $\mel{\xi'' \gamma''}{{Q}_{ij}(r)}{\xi' \gamma'}$ ($\gamma \in [\alpha, \beta]$) to their tensorial (`\Duo') representation $\mel{\xi'' \Lambda''}{Q^{(k)}_{m}(r)}{\xi' \Lambda'}$  in Eq.~\eqref{eq:electronic_quadrupole_moment}.

To construct this transformation and also to keep track of the relative phases of  `electronic structure' wavefunctions \Duo\ makes the use of  the Cartesian matrix elements  of the electronic angular momentum operator $\hat{L}_z$. We choose the Cartesian components $\ket{\alpha}, \ket{\beta}$ such that for wavefunctions with $|\Lambda|>0$ the $\hat{L}_z$ matrix is given (up to an arbitrary phase factor) by: 
\begin{equation}
	\mathbf{L}_z = 
	\begin{pmatrix}
		\mel{\ \alpha}{\hat{L}_z}{\ \alpha} & \mel{\ \alpha}{\hat{L}_z}{\ \beta} \\
		\mel{\ \beta}{\hat{L}_z}{\ \alpha} & \mel{\ \beta}{\hat{L}_z}{\ \beta}
	\end{pmatrix} =
	\begin{pmatrix}
		0 & -i \hbar |\Lambda| \\
		i \hbar|\Lambda| & 0
	\end{pmatrix},
    \label{eq:Lz_cartesian_element}
\end{equation}
where $\mathbf{L}_z$ is the Cartesian matrix representation of $\hat{L}_z$ with the elements $\mel{\xi \gamma''}{{L}_z}{\xi \gamma'}$ and the index $\xi$ is dropped for simplicity. The wavefunctions  $\ket{\xi, \pm |\Lambda|}$ in Eq.~\eqref{eq:lambda_from_cartesian} can be formed as eigenfunctions of the operator $\hat{L}_z$ in the Cartesian representation by diagonalizing the 2$\times$2 matrix  matrix $\mathbf{L}_z$ with the eigenvalues $\hbar |\Lambda|$ and $-\hbar |\Lambda|$.\citep{Duo} The corresponding unitary matrix that diagonalizes $\mathbf{L}_z$, 
\begin{equation}
U= 	\begin{pmatrix}
		\frac{1}{\sqrt{2}} & \frac{i}{\sqrt{2}} \\
		\frac{1}{\sqrt{2}} & \frac{-i}{\sqrt{2}}
	\end{pmatrix}
\end{equation}
provides the transformation between the Cartesian and tensorial representations for any electronic structure property, including the electric quadrupole  
\begin{equation}
 Q^\text{tens.} = U^{-1}Q^\text{Cart.}U. 
\end{equation}

\begin{table}
	\caption{Product table for the quadratic functions of Cartesian components and the isotropic function $s$, that transform as the product of different irreducible representations for the $C_{2v}$ point group.}
	\label{tab:C2v_product_table}
	\begin{tabular}{|c|cccc|}
		\hline
		      & $A_1$ & $A_2$ & $B_1$ & $B_2$ \\ 
		\hline
		$A_1$ & $s$   & $xy$  & $xz$  & $yz$  \\
		$A_2$ & $xy$  & $s$   & $yz$  & $xz$  \\
		$B_1$ & $xz$  & $yz$  & $s$   & $xy$  \\
		$B_2$ & $yz$  & $xz$  & $xy$  & $s$   \\
		\hline
	\end{tabular}
\end{table}

\begin{table}
	\caption{Product table for the quadratic functions of Cartesian components and the isotropic function $s$, that transform as the product of different irreducible representations for the $D_{2h}$ point group.}
	\label{tab:D2h_product_table}
	\begin{tabular}{|c|cccccccc|}
		\hline
		         & $A_g$ & $B_{1g}$ & $B_{2g}$ & $B_{3g}$ & $A_u$    & $B_{1u}$ & $B_{2u}$ & $B_{3u}$ \\
		\hline
		$A_g$    & $s$   & $xy$     & $xz$     & $yz$     &          &          &          &          \\
		$B_{1g}$ & $xy$  & $s$      & $yz$     & $xz$     &          &          &          &          \\
		$B_{2g}$ & $xz$  & $yz$     & $s$      & $xy$     &          &          &          &          \\
		$B_{3g}$ & $yz$  & $xz$     & $xy$     & $s$      &          &          &          &          \\
		$A_u$    &       &          &          &          & $s$      & $xy$     & $xz$     & $yz$     \\
		$B_{1u}$ &       &          &          &          & $xy$     & $s$      & $yz$     & $xz$     \\
		$B_{2u}$ &       &          &          &          & $xz$     & $yz$     & $s$      & $xy$     \\
		$B_{3u}$ &       &          &          &          & $yz$     & $xz$     & $xy$     & $s$      \\ 
		\hline
	\end{tabular}
\end{table}

Together with the 3-$j$ symbol in Eq.~\eqref{eq:duo_quadrupole_linestrength}, which implies that each component $Q^{(2)}_{m'}$ couples electronic states with $\Lambda_f - \Lambda_i = m'$, this allows for the following additional relations to be made
\begin{align}
    \begin{split}
        \mel{\pm|\Lambda|}{Q^{(2)}_{0}}{\pm|\Lambda|} &= \frac{3}{2\sqrt{6}}\Big[ \mel{\alpha}{Q_{zz}}{\alpha} + \mel{\beta}{Q_{zz}}{\beta}\Big] \\
        &= \frac{3}{\sqrt{6}} \mel{\alpha}{Q_{zz}}{\alpha},
    \end{split} \label{QMTC1}\\
    \begin{split}
        \mel{\Sigma^+}{Q^{(2)}_{\pm 1}}{\mp \Pi} &= \mp \frac{1}{\sqrt{2}}\Big[\mel{\Sigma^+}{Q_{xz}}{\Pi_x} +\mel{\Sigma^+}{Q_{yz}}{\Pi_y}\Big] \\
    	&= \mp \sqrt{2}\mel{\Sigma^+}{Q_{xz}}{\Pi_x},
    \end{split} \label{QMTC2}\\
    \begin{split}
        \mel{\Sigma^-}{Q^{(2)}_{\pm 1}}{\mp \Pi} &= - \frac{i}{\sqrt{2}}\Big[\mel{\Sigma^-}{Q_{xz}}{\Pi_y} + \mel{\Sigma^-}{Q_{yz}}{\Pi_x}\Big] \\
        &= -i\sqrt{2}\mel{\Sigma^-}{Q_{xz}}{\Pi_y},
    \end{split} \label{QMTC3}\\
    \begin{split}
        \mel{\Sigma^+}{Q^{(2)}_{\pm 2}}{\mp \Delta} &= + \frac{1}{\sqrt{2}}\Big[\mel{\Sigma^+}{Q_{xx}}{\Delta_{xx}} +\mel{\Sigma^+}{Q_{xy}}{\Delta_{xy}}\Big] \\
        &= +\sqrt{2}\mel{\Sigma^+}{Q_{xx}}{\Delta_{xx}},
    \end{split} \label{QMTC4}\\
    \begin{split}
        \mel{\Sigma^-}{Q^{(2)}_{\pm 2}}{\mp \Delta} &= \pm \frac{i}{\sqrt{2}}\Big[\mel{\Sigma^-}{Q_{xx}}{\Delta_{xy}} + \mel{\Sigma^-}{Q_{xy}}{\Delta_{xx}}\Big] \\
        &= \pm i\sqrt{2}\mel{\Sigma^-}{Q_{xx}}{\Delta_{xy}},
    \end{split} \label{QMTC5}\\
	\begin{split}
	    \mel{\mp\Pi}{Q^{(2)}_{\pm 1}}{\mp\Delta} &= \mp \frac{1}{2} \Big[ \mel{\Pi_x}{Q_{xz}}{\Delta_{xx}} + \mel{\Pi_x}{Q_{yz}}{\Delta_{xy}} - \mel{\Pi_y}{Q_{yz}}{\Delta_{xx}} + \mel{\Pi_y}{Q_{xz}}{\Delta_{xy}} \Big] \\
	    &= \mp 2 \mel{\Pi_x}{Q_{xz}}{\Delta_{xx}},
	\end{split}  \label{QMTC6}
\end{align}
The initial expressions in Eqs.~(\ref{QMTC1}-\ref{QMTC6}) are obtained from Eqs.~(\ref{QM20}-\ref{QM22}) by substituting the symmetric components $Q_{zx} = Q_{xz}$, $Q_{zx} = Q_{yz}$, $Q_{xy} = Q_{yx}$, and $Q_{xx} = -Q_{yy}$. The second line in each expression is obtained by setting matrix elements that do not satisfy the selection rule in Eq.~\eqref{eq:lambda_selection_rule} (e.g $\mel{\Sigma^+}{Q^{(2)}_{\mp 1}}{\mp \Pi}$, $\mel{\Sigma^+}{Q^{(2)}_{\mp 2}}{\mp \Delta}$, etc.) equal to zero and rearranging to obtain relations between different Cartesian components of the matrix elements. In the case of $D_{2h}$ symmetry, the corresponding equations  (\ref{QMTC1}-\ref{QMTC6}) are identical except for the addition of the relevant $g/u$ parity label.

\section{Demonstrations}
\label{sec:demonstrations}

In this section we provide a demonstration of the \Duo\ electric quadrupole program for the simple $^1\Sigma$ systems of H\textsubscript{2}, CO and HF.  In particular, we choose H\textsubscript{2} as the initial proof of the program due to the highly accurate spectroscopic data available for this molecule, which we aim to reproduce. The demonstrations for CO and HF exemplify heteronuclear systems with large molecular quadrupole moments in which the consideration of E2 transitions is necessary to obtain accurate cross-sections. An application to an more complex system involving interstate transitions with a non-$\Sigma$ electronic state is illustrated by way of simulating the Noxon electronic (E2) band $a^1\Delta_g$ -- $b^1\Sigma_g^+$  of the O\textsubscript{2} molecule. The spectroscopic models detailed in this section are provided as supplementaries in the form of \Duo\ input files, the \Duo\ program itself is open-source and can be obtained from the ExoMol public repository at \href{https://github.com/Exomol/Duo)}{github.com/Exomol}.

\subsection{Molecular Hydrogen}
\label{sec:molecular_hydrogen}

Molecular hydrogen is the simplest diatomic molecule, containing just two electrons and two protons. It is the most abundant molecule in the universe and plays an important role in star formation, \citep{83JoGoGo.coolstars,02FrMcWi.H2,10KrBrCi.H2,10NiGiNe.H2} interstellar physics, \citep{10IsCeVi.is, 99HoTixx.is, 00Dalgarno.H2} (exo)planetary atmospheres, \citep{10BoLiDu.exo,12HuSiVi.exo,14StBeSe.exo,17ArRiWa.exo} and investigations of fundamental physics.\citep{16UbKoEi.H2,16UbBaSa.H2}

Owing to its molecular symmetry, the homonuclear H\textsubscript{2} molecule has no permanent electric dipole moment, and thus rovibrational transitions are forbidden in the electric dipole approximation. The availability of highly accurate electronic potential energy curves (PECs) and electric quadrupole moment curves (QMCs) makes H\textsubscript{2} an ideal candidate for validating the implementation of E2 transitions in \Duo. The simplicity of the H\textsubscript{2} molecule makes it an extremely tractable quantum mechanical problem - indeed, it was the model molecule for many early calculations of molecular dynamics, on the world's first mass-produced computers.\citep{63KoWoxx.H2,65KoWoxx.H2,68LeBexx.H2} Even for these early calculations linestrength accuracies within a few percent were attainable.\citep{69DaAlBr.H2,72Truhla.H2,77TuKiDa.H2} As a result, there is a wealth of accurate spectroscopic data available with which the \Duo\ implementation can be validated. Most recently, \citet{19RoAbCz.H2} calculated a highly accurate (order \SI{e-6}{\per\cm}) infrared spectrum for the H\textsubscript{2} molecule including several higher order correction terms.\citep{H2SPECTRE}

The calculations of \citet{19RoAbCz.H2} are based on an extensive series of earlier works by \citet{10Pachucki.H2,09PaKoxx.H2,14PaKoxx.H2,15PaKoxx.H2}, in which the H\textsubscript{2} Born-Oppernheimer PEC was obtained with \SI{e-15}{} relative numerical precision using ~22,000 exponential basis functions and explicit electron correlation calculations.\citep{12Pachuc.H2, 10Pachucki.H2} They also compute non-adiabatic,\citep{09PaKoxx.H2, 15PaKoxx.H2} adiabatic\citep{14PaKoxx.H2} and high-order relativistic \citep{17PuKoPu.H2} corrections to the Born-Oppenheimer potential energy. The quadrupole moment function employed in their calculations is obtained using the Born-Oppenheimer wavefunction, and is in agreement with the values reported by \citet{98WoSiDa.H2}, who employ a 494-term correlated basis representation of the wavefunction to obtain the quadrupole moment function with an estimated accuracy on the order of 0.001\%.

\begin{figure}
	\centering
	\includegraphics[width=.49\linewidth]{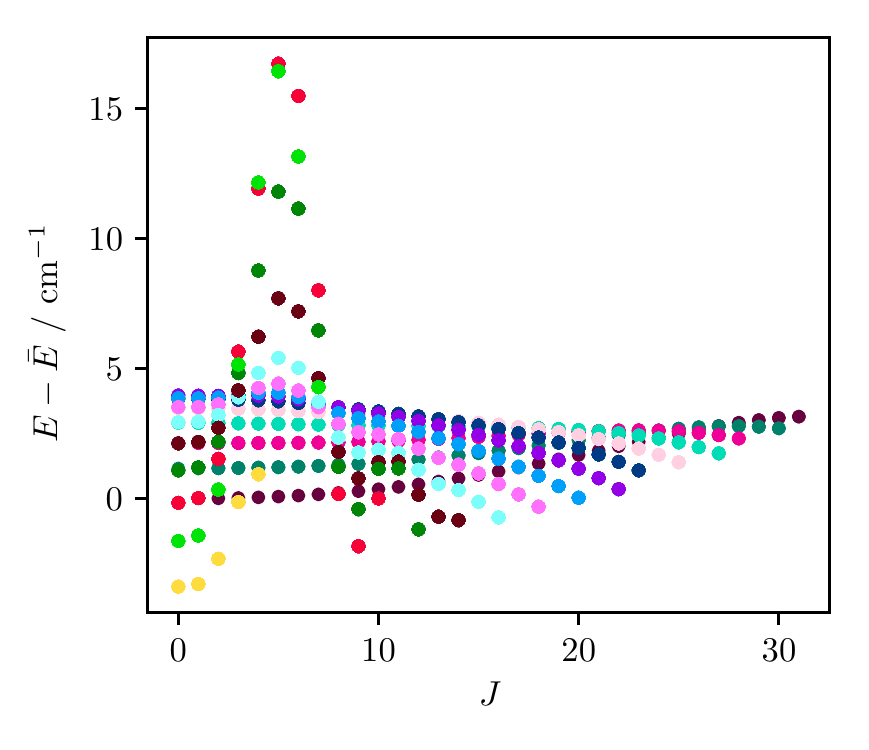}
	\includegraphics[width=.49\linewidth]{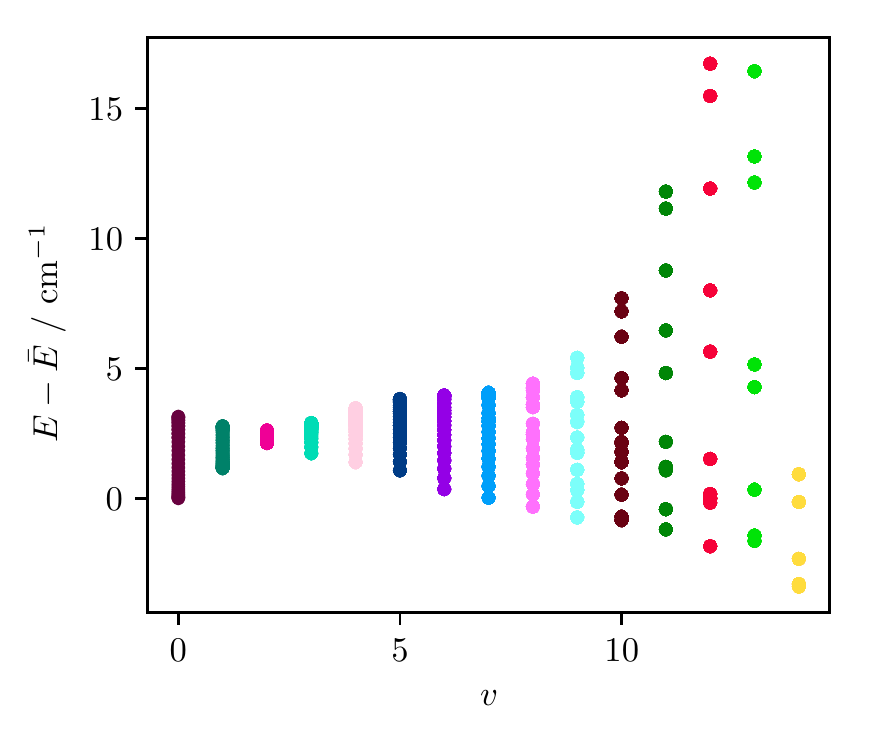}
	\includegraphics[width=.49\linewidth]{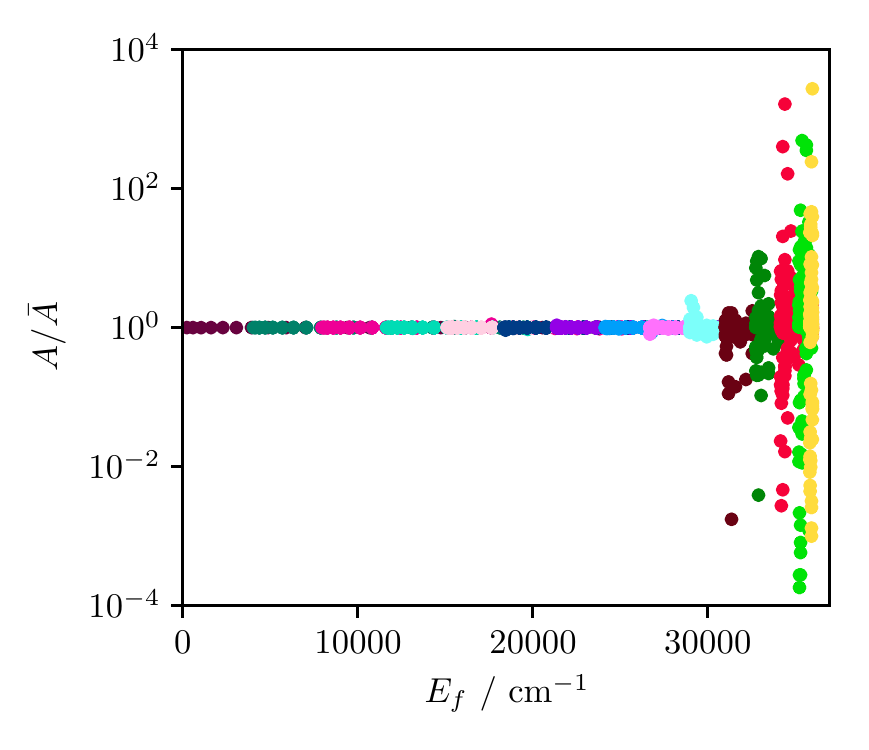}
	\includegraphics[width=.49\linewidth]{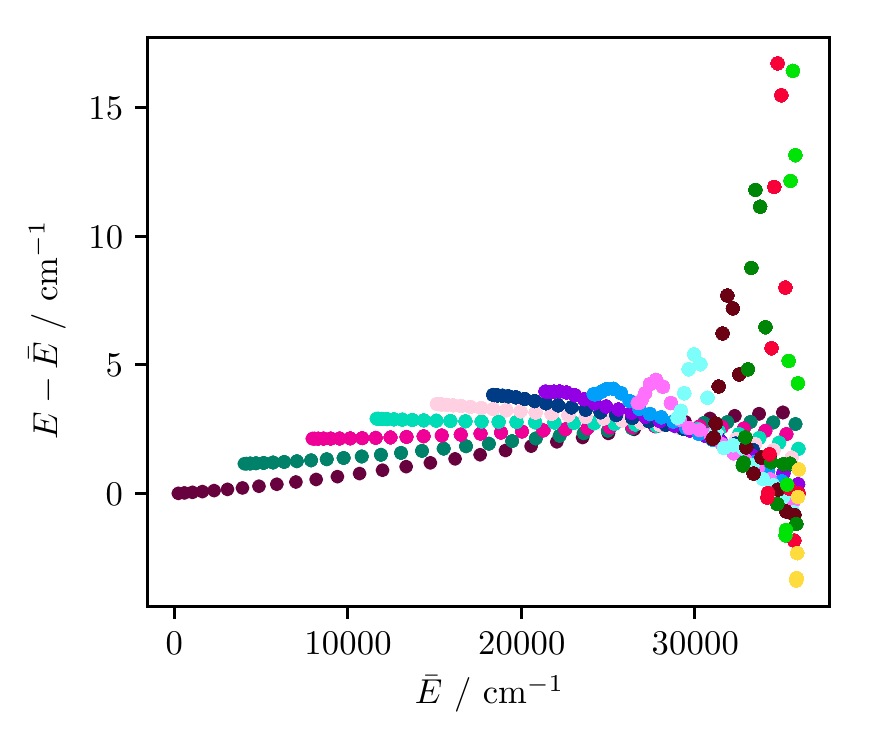}
	\caption{Agreement between the Einstein-$A$ coefficients (bottom-left) and state energies of H\textsubscript{2} calculated by \Duo\ ($A_\mathrm{if}, E$), and by \citet{19RoAbCz.H2} ($\bar{A}_\mathrm{if}, \bar{E}$). The energy differences $E-\bar{E}$ in the upper panels are plotted as functions of the level of rotational $J$ and vibrational $v$ excitations. The energy and $A$-coefficient differences in the lower panels are plotted as functions of (upper) state energy.  The colors in each plot correspond to the (upper) vibrational quantum number of the state.}
	\label{fig:H2_19RoAbCz_comparison} 
\end{figure}

For the validation of the \Duo\ implementation, their original Born-Oppenheimer potential is retrieved using the \verb|V(DR)| function made available via the H2SPECTRE program.\citep{H2SPECTRE} The contribution of the adiabatic and non-adiabatic corrections computed by \citet{19RoAbCz.H2} are in the range \SIrange{5}{20}{\per\cm} and \SIrange{0.4}{4.0}{\per\cm}, respectively, increasing the total state energy. The higher order relativistic corrections are on the order of \SI{0.01}{\per\cm} or less. Since the Born-Oppenheimer PEC provided does not include adiabatic or non-adiabatic corrections, significant deviation is expected between the calculated state energies for high $v$ and $J$ states. Typically these deviations could be corrected in \Duo\ via an empirical fit to experimentally accurate state energies. Such refinement is not performed in this work, as the aim here is to illustrate the implementation of E2 transition strengths, rather than provide an accurate or improved line list for H\textsubscript{2}. The quadrupole moment function of \citet{98WoSiDa.H2} is also employed, given as a grid of 253 E2 moment values between \SIrange{0.2}{20.0}{\bohr}, which \Duo\ interpolates using quintic splines.

The vibrational grid is defined by 501 equally spaced points in the range \SIrange{0.38}{18.90}{\bohr}. After solving the vibrational Schr\"{o}dinger using the sinc-DVR method, the first 30 vibrational states are selected to form the contracted vibrational basis and the rovibrational Schr\"{o}dinger equation is solved for rotational states with total angular momentum quantum numbers $0 \leq J \leq 200$ at \SI{296}{\kelvin}.

Fig. \ref{fig:H2_19RoAbCz_comparison} illustrates the results of a line-by-line comparison of the \Duo\ results to the accurate line list of \citet{19RoAbCz.H2} (including all corrections). As expected, significant differences between the energies calculated by \Duo\ ($E$) and the accurate energies provided by H2SPECTRE ($\bar{E}$) for high $v$, $J$ states are observed. We also expect to see significant deviation in the Einstein coefficients obtained for transitions involving these states, due to the factor of $\nu_{fi}^5$ present in Eq. \ref{eq:einstein_coefficient} coupled with vanishingly small Einstein coefficients for transitions to states with large $v$ quantum number. Thus states with $v \ge 10$ are excluded from the analysis.

For the 3,027 remaining transitions between the remaining vibrational levels, 99.0\% of Einstein coefficients ($A_{fi}$) lie within 1\% of the values calculated by \citet{19RoAbCz.H2} ($\bar{A}_{fi}$). The 99-th percentile is $|1 - A_{fi}/\bar{A}_{fi}| = 0.0672$. Note that all Einstein coefficients with errors greater than 5\% correspond to weak transitions with absorption intensities $I_{fi} <$ \SI{1e-35}{\cm\per\molecule}. For example, the largest discrepancy $A_{fi}/\bar{A}_{fi}$ = \SI{2.45} corresponds to the $v=9\leftarrow 0$ transition with $A_{fi}$ = \SI{5.27e-15}{\per\second} and $I_{fi}$ = \SI{5.45e-36}{\cm\per\molecule}.

\begin{figure*}
	\centering
	\includegraphics[width=.49\linewidth]{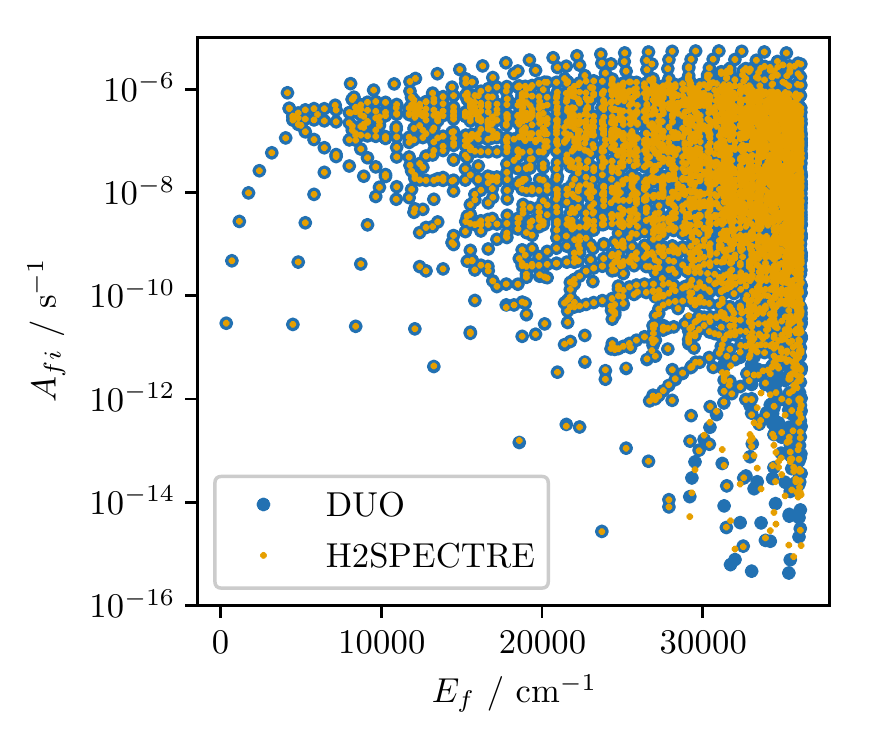}
	\includegraphics[width=.49\linewidth]{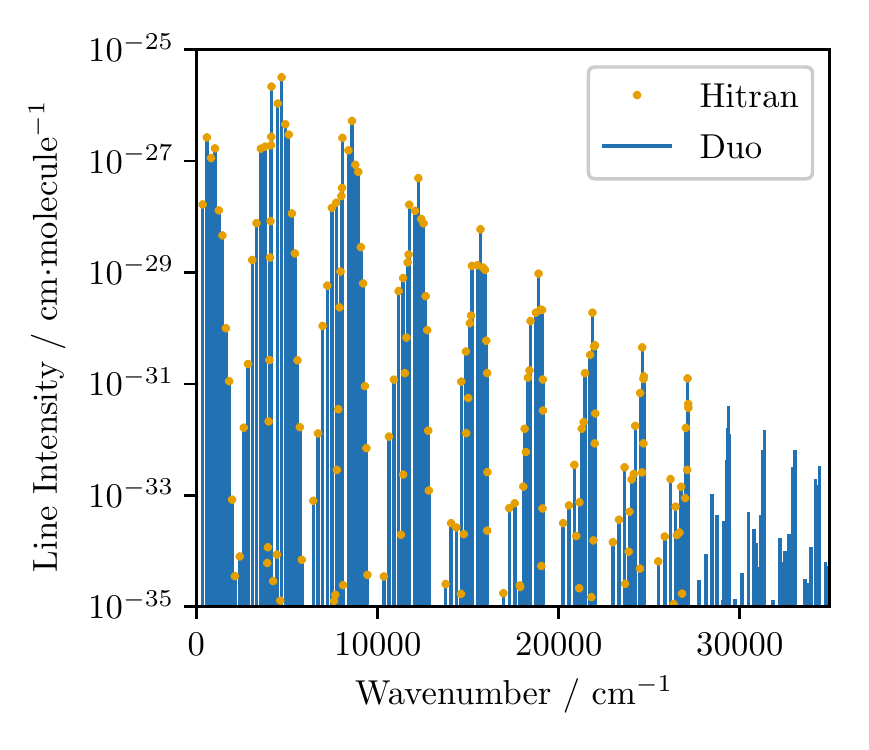}
	\caption{
		Comparison of the \Duo\ calculated Einstein-$A$ coefficients with the target values predicted by \citet{19RoAbCz.H2} (left), and of the \Duo\ calculated absorption intensities ($T= 296$~K) with the intensities listed in the HITRAN database\citep{HITRAN2016, 11KoPiLa.H2, 98WoSiDa.H2} (right).}
	\label{fig:H2XQM-19RoAbCz-98WoSiDa-Hitran}
\end{figure*}

Table~\ref{tab:H2_Bragg_compare} compares the results of the calculation to the experimentally measured intensities and line position of \citet{82BrBrSm.H2} ($T= 296$~K), and in Table~\ref{tab:H2_Campargue_compare} to more recent measurements of \citet{12CaKaPa.H2}, as well as their theoretical predictions based on the effective quadrupole moment method. The Duo calculated intensities reproduce closely the accurate experimental measurements of \citet{12CaKaPa.H2}, and match their theoretical predicted values to within 0.1\%. Agreement with the older measurements of \citet{82BrBrSm.H2} are less consistent but generally agree, particularly for the $Q$-branch transitions of the first overtone band. In both cases the line positions differ considerably, but by a roughly constant value across each vibrational band. This is due to the fact that no \Duo\ refinement procedure is performed and no adiabatic or non-adiabatic corrections are included in the calculations. Also illustrated, in Fig. \ref{fig:H2XQM-19RoAbCz-98WoSiDa-Hitran} are (left) direct comparisons of the Einstein coefficients obtained via \Duo\ to those of \citet{19RoAbCz.H2}, and (right) the absorption intensities via the \Exocross\ program, as compared to transitions listed in the HITRAN\citep{HITRAN2016} database. Here and in the following we use the HITRAN intensity units cm$/$molecule.

\begin{table}
    \centering
    \caption{Comparison of various H\textsubscript{2} $v'\gets 0$ transitions (positions and intensities), measured experimentally by \citet{82BrBrSm.H2}, to the values predicted by \Duo\ calculations at $T$ = 296~K. The line positions are in \si{\per\cm}.}
    \begin{tabular}{cccc}
        \hline
        $v'$ & Branch & $\tnu_\text{obs.} - \tnu_\text{calc.}^\text{\Duo}$ & $I_\text{obs.}/I_\text{calc.}^\text{\Duo}$ \\[2pt] 
        \hline\hline
        1 & Q(3) & -1.158 & 1.080 \\[-.9em]
        1 & Q(2) & -1.165 & 1.027 \\[-.9em]
        1 & Q(1) & -1.171 & 1.040 \\[-.9em]
        1 & S(0) & -1.181 & 1.158 \\[-.9em]
        1 & S(1) & -1.185 & 1.648 \\[-.9em]
        1 & S(2) & -1.185 & 1.594 \\[-.9em]
        1 & S(3) & -1.187 & 1.013 \\[-.9em]
        2 & O(3) & -2.121 & 0.852 \\[-.9em]
        2 & O(2) & -2.138 & 0.915 \\[-.9em]
        2 & Q(3) & -2.121 & 0.949 \\[-.9em]
        2 & Q(2) & -2.136 & 0.973 \\[-.9em]
        2 & Q(1) & -2.147 & 1.624 \\[-.9em]
        2 & S(0) & -2.152 & 0.984 \\[-.9em]
        2 & S(1) & -2.147 & 0.988 \\[-.9em]
        3 & S(0) & -2.923 & 0.816 \\[-.9em]
        3 & S(1) & -2.912 & 0.911 \\[-.9em]
        3 & S(2) & -2.887 & 1.017 \\[-.9em]
        3 & S(3) & -2.858 & 0.878 \\[-.9em]
        4 & S(0) & -3.480 & 0.606 \\[-.9em]
        4 & S(1) & -3.469 & 0.874 \\[-.9em]
        4 & S(2) & -3.432 & 0.727 \\[-.9em]
        4 & S(3) & -3.382 & 0.831 \\
        \hline
    \end{tabular}
    \label{tab:H2_Bragg_compare}
\end{table}

\begin{table}
    \centering
    \caption{Comparison of various H\textsubscript{2} $v'=2\leftarrow0$ overtone lines, measured experimentally and computed via an effective quadrupole moment by \citet{12CaKaPa.H2} ($\tnu_\text{calc.}$), and the values predicted by \Duo\ calculations ($\tnu_\text{calc.}^\text{\Duo}$) for $T=296$~K. The line positions are in \si{\per\cm}.}
    \begin{tabular}{ccccc}
        \hline
        Branch & $\tnu_\text{obs.} - \tnu_\text{calc.}$ & $I_\text{obs.}/I_\text{calc.}$ & $\tnu_\text{obs.} - \tnu_\text{calc.}^\text{\Duo}$ & $I_\text{obs.}/I_\text{calc.}^\text{\Duo}$ \\[2pt] 
        \hline\hline
        O(5) & -0.0019 & 0.924 & -2.061 & 0.924 \\[-.9em]
        O(4) & -0.0040 & 0.931 & -2.093 & 0.930 \\[-.9em]
        O(3) & -0.0033 & 1.008 & -2.115 & 1.007 \\[-.9em]
        O(2) & -0.0031 & 1.001 & -2.132 & 1.000 \\[-.9em]
        O(5) & -0.0030 & 1.020 & -2.067 & 1.020 \\
        \hline
    \end{tabular}
    \label{tab:H2_Campargue_compare}
\end{table}

\subsection{Carbon Monoxide}
\label{sec:carbon_monoxide}

Carbon monoxide is a heteronuclear diatomic molecule, and thus electric dipole transitions are allowed within its ground $X$~$^1\Sigma^+$ state. However, it also possesses a strong electric quadrupole moment,\citep{11ChCoxx.CO} and as a result, the electric dipole infrared spectrum is accompanied by weaker electric quadrupole lines. We show that many of the E2 spectral lines at room temperature lie higher in intensity than the minimum spectroscopic cutoff of $10^{-30}$ cm/molecule at the HITRAN reference temperature of $T= 296$~K, typically applied to E1 spectra. As a result, their inclusion or emission in spectroscopic databases has significant implications for applications where accurate cross-sections are required.

Numerous experimental and \ai\ studies have been performed of the electric dipole moment spectra for the CO molecule, including recent accurate calculations by \citet{15LiGoRo.CO}. Li \etal\ seek to resolve a long-standing uncertainty in the line intensities of CO E1 spectra, namely significant differences observed between the intensities predicted by the calculations of \citet{94Goorvich.CO} and those of \citet{96HuRoxx.CO}. The former uses Chackerian's \citep{84ChFaGu.CO} semi-empirical dipole moment function, obtained from a nonlinear least-squared fit to vibrational states up to $v=38$. The latter uses a purely \textit{ab initio} electric dipole moment curve (DMC), computed by Langhoff and Bauschlicher via ACPF calculations on a 5Z basis set.\citep{95LaBaxx.CO} Li \etal\ perform new CRDS measurements in order to produce an accurate DMC via a direct-fit. At long bond lengths, where experimental data is not attainable, they reproduce the calculations of \citet{95LaBaxx.CO} but with a finer grid, and determine that the interpolation used on the original grid was insufficient to capture the full shape of the DMC. Their PEC of choice is the analytical MLR3 function obtained by \citet{04CoHaxx.CO} via a direct fit to 21559 spectroscopic lines.\citep{04CoHaxx.CO}

Studies of the quadrupole moment of CO are somewhat sparser. Although several experimental measurements exist for the equilibrium molecular quadrupole moment, only a single study presents a QMC across a range of geometries. The early work by \citet{72Truhlar.CO} presents simple Hartree-Fock calculations of the quadrupole moment at just 6 internuclear geometries. The accuracy of the vibrational matrix elements calculated is low, particularly for weaker transitions corresponding to higher vibrational quantum numbers. In particular the methodology struggles to accurately describe the quadrupole moment at intermediate and long internuclear distances, which are necessary for calculating the vibrational overtones. \citet{03CoHaJo.CO} compares the results of CCSD and CC3 calculations on the CO molecule with a variety of basis sets. The results show that the CCSD level of theory is insufficient to correctly describe the electric properties of the CO molecule, and that consideration of triple excitations is vital. They also study the convergence of such calculations with increasing basis set size, and find the results converge quickly for bases larger than DZ.

\begin{figure}
	\centering
	\includegraphics{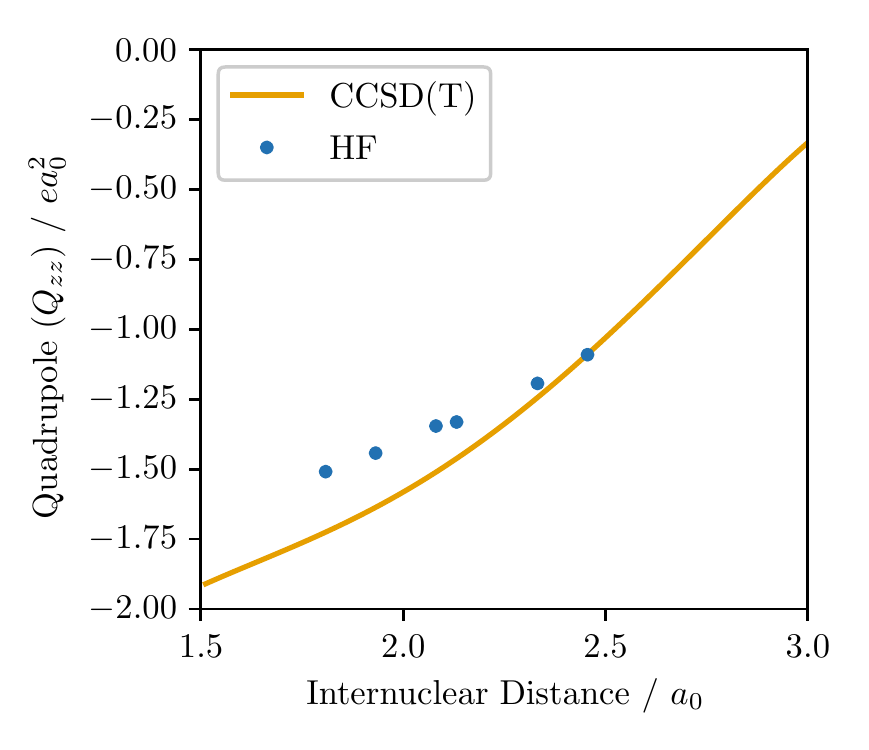}
	\caption{Electric quadrupole moments  in \si{\au} ($e a_0^2$)  for CO obtained in this work via CCSD(T) calculations compared to Hartree-Fock calculations by \citet{72Truhlar.CO}.}
	\label{fig:CO_EQC}
\end{figure}

In the present work, following the success of \citet{03CoHaJo.CO}, the CCSD(T) method is employed with an aug-cc-pwCVQZ basis as implemented in the CFOUR program \cite{CFOUR} to calculate the strength of the non-zero quadrupole component $Q_{zz}$ for 100 nuclear geometries in the range \SIrange{1.50}{3.78}{\bohr}. Divergent behaviour at large internuclear separations is attributed to CCSD(T)'s inability to account for multireference effects. The curve is therefore truncated at \SI{3.0}{\bohr}. The QMC obtained from these calculations is shown in Fig. \ref{fig:CO_EQC}. 

The value of the electric quadrupole moment curve at equilibrium separation $Q_{zz}$ = \SI{-1.45}{\au} (\si{\au} =  $e a_0^2$) agrees reasonably well with the Hartree-Fock calculations of \citet{72Truhlar.CO}, which obtain $Q_{zz}$ = \SI{-1.33}{\au}. Note that \citet{72Truhlar.CO} chooses a definition of the quadrupole moment which is a factor of two larger than the definition employed by \textsc{MOLPRO} and \Duo, the value quoted here is adjusted accordingly. Importantly, we obtain very good agreement with experimental values of the ZPE-averaged quadrupole moment from the literature. From the CCSD(T) quadrupole moment shown in Fig. \ref{fig:CO_EQC}, \Duo\ calculates $\mel{v=0}{Q_{zz}}{v=0}$ = \SI{-1.4522}{\au} which agrees closely with the accurate MBERS measurement of \citet{77MeLeDy.CO}, the CC3 calculations of \citet{03CoHaJo.CO}, and EFGIB measurements from other sources. These comparisons are presented in Table \ref{tab:CO_quadrupole_comparison}. 

\begin{table}
	\caption{A comparison of various electric quadrupole moment values for CO in \si{\au} ($e a_0^2$ = \SI{4.486484(28)e-40}{\coulomb\metre\squared}\; \citep{CODATA}) from the literature. All values are averaged over the vibrational ZPE and are given in the molecular centre of mass reference frame, $Q_{zz}^\text{(CM)} = 2R_z\mu + Q_{zz}^\text{(EQC)}$ with the displacement between the centre of mass and the electric quadrupole centre given by $R_z$ = \SI{-5.96}{\au} and a dipole moment $\mu$ = \SI{-0.043159}{\au}\citep{03CoHaJo.CO, 11ChCoxx.CO}}
		\begin{center}
		\begin{tabular}{lll}
			\hline
			$Q_{zz}$ / \si{\au}\ & Method & Ref.\\
			\hline\hline
			-1.4522    & CCSD(T)     & This work \\
			-1.445(2)  & CC3         & \onlinecite{03CoHaJo.CO}\\
			-1.43(3)   & MBERS        & \onlinecite{77MeLeDy.CO}\\
			-1.440(69) & EFGIB       & \onlinecite{11ChCoxx.CO}\\
			-1.382(31) & EFGIB       & \onlinecite{98GrImRa.CO,11ChCoxx.CO}\\
			-1.18(22)  & EFGIB       & \onlinecite{68BuDiDu.CO,11ChCoxx.CO}\\
			\hline
		\end{tabular}
	\end{center}
	\label{tab:CO_quadrupole_comparison}
\end{table}

%, which has the following analytical form:
%\begin{equation}
%	U(\beta, r) = Y(r) \Lambda(\beta, z(r)) \\
%\end{equation}
%where $Y(r)$ accounts for the asymptotic behaviour of the potential in both the short- and %long-range limits;
%\begin{align}
%	Y(r) &= Y_0(r) + Y_\infty(r), \\
%	Y_0(r) &= \left[ \frac{K}{r} + Kd_0 + E_0 + \left(\frac{1}{2}Kd_0 + E_0\right) d_0r \right] \exp(-d_0r), \\
%	Y_\infty(r) &= \sum_{n=5,6,8} D_n(r) \frac{C_n}{r^n},
%\end{align}
%where $D_n(r)$ is a modified Douketis damping function\cite{82DoScMa.ai}
%\begin{equation}
%	D_n(r) = \left[1 - \exp\left(-\frac{d_1}{n}r - \frac{d_2}{\sqrt{n}}r^2 \right) \right]^{n+2}
%\end{equation}
%and the mapping function $z(r)$ is chosen as
%\begin{equation}
%	z(r) = \tanh(c_1r - c_2r^{-1})	
%\end{equation}
%with $d_{0,1}$ and $c_{1,2}$ constants. The function $\Lambda(\beta, z)$ is non-singular in the %domain $z \in [-1, +1]$, and is represented as a Chebyshev polynomial of the first kind,
%\begin{equation}
%	\Lambda(\beta, z) = \sum^{N}_{n=0} \beta_n T_n(z).
%\end{equation}
%The details of their  analytical form and the fitting procedure are given by \citet{18MeStEr.CO}. 
Nuclear motion calculations are performed using the semi-empirical PEC of \citet{18MeStEr.CO}. This accurate analytical representation of the PEC is chosen for the \Duo\ solutions in order to improve the quality of the wavefunctions used to calculate the linestrengths. 
The \Duo\  vibrational grid used for the calculation consists of 501 equally spaced points in the range \SIrange{1.50}{3.00}{\bohr}, and the first 21 vibrational states are selected to form the contracted basis. These excitations correspond to energies within the spectroscopically relevant region ($E/hc<$ \SI{40,000}{\per\cm}) for the room temperature applications.  

%The functional form of the PEC allows us to solve the vibrational Schr\"{o}dinger equation over a greater range of internuclear distances than the CCSD(T) quadrupole moment is obtained for and, more importantly, outside the spectroscopically relevant region, in turn improving the quality of the vibrational basis set. 

%domain of the CCSD(T) electric quadrupole moment. These vibrational levels are discarded during the intensity calculations after solving the vibrational problem by selecting $v_\text{max} < 21$ in the contracted basis set.

%The following functional forms were used for extrapolation outside the original \ai\ range:\citep{Duo}
%\begin{eqnarray}
%\nonumber
%  f_{\rm short}(r) &=& A + B r, \\
%\nonumber
% f_{\rm long}(r)  &=& A/r^2 + B/r^3
%\end{eqnarray}
%for  long range, where $A$ and $B$ are stitching parameters.

After solving the Schr\"{o}dinger equation for rotational quantum numbers $0 \leq J \leq 50$, with a vibrational transition quadrupole moment  $\mel{\xi_f v_f}{Q^{(2)}_0}{\xi_i v_i} <$ \SI{1e-5}{\au} are discarded. It was found by \citet{15MeMeSt.CO} numerically computed transition dipole moments of high overtones corresponding to large changes in vibrational quanta can suffer  from numerical instabilities  and lead to unphysically large intensities. 
%In particular, representing an electric dipole moment as a grid of \ai\ points, and applying spline interpolation to obtain its on the vibrational grid, results in discontinuities in the derivative of the dipole. 
%For electric dipole transitions, performing a simultaneous fit to both the empirical and \ai\ values can be an effective solution, using long-range \ai\ data to constrain the asymptotic behaviour of the curves. 
In the case of electric quadrupole transitions however, the intensity of these high overtone vibrational bands is sufficiently weak that absorption lines with transition quadrupole moments $\mel{\xi_f v_f}{Q^{(2)}_0}{\xi_i v_i} <$ \SI{1e-5}{\au} (corresponding to high overtone bands) can simply be excluded from the line list altogether.

\begin{figure}
	\centering
	\includegraphics[width=.49\linewidth]{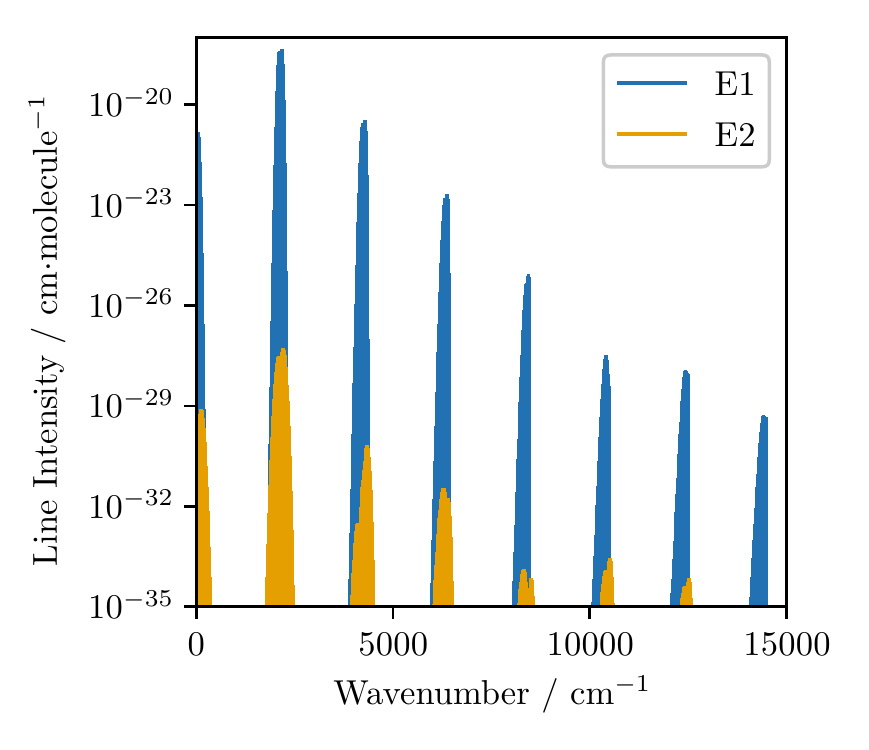}
	\includegraphics[width=.49\linewidth]{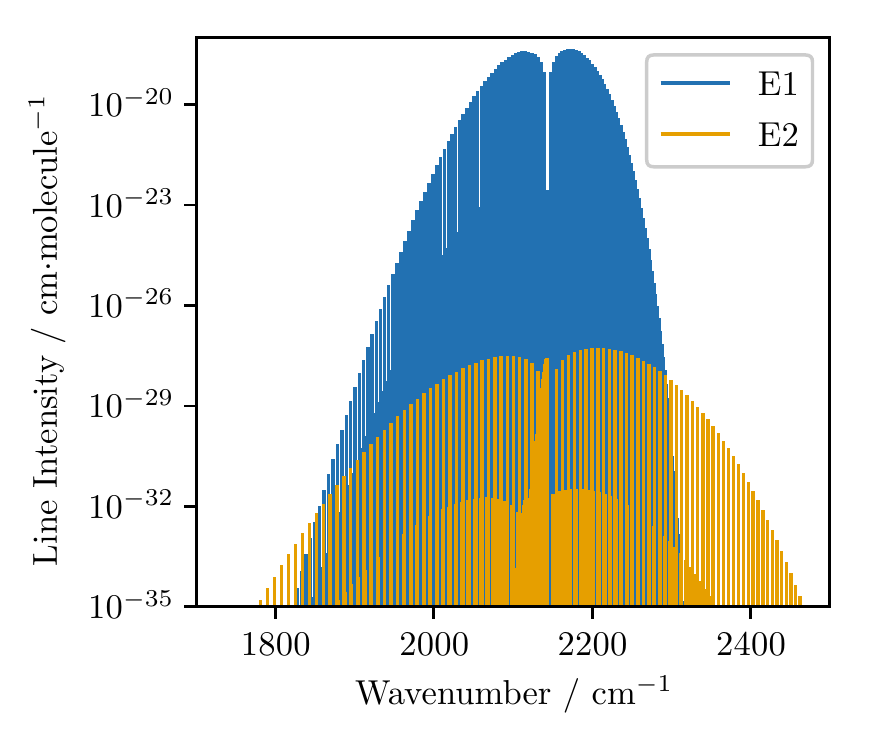}
	\caption{Vibrational bands (left) and rotational $v=0-1$ transitions (right) of the E1 and E2 rovibrational spectra in the ground $X$~$^1\Sigma^+$ state of the $^{12}$C$^{16}$O molecule. The E1 intensities are those of \citet{15LiGoRo.CO}, via the ExoMol database.}
	\label{fig:CO_li2015_CCSDT_hitran}
\end{figure}

The calculated state energies are  substituted for those obtained by \citet{15LiGoRo.CO} in a simultaneous direct-fit to experimentally determined energy levels. This improves the accuracy in the line positions of the final stick spectrum, obtained via \textsc{ExoCross}\citep{ExoCross}, but has no effect on the quadrupole Einstein coefficients or linestrengths. The energy level data of \citet{15LiGoRo.CO} is made available through the HITRAN or ExoMol (\href{http://www.exomol.com)}{exomol.com}) databases.\citep{20TeYuAl} 

The resultant room temperature ($T = 296$~K) line list for $^{12}$C$^{16}$O with a cut-off intensity of \SI{e-35}{\cm\per\molecule} consists of 6474 electric quadrupole transitions between rotational states up to $J_{\rm max} = 48$, and vibrational states $v=7$. A synthetic room temperature E2 spectrum is illustrated in Fig. \ref{fig:CO_li2015_CCSDT_hitran}, where it is compared to the E1 spectrum of \citet{15LiGoRo.CO}. The difference is approximately eight orders of magnitude. Nonetheless, many E2 lines - particularly for the $v=0\leftarrow0$ and $v=1\leftarrow0$ bands - lie above the typical cutoff intensity used in many spectroscopic databases ($\sim$\SI{e-30}{\centi\metre\squared\per\molecule} at $T= 296$~K).

The computed electric quadrupole Einstein~A coefficients of $^{12}$C$^{16}$O are combined with the ExoMol E1 line list Li2015 for CO in a form of an E2 Transition file, see an extract in Table~\ref{tab:CO_trans_extract}. Apart from the Einstein~A E2 coefficients (s$^{-1}$), the Transition file contains the upper and lower state counting numbers of the  Li2015 State file, 
as illustrated in Table~\ref{tab:CO_states_extract}, which presents an extract from the ExoMol State file of the $^{12}$C$^{16}$O line list Li2015. For more details on the ExoMol line list structure see \citet{20TeYuAl}.

\begin{table}[]
    \centering
    \begin{tabular}{R{3em}R{2em}R{6em}R{7em}}
        \hline
        $f$ & $i$ & $A_{fi}$ & $\tnu_{fi}$ \\ 
        \hline\hline 
        \texttt{ 94} & \texttt{10} & \texttt{1.0587E-17} & \texttt{10.591935} \\
        \texttt{ 93} & \texttt{ 9} & \texttt{1.1546E-17} & \texttt{10.696876} \\
        \texttt{ 92} & \texttt{ 8} & \texttt{1.2569E-17} & \texttt{10.801832} \\
        \texttt{ 91} & \texttt{ 7} & \texttt{1.3657E-17} & \texttt{10.906802} \\
        \texttt{ 90} & \texttt{ 6} & \texttt{1.4815E-17} & \texttt{11.011786} \\
        \texttt{ 89} & \texttt{ 5} & \texttt{1.6043E-17} & \texttt{11.116781} \\
        \texttt{ 88} & \texttt{ 4} & \texttt{1.7346E-17} & \texttt{11.221787} \\
        \texttt{ 87} & \texttt{ 3} & \texttt{1.8725E-17} & \texttt{11.326802} \\
        \texttt{ 86} & \texttt{ 2} & \texttt{2.0183E-17} & \texttt{11.431825} \\
        \texttt{ 85} & \texttt{ 1} & \texttt{2.1722E-17} & \texttt{11.536856} \\
        \texttt{136} & \texttt{52} & \texttt{1.7502E-16} & \texttt{17.652735} \\
        \hline
    \end{tabular}
    \caption{Extract from the $^{12}$C$^{16}$O electric quadrupole Transition file. It contains the upper (f) and lower (i) states counting numbers,   Einstein A coefficients (s$^{-1}$) and transition wavenumbers (\si{\per\cm}).}
    \label{tab:CO_trans_extract}
\end{table}

\begin{table}[]
    \centering
    \caption{Extract from the Li2015 States file for $^{12}$C$^{16}$O.}
    \label{tab:CO_states_extract}
    \begin{tabular}{R{2em}R{8em}R{2em}R{2em}R{3em}R{3em}R{6em}R{2em}R{2em}R{2em}R{2em}}
        \hline
        $i$ & $E$ & $g$ & $J$ & $v$& $\tau$  \\ 
        \hline\hline 
        \texttt{ 1} & \texttt{    0.000000} & \texttt{1} & \texttt{0} & \texttt{ 0}& \texttt{e} \\
        \texttt{ 2} & \texttt{ 2143.271100} & \texttt{1} & \texttt{0} & \texttt{ 1}& \texttt{e} \\
        \texttt{ 3} & \texttt{ 4260.062200} & \texttt{1} & \texttt{0} & \texttt{ 2}& \texttt{e} \\
        \texttt{ 4} & \texttt{ 6350.439100} & \texttt{1} & \texttt{0} & \texttt{ 3}& \texttt{e} \\
        \texttt{ 5} & \texttt{ 8414.469300} & \texttt{1} & \texttt{0} & \texttt{ 4}& \texttt{e} \\
        \texttt{ 6} & \texttt{10452.222200} & \texttt{1} & \texttt{0} & \texttt{ 5}& \texttt{e} \\
        \texttt{ 7} & \texttt{12463.768600} & \texttt{1} & \texttt{0} & \texttt{ 6}& \texttt{e} \\
        \texttt{ 8} & \texttt{14449.181300} & \texttt{1} & \texttt{0} & \texttt{ 7}& \texttt{e} \\
        \texttt{ 9} & \texttt{16408.534600} & \texttt{1} & \texttt{0} & \texttt{ 8}& \texttt{e} \\
        \texttt{10} & \texttt{18341.904400} & \texttt{1} & \texttt{0} & \texttt{ 9}& \texttt{e} \\
        \texttt{11} & \texttt{20249.368200} & \texttt{1} & \texttt{0} & \texttt{10}& \texttt{e} \\
        \hline
    \end{tabular}
\mbox{}\\
{\flushleft
$i$:   State counting number.     \\
$\tilde{E}$: State energy in \si{\per\cm}. \\
$g_i$:  Total statistical weight, equal to ${g_{\rm ns}(2J + 1)}$.     \\
$J$: Total angular momentum.\\
State: Electronic state.\\
$v$:   State vibrational quantum number. \\
$\tau$:   Rotationless parity $e$/$f$.\citep{75BrHoHu.diatom}\\
}
\end{table}

\subsection{Hydrogen Fluoride}
\label{sec:hydrogen_fluoride}

Like the CO molecule, HF possesses a strong permanent electric dipole moment\citep{63Weissx.HF}, it also possesses a strong permanent electric quadrupole moment.\citep{73deDyxx.HF} Numerous \ai\ studies have been performed for HF, including several which produce QMCs for the ground $X$~$^1\Sigma^+$ electronic state.\citep{96PiKoSp.HF,03Maroul.HF,08Harris.HF} \citet{96PiKoSp.HF} use the orthogonally spin-adapted linear-response coupled-cluster (LRCC) theory with singly and doubly excited clusters (CCSD) and obtain quadrupole moments at 15 internuclear geometries in the range \SIrange{1.12632}{12.1296}{\bohr} Their basis set of choice is that introduced by Sadlej for correlated calculations of molecular electric properties,\citep{88Sadlej.ai} which they compare to standard basis sets at the TZ level. They also provide the results of full CI calculations on a DZ basis set. \citet{03Maroul.HF} presents all-electron CCSD(T) calculations of the quadrupole moment at nine internuclear geometries in the range \SIrange{0.9328}{2.5328}{\bohr} For comparison, the quadrupole moment for the $X^1\Sigma^+$ state is computed via the MRCI method and an aug-cc-pVQZ basis set at 501 internuclear geometries in the range \SIrange{1.32}{6.99}{\bohr} using \textsc{Molpro}.

\begin{table}
	\caption{A comparison of various \ai\ electric quadrupole moment values for HF in \si{\au} ($e a_0^2$). All values are given in the molecular centre of mass reference frame, and at the equilibrium nuclear geometry.}
	\begin{center}
		\begin{tabular}{rlr}
			\hline
			$Q_{zz}$ / \si{\au} & Method & Ref.\\
			\hline\hline
			1.706 & MRCI    & This work \\
			1.72  & CCSD     & \onlinecite{96PiKoSp.HF} \\
			1.72  & CCSD(T) & \onlinecite{03Maroul.HF} \\
			1.66  & CI      & \onlinecite{96PiKoSp.HF} \\
			\hline
		\end{tabular}
	\end{center}
	\label{tab:HF_quadrupole_comparison}
\end{table}

\begin{figure}
	\centering
	\includegraphics{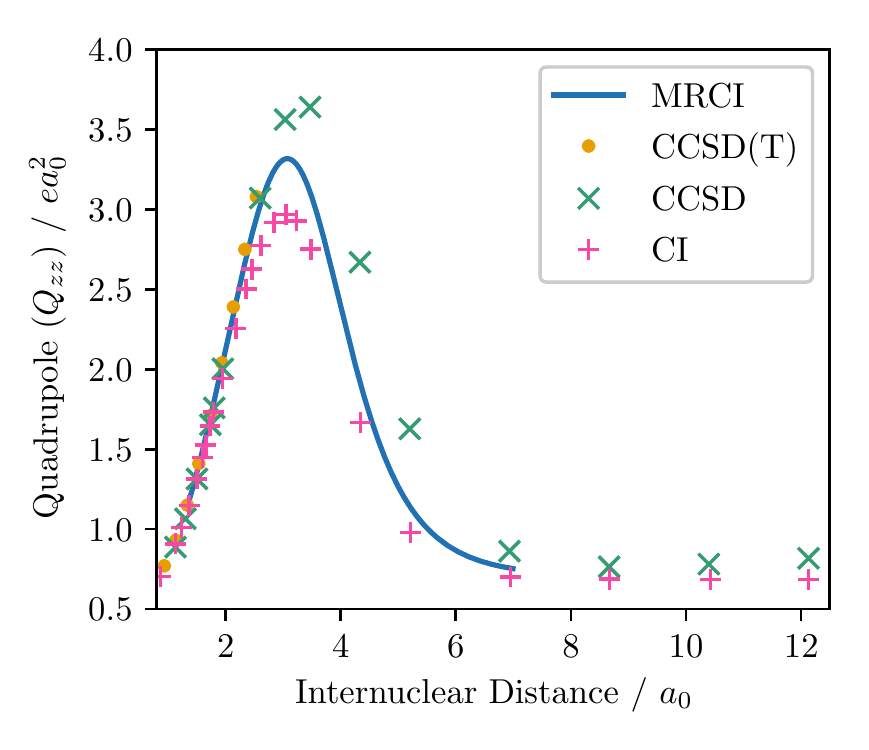}
	\caption{Comparison of the quadrupole moment curves  in \si{\au} ($e a_0^2$)  for HF obtained via various \textit{ab initio} methods. The MRCI calculations presented in this work, CCSD(T) calculations of \citet{03Maroul.HF}, and CCSD and full-CI calculations of \citet{96PiKoSp.HF}.}
	\label{fig:HF_EQC_method_comparison}
\end{figure}

The electric quadrupole moments of HF obtained via these various methods are illustrated in Fig. \ref{fig:HF_EQC_method_comparison}. Although the four curves have the same general shape, significant variation is apparent between the value of $Q_{zz}$ computed at intermediate bond lengths close to \SI{3.8}{\bohr}. Here the strength of the quadrupole moment is greatest, and difference of more than \SI{0.5}{\au} is apparent between the full CI and CCSD methods. Table \ref{tab:HF_quadrupole_comparison} shows the differences in the value of the quadrupole moment at the equilibrium internuclear distance for the four \ai\ methods presented. All four calculations produce similar values for $Q_{zz}(R_e)$, but the coupled-cluster methods systematically overestimate the strength relative to experimental measurements. Importantly, when averaged over the vibrational ZPE, the MRCI results obtained in the present work give good agreement with the experimental MBERS measurement of \citet{73deDyxx.HF}. They obtain $\mel{v=0}{Q_{zz}(r)}{v=0}$ = \SI{1.75(2)}{\au}, whilst \Duo\ calculates a value of \SI{1.747}{\au}, which is within the range of experimental uncertainty.

For the PEC, \citet{15CoHaxx.HF} provide a very accurate RKR-style analytical expression for the potential energy and Born-Oppenheimer breakdown functions of the $X^1\Sigma^+$ ground electronic state of various hydrogen halide isotopologues, including \textsuperscript{1}H\textsuperscript{19}F. They devise a novel analytical form (MLR3) of the diatomic electronic potential and perform a non-linear least squares fit to experimental energies. 

%The MLR3 function is given by \citep{10CoHaxx.ai}:
%\begin{equation}
%	V(r) = D_e \left( 1 - \frac{u_\text{LR}(r)}{u_\text{LR}(r_e)} \exp\left[ %-\phi_\text{MLR3}(r) y_{p,a}(r, r_e) \right] \right)^2,
%\end{equation}
%where
%\begin{equation}
%	y_{p,a}(r,r_e) = \frac{r^p - r_e^p}{r^p - ar_e^p}
%\end{equation}
%and the long-range potential function $u_\text{LR}(r)$ is given by:
%\begin{equation}
%	u_\text{LR}(r) = \sum_n D_n(r) \frac{C_n}{r^n},
%\end{equation}
%\citet{10CoHaxx.ai} propose generalised Douketis damping \citep{82DoScMa.ai} or %Tang-Toennies \citet{84TaToxx.ai} damping functions for $D_n(r)$, the \Duo\ %implementation of MLR3 employs Douketis damping functions defined as:
%\begin{equation}
%	D_n(r) = \left( 1 - \exp\left[ -\frac{b(s)(\rho r)}{n} -\frac{c(s)(\rho r)^2}{\sqrt{n}} \right] \right)^{m+s},
%\end{equation}
%with $\rho = 2\rho_A\rho_B/(\rho_A+\rho_B)$ for atoms $A, B$ where $\rho_{A,B} = (I_p^{A, B}/I_p^{A, B})^{2/3}$ and $I_p^H$ is the ionisation potential of the hyrogen atom. Lastly the function $\phi_\text{MLR3}(r)$ is given by:
%\begin{equation}
%	\phi_\text{MLR3}(r) = y_m(r,r_\text{ref}) \phi_\text{MLR3}(\infty) + \left[ 1 - y_m(r,r_\text{ref})\right] \sum_{i=0}^{N_\phi} \phi_i y_q(r, r_\text{ref})^i,
%\end{equation}
%where
%\begin{gather}
%	y_{m,q}(r, r_\text{ref}) = \left( \frac{r^{m,q} - r_\text{ref}^{m,q}}{r^m + r_\text{ref}^{m,q}} \right) ,\\
%	\phi_\text{MLR3}(\infty) = \ln\left(\frac{2D_e}{u_\text{LR}(r_e)}\right).
%\end{gather}
Their analytical representation of the MLR3 potential has been newly implemented in \Duo\, and for the present calculations, the HF MLR3 parameters obtained by \citet{15CoHaxx.HF} are employed, as well as their Born-Oppenheimer breakdown (BOB) function which is obtained from the Fortran source code provided in the supplementary material of \citet{15CoHaxx.HF}. 

\begin{figure}
	\centering
	\includegraphics[width=.49\linewidth]{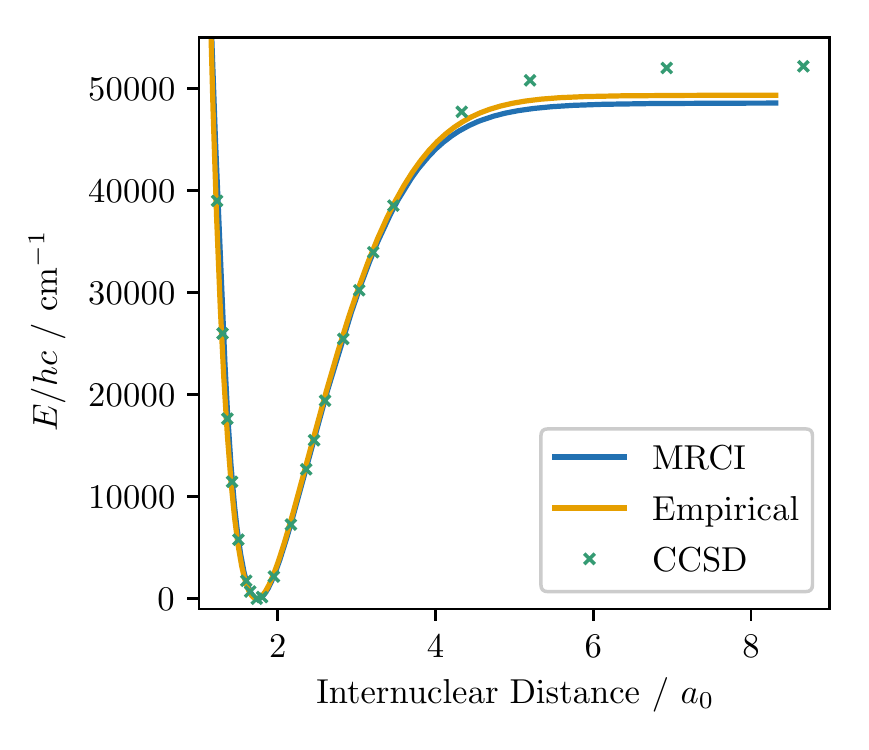}
	\caption{Comparison of the potential energy curves for the $X^1\Sigma^+$ ground state of HF. The MRCI calculations are from this work, empirically fitted MLR3 potential of \citet{15CoHaxx.HF}, and the CCSD calculations of \citet{96PiKoSp.HF}.}
	\label{fig:HF_PEC_method_comparison}
\end{figure}

\begin{figure}
	\centering
	\includegraphics[width=.49\linewidth]{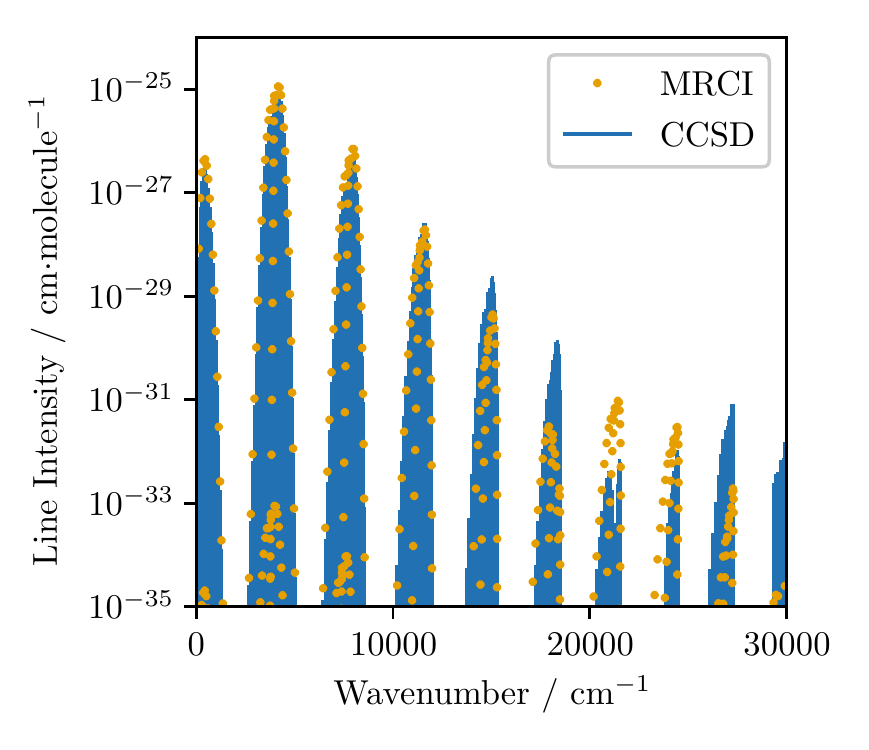}
	\caption{Comparison of the electric quadrupole absorption spectrum for H$^{19}$F obtained via spectroscopic models using the CCSD and MRCI quadrupole moment curves illustrated in Fig. \ref{fig:HF_EQC_method_comparison}.}
	\label{fig:HF_method_comparison}
\end{figure}

Fig. \ref{fig:HF_PEC_method_comparison}, shows a comparison of the potential energy curves obtained from our MRCI calculations, the CCSD calculations of \citet{96PiKoSp.HF} and the MLR3 potential of \citet{15CoHaxx.HF}. All three methods give similar results at short and intermediate bond lengths. The CCSD calculations overestimate the dissociation energy, relative to the empirical MLR3 potential, and the MRCI results predict a slightly lower dissociation energy. Fig. \ref{fig:HF_method_comparison} illustrates the results of calculations from two spectroscopic models. In each case the potential energies are the same; the MLR3 and BOB curves of \citet{15CoHaxx.HF}; but one model uses the MRCI quadrupole moment presented in this work, and the other uses Piecuch's CCSD quadrupole moment. 
In both cases nuclear motion calculations are performed for rotational states $0 \leq J \leq 41$, the vibrational grid is defined for 501 equally spaced points in the range \SIrange{0.76}{4.40}{\bohr}, and the first 20 vibrational states are chosen for the contracted basis. 

For the first three vibrational bands, the absorption intensities predicted by both spectroscopic models are nearly identical. Higher order vibrational bands, however, exhibit significant discrepancies. The CCSD intensities begin to plateau above \SI{20000}{\per\cm}, we propose that this intensity plateau arises as a result of the same effect encountered in section \ref{sec:carbon_monoxide} and detailed by \citet{15MeMeSt.CO}. Comparatively, the MRCI spectrum shows no such intensity plateau, indeed the MRCI quadrupole moment is obtained on a considerably finer grid spacing, which aids in smoothing the interpolation. 

\begin{figure}
	\centering
	\includegraphics[width=.49\linewidth]{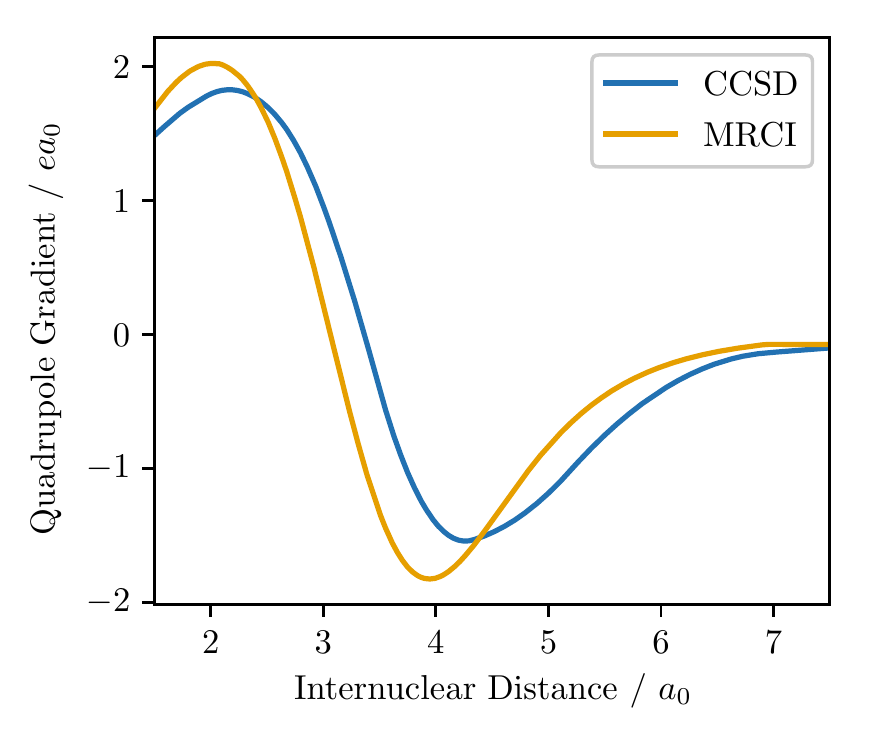}
	\caption{Central finite difference gradients of the HF quadrupole moment obtained via MRCI and CCSD methods with respect to internuclear distance.}
	\label{fig:HF_Qgrad_method_comparison}
\end{figure}

A second possible cause proposed by \citet{15MeMeSt.CO} is the asymptotic behaviour of the quadrupole moment curves at longer internuclear distances. Here the magnitude of the coupling becomes exponentially smaller, and significant relative variations in the gradient of $Q_{zz}$ are observed between the two methods. The gradient of CCSD quadrupole moment curve at distances $R$ $>$ \SI{3}{\bohr} decays considerably slower than that obtained via MRCI calculations. Fig. \ref{fig:HF_Qgrad_method_comparison} shows the gradient of the two quadrupole moment functions computed using a central finite difference scheme on the Duo integration grid. 

The MRCI spectrum exhibits a local minimum in intensity for the $v=5\leftarrow0$ band. A similar abnormal intensity was observed by \citet{15MeMeSt.CO} for the same vibrational band of the electric dipole spectrum. Regardless, the expected E2 absorption intensities for the $v=5\leftarrow0$ band is extremely weak, far weaker than typical spectroscopic cutoff intensity ($~10^{-30}$ cm/molecule at $T= 296$~K).

\begin{figure}
	\centering
	\includegraphics[width=.49\linewidth]{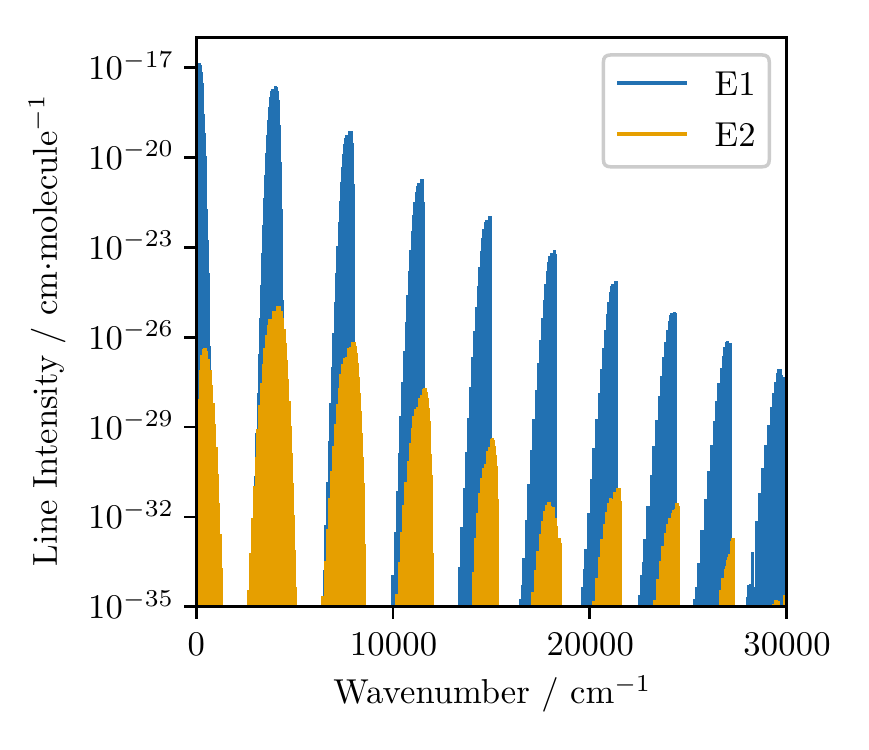}
	\includegraphics[width=.49\linewidth]{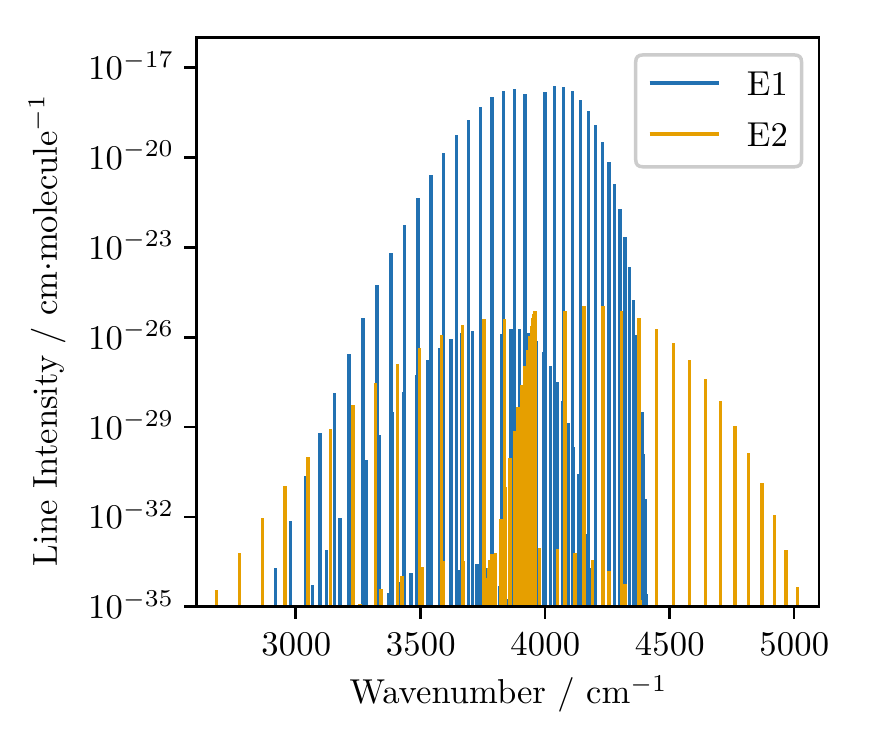}
	\caption{Vibrational bands (left) and rotational $v=0-1$ transitions (right) of the E1 and E2 rovibrational spectra in the ground $X^1\Sigma^+$ state of the H$^{19}$F molecule as line intensities (cm/molecule). The E1 spectrum is that of \citet{15CoHaxx.HF}, via the ExoMol database.}
	\label{fig:HF_coxon_MRCI_hitran}
\end{figure}

Intensities obtained using the MRCI quadrupole moment are chosen for the final \textsuperscript{1}H\textsuperscript{19}F spectroscopic model and line list. This is combined with the ExoMol E1 line list Coxon-Hajig in the form of an E2 Transition file. Fig.~\ref{fig:HF_coxon_MRCI_hitran} compares the E2 intensities obtained for room temperature calculations to the E1 intensities of \citet{15CoHaxx.HF}. It consists of 2716 electric quadrupole transitions between rotational states up to $J=18$ and vibrational states up to $v=9$ with a cutoff intensity of \SI{e-35}{\cm\per\molecule} ($T= 296$~K) and is included into the supplementary material of this work.

\subsection{Oxygen Noxon Band}
\label{sec:oxygen_noxon_band}

Owing to its molecular symmetry, the homonuclear O\textsubscript{2} molecule possesses no permanent dipole moment. Additionally, the three lowest lying electronic states, $X^3\Sigma_g^-$, $a^1\Delta_g$ and $b^1\Sigma_g^+$ all have {\it gerade} symmetry. The $\Sigma$ spin-orbit mixing results in electric quadrupole transitions in the  $a^1\Delta_g$ -- $X^3\Sigma_g^-$ system, which borrow strength from the direct $a^1\Delta_g$ -- $b^1\Sigma_g^+$ transitions of the so-called Noxon band.\citep{61Noxonx.O2, 11MiBaSh.O2}
\begin{equation}
    \mel{a ^1 \Delta_g}{Q^{(2)}_{\pm 2}}{X ^3\Sigma_g^-} \propto \mel{a ^1 \Delta_g}{Q^{(2)}_{\pm 2}}{b ^1\Sigma_g^+}
\end{equation}

% The well-known IR and Vis spectra of O\textsubscript{2} are therefore due to the electric quadrupole and magnetic dipole moments. The strongest transitions belong to the magnetic dipole bands $b^1\Sigma_g^+$ -- $X^3\Sigma_g^-$ A, B, $\gamma$ and $\delta$. In fact these electronic bands are directly forbidden by the magnetic dipole selection rule $\Delta S = 0$, but arise via a spin-orbit mixing of the singlet and triplet $\Sigma$ states. Although weaker, electric quadrupole transitions for this system are also present and arise via the same spin-orbit mixing. In this case, two electric quadrupole moments contribute to the intensity of the $b^1\Sigma_g^+$ -- $X^3\Sigma_g^-$ bands:
% \begin{align}
%     \mel{\tilde X ^3\Sigma_g^-}{Q^{(2)}_0}{\tilde b ^1\Sigma_g^+} &\propto \mel{X ^3\Sigma_g^- (\Omega=0)}{Q^{(2)}_0}{X ^3\Sigma_g^- (\Omega=0)} - \mel{b ^1\Sigma_g^+}{Q^{(2)}_0}{b ^1\Sigma_g^+} \\
%     \mel{\tilde X ^3\Sigma_g^-}{Q^{(2)}_{\pm 1}}{\tilde b ^1\Sigma_g^+} &\propto \mel{d ^1\Pi_g}{Q^{(2)}_{\pm1}}{b ^1\Sigma_g^+},
% \end{align}
% where the tilde indicates the spin-orbit perturbed eigenstates. This leads to four absorption bands corresponding to $\Delta N = \pm 3, \pm 1$ with $\Delta J = \pm 2$.

Although weak, with intensities on the order of \SI{e-45}{\centi\metre\squared\per\molecule}, rotational lines in both the $(1-0)$ and $(0-0)$ Noxon bands have been measured experimentally.\citep{09FoCeMa.O2, 86FiKrRa.O2} This electronic band is forbidden by the magnetic dipole $\Delta \Lambda = 0, \pm 1$ selection rule, and consequently the Noxon band is purely quadrupolar in nature. This makes the Noxon band ideal for validations of the electric quadrupole methodology applied to open-shell molecules.

\begin{figure}
    \centering
    \includegraphics{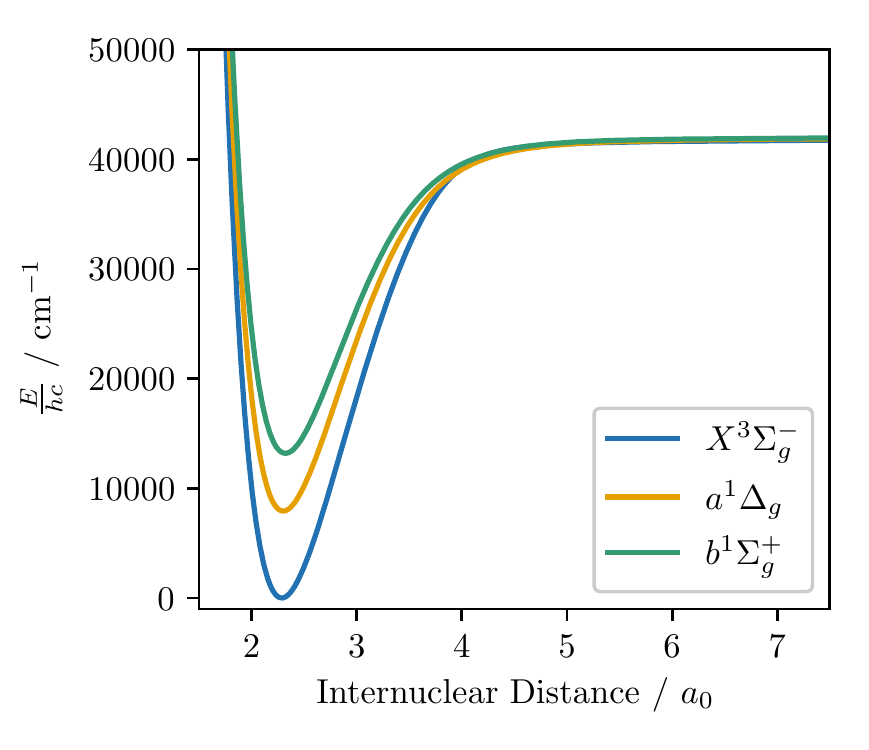}
    \caption{Potential energy curves for the three lowest lying electronic states of O\textsubscript{2}, obtained via MRCI calculations with an aug-cc-pV6Z basis set.}
    \label{fig:O2_DUO_V}
\end{figure}

\begin{figure}
    \centering
    \includegraphics{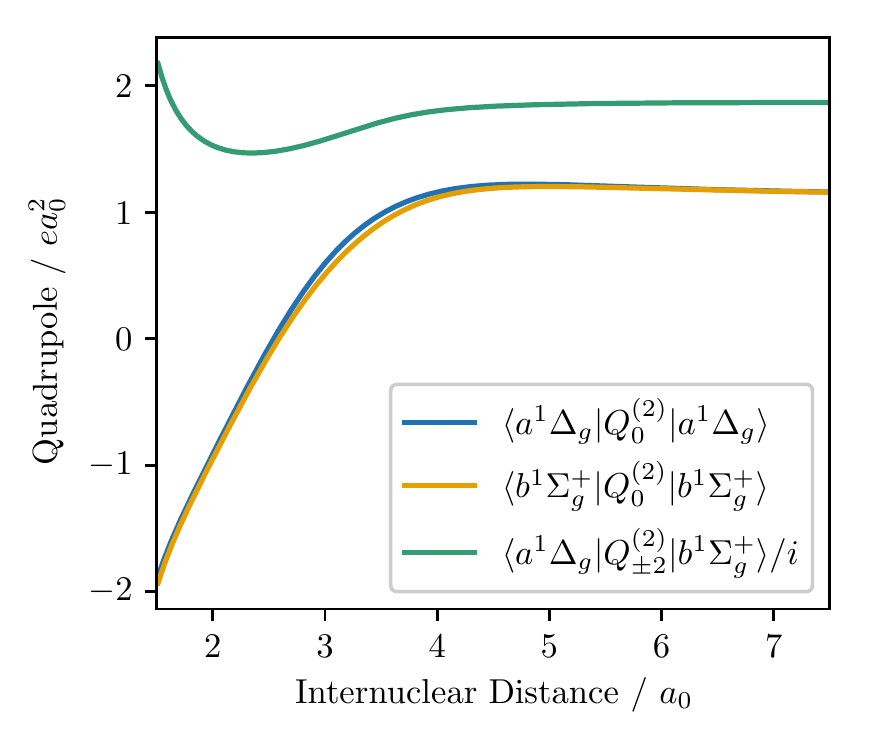}
    \caption{Diagonal quadrupole moment curves in \si{\au} ($e a_0^2$) for the $a ^1\Delta_g$ and $b ^1\Sigma_g^+$ electronic states of O\textsubscript{2} obtained via MRCI calculations with a aug-cc-pV6Z basis set.}
    \label{fig:O2_DUO_Q}
\end{figure}

The emission spectrum of the fundamental Noxon band was measured at \SI{313 \pm 10}{\kelvin} by \citet{86FiKrRa.O2} with an estimated precision of \SIrange{0.010}{0.020}{\per\centi\metre}. This measurement is replicated computationally with \Duo\ calculated Einstein coefficients and the \Exocross\ program. The \ai\ data for the \Duo\ calculations was produced using \textsc{Molpro} \citep{MOLPRO2015} with the MRCI program and an aug-cc-pV6Z basis set. The calculation includes PECs for the three lowest lying electronic states $X ^3\Sigma_g^-$, $a ^1\Delta_g$ and $b ^1\Sigma_g^+$ (Fig. \ref{fig:O2_DUO_V}), as well as diagonal quadrupole moment curves $Q^{(2)}_0(r) = 3Q_{zz}(r)/\sqrt{6}$ for the $a ^1\Delta_g$ and $b ^1\Sigma_g^+$ electronic states, and the off-diagonal $a ^1\Delta_g$ -- $b ^1\Sigma_g^+$ quadrupole $Q^{(2)}_{\pm 2}(r) = \sqrt{2} Q_{xx}(r)$ (Fig. \ref{fig:O2_DUO_Q}). Calculations are performed on a grid of 116 internuclear distances in the range \SIrange{1.5}{7.5}{\bohr}. The contracted vibrational basis set consists of the first 25 vibrational states for each electronic state, and calculations are performed for rotational states $0 \leq J \leq 50$.

\begin{figure}
    \centering
    \includegraphics{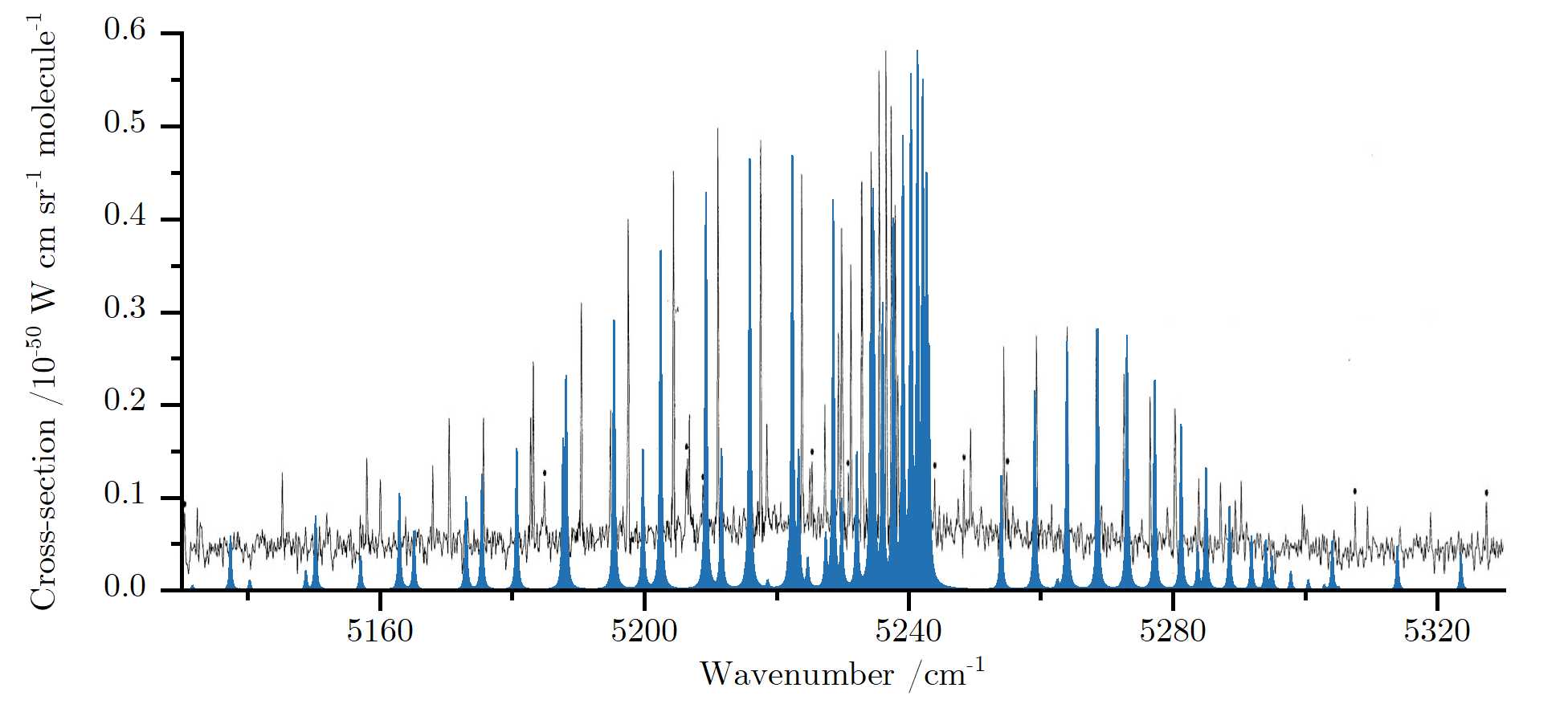}
    \caption{Overlay of the \Duo\ calculated O\textsubscript{2} Noxon emission cross-sections with the measured spectrum from \citet{86FiKrRa.O2}, scaled relative to the peak intensity of the $Q(8)$ transition. Cross-sections are calculated at $T$ = \SI{313}{\kelvin} with a Voigt profile (HWHM = \SI{0.15}{\per\cm})}
    \label{fig:Fink-Noxon-O2_duo-comparison} 
\end{figure}

Fig. \ref{fig:Fink-Noxon-O2_duo-comparison} shows an overlay of the experimental spectrum by  \citet{86FiKrRa.O2} with the calculated emission cross-section for the fundamental Noxon band, obtained via \Exocross\ using the \Duo\ calculated Einstein coefficients at \SI{313}{\kelvin} with a Voigt line profile (HWHM = \SI{0.15}{\per\centi\metre}). The intensities have been scaled relative to the most intense $Q(8)$ transition. There is a systematic error in the line positions calculated by \Duo\ \SI{\sim 7}{\per\centi\metre}, which is attributed primarily to the fact that the calculations do not include the strongly coupled excited $C{}^3\Pi_g$ state.\cite{94BaNaxx.O2} Due to the number of couplings required for a complete treatment of the open-shell O\textsubscript{2} molecule the full rovibronic spectrum, including such highly excited states will be the focus of a future publication. Consequently, and for the sake of simplicity, no empirical refinement of the PECs is performed in the present work. Nonetheless, the relative line positions and intensities are in good agreement with those measured by \citet{86FiKrRa.O2}, and demonstrate the validity of the approach for open-shell diatomic systems and excited electronic states.

\section{Conclusions}
\label{sec:conclusions}

Generic expressions for the electric quadrupole Einstein coefficients and matrix elements between arbitrary electronic states of (open-shell) diatomic molecules have been derived and implemented in the \Duo\ spectroscopic code. The implementation is general, and allows for the creation of highly accurate \ai\ and empirical spectroscopic models and line lists for an array of astrophysically important molecules. The work has been validated by reproducing highly accurate literature data for the homonuclear H\textsubscript{2} molecule, as well by comparison to the electronic emission spectrum of the O\textsubscript{2} Noxon band, and further demonstrated by the calculation of novel electric quadrupole spectra for the heteronuclear CO and HF molecules. The line lists for CO and HF have been included into the ExoMol database.

Through this calculation, we have shown that even for electric dipole-allowed systems, electric-quadrupole line intensities can often lie above the typically cutoff intensities used in spectroscopic databases, atmospheric retrievals and remote-sensing applications. For many homonuclear systems where rovibrational, and many electronic, transitions are forbidden in the electric dipole approximation, calculation of the quadrupole intensities is crucial for producing accurate rovibronic line lists. Our goal is to provide accurate E2 and M1 line lists for electronic transitions of (open-shell) diatomic molecules such as O\textsubscript{2}, N\textsubscript{2}, S\textsubscript{2}, SO etc.

\section{Supplementary material}

The  supplementary material includes the spectroscopic models for H\textsubscript{2}, HF, CO and O\textsubscript{2} in the form of \Duo\ input files;  E2 line lists for H\textsubscript{2}, HF, CO and O\textsubscript{2} using the ExoMol format; examples of E2 room temperature spectra of these molecules with the upper and lower states fully assigned.

\begin{acknowledgments}
This work was supported by the STFC Projects No. ST/R000476/1 and ST/S506497/1. The authors acknowledge the use of the UCL Legion High Performance Computing Facility (Myriad@UCL) and associated support services in the completion of this work. This work was also supported by the European Research Council (ERC) under the European Union’s Horizon 2020 research and innovation programme through Advance Grant number 883830.
The work of A.Y. has been supported by Deutsches Elektronen-Synchtrotron DESY, a member of the 
Helmholtz Association (HGF), and by the Deutsche Forschungsgemeinschaft (DFG) through the
cluster of excellence “Advanced Imaging of Matter” (AIM, EXC 2056, ID 390715994).
\end{acknowledgments}

\section{Data Availability}

The data that supports the findings of this study are available within the article, its supplementary material and  are also openly available at \url{www.exomol.com}.

\section{Conflict of interest}

The authors have no conflicts to disclose.

\vfill

\newpage
\clearpage
\appendix

%\counterwithin{figure}{section}

%\setcounter{table}{0}
%\renewcommand{\thetable}{A\arabic{table}}

\section{Correlation of \textsc{Molpro} enumeration to term symbols}
\label{ap:molpro_enumeration}

Tables \ref{t:A1} and \ref{t:A2}  are versions of Tables~\ref{tab:D2h_irreps} and \ref{tab:C2v_irreps} with the addition of \textsc{Molpro} enumerations for the irreducible representations. Which can be used to simplify the conversion of \textsc{Molpro} output data to Duo input. 

\begin{table}[h!]
    \caption{Irreducible representations for homonuclear symmetry groups, the functions that transform according to the irreducible representations, their \textsc{Molpro} enumeration, and corresponding components of electronic states.}
    \label{t:A1}
    \begin{tabular}{c|c|c|r}
        Symmetry & Function & \textsc{Molpro} No. & Components \\
        \hline
        $A_g$    & s   & 1 & $\Sigma_g^+$, $(\Delta_g)_{xx}$ \\
        $B_{1g}$ & xy  & 4 & $\Sigma_g^-$, $(\Delta_g)_{xy}$ \\
        $B_{2g}$ & xz  & 6 & $(\Pi_g)_x$ \\
        $B_{3g}$ & yz  & 7 & $(\Pi_g)_y$ \\
        $A_u$    & xyz & 8 & $\Sigma_u^-$, $(\Delta_u)_{xy}$ \\
        $B_{1u}$ & z   & 5 & $\Sigma_u^+$, $(\Delta_u)_{xx}$ \\
        $B_{2u}$ & y   & 3 & $(\Pi_u)_y$ \\
        $B_{3u}$ & x   & 2 & $(\Pi_u)_x$
    \end{tabular}
\end{table}

\begin{table}[h!]
    \caption{Irreducible representations for heteronuclear symmetry groups, the functions that transform according to the irreducible representations, their \textsc{Molpro} enumeration, and corresponding components of electronic states.}
    \label{t:A2}
    \begin{tabular}{c|c|c|r}
        Symmetry & Function(s) & \textsc{Molpro} No. & Components \\
        \hline
        $A_1$ & $s$, $z$  & 1 & $\Sigma^+$, $\Delta_{xx}$ \\
        $A_2$ & $xy$      & 4 & $\Sigma^-$, $\Delta_{xy}$ \\
        $B_1$ & $x$, $xz$ & 2 & $\Pi_x$ \\
        $B_2$ & $y$, $yz$ & 3 & $\Pi_y$ \\
    \end{tabular}
\end{table}

\newpage
%\renewcommand\bibname{References}
%\printbibliography

\bibliographystyle{aipnum4-1}

\bibliography{journals_phys,abinitio,linelists,programs,methods,H2,O2,H2O,HF,quadrupole,CO,stars,coolstars,exoplanets,sy,Books,S2,N2,CO2,He2+,interstellar,codata,diatomic}% Produces the bib

%merlin.mbs aipnum4-1.bst 2010-07-25 4.21a (PWD, AO, DPC) hacked
%Control: key (0)
%Control: author (8) initials jnrlst
%Control: editor formatted (1) identically to author
%Control: production of article title (-1) disabled
%Control: page (0) single
%Control: year (1) truncated
%Control: production of eprint (0) enabled
\begin{thebibliography}{103}%
\makeatletter
\providecommand \@ifxundefined [1]{%
 \@ifx{#1\undefined}
}%
\providecommand \@ifnum [1]{%
 \ifnum #1\expandafter \@firstoftwo
 \else \expandafter \@secondoftwo
 \fi
}%
\providecommand \@ifx [1]{%
 \ifx #1\expandafter \@firstoftwo
 \else \expandafter \@secondoftwo
 \fi
}%
\providecommand \natexlab [1]{#1}%
\providecommand \enquote  [1]{``#1''}%
\providecommand \bibnamefont  [1]{#1}%
\providecommand \bibfnamefont [1]{#1}%
\providecommand \citenamefont [1]{#1}%
\providecommand \href@noop [0]{\@secondoftwo}%
\providecommand \href [0]{\begingroup \@sanitize@url \@href}%
\providecommand \@href[1]{\@@startlink{#1}\@@href}%
\providecommand \@@href[1]{\endgroup#1\@@endlink}%
\providecommand \@sanitize@url [0]{\catcode `\\12\catcode `\$12\catcode
  `\&12\catcode `\#12\catcode `\^12\catcode `\_12\catcode `\%12\relax}%
\providecommand \@@startlink[1]{}%
\providecommand \@@endlink[0]{}%
\providecommand \url  [0]{\begingroup\@sanitize@url \@url }%
\providecommand \@url [1]{\endgroup\@href {#1}{\urlprefix }}%
\providecommand \urlprefix  [0]{URL }%
\providecommand \Eprint [0]{\href }%
\providecommand \doibase [0]{http://dx.doi.org/}%
\providecommand \selectlanguage [0]{\@gobble}%
\providecommand \bibinfo  [0]{\@secondoftwo}%
\providecommand \bibfield  [0]{\@secondoftwo}%
\providecommand \translation [1]{[#1]}%
\providecommand \BibitemOpen [0]{}%
\providecommand \bibitemStop [0]{}%
\providecommand \bibitemNoStop [0]{.\EOS\space}%
\providecommand \EOS [0]{\spacefactor3000\relax}%
\providecommand \BibitemShut  [1]{\csname bibitem#1\endcsname}%
\let\auto@bib@innerbib\@empty
%</preamble>
\bibitem [{\citenamefont {Reid}\ \emph {et~al.}(1981)\citenamefont {Reid},
  \citenamefont {Sinclair}, \citenamefont {Robinson},\ and\ \citenamefont
  {McKellar}}]{44ReSiRo.O2}%
  \BibitemOpen
  \bibfield  {author} {\bibinfo {author} {\bibfnamefont {J.}~\bibnamefont
  {Reid}}, \bibinfo {author} {\bibfnamefont {R.~L.}\ \bibnamefont {Sinclair}},
  \bibinfo {author} {\bibfnamefont {A.~M.}\ \bibnamefont {Robinson}}, \ and\
  \bibinfo {author} {\bibfnamefont {A.~R.~W.}\ \bibnamefont {McKellar}},\
  }\href {\doibase 10.1103/PhysRevA.24.1944} {\bibfield  {journal} {\bibinfo
  {journal} {Phys. Rev. A}\ }\textbf {\bibinfo {volume} {24}},\ \bibinfo
  {pages} {1944} (\bibinfo {year} {1981})}\BibitemShut {NoStop}%
\bibitem [{\citenamefont {Brault}(1980)}]{80Brault.O2}%
  \BibitemOpen
  \bibfield  {author} {\bibinfo {author} {\bibfnamefont {J.~W.}\ \bibnamefont
  {Brault}},\ }\href {\doibase 10.1016/0022-2852(80)90149-6} {\bibfield
  {journal} {\bibinfo  {journal} {J. Mol. Spectrosc.}\ }\textbf {\bibinfo
  {volume} {80}},\ \bibinfo {pages} {384} (\bibinfo {year} {1980})}\BibitemShut
  {NoStop}%
\bibitem [{\citenamefont {Goldman}, \citenamefont {Reid},\ and\ \citenamefont
  {Rothman}(1981)}]{81GoReRo.O2}%
  \BibitemOpen
  \bibfield  {author} {\bibinfo {author} {\bibfnamefont {A.}~\bibnamefont
  {Goldman}}, \bibinfo {author} {\bibfnamefont {J.}~\bibnamefont {Reid}}, \
  and\ \bibinfo {author} {\bibfnamefont {L.~S.}\ \bibnamefont {Rothman}},\
  }\href {\doibase 10.1029/GL008i001p00077} {\bibfield  {journal} {\bibinfo
  {journal} {Geophys. Res. Lett.}\ }\textbf {\bibinfo {volume} {8}},\ \bibinfo
  {pages} {77} (\bibinfo {year} {1981})}\BibitemShut {NoStop}%
\bibitem [{\citenamefont {Rothman}\ and\ \citenamefont
  {Goldman}(1981)}]{81RoGoxx.O2}%
  \BibitemOpen
  \bibfield  {author} {\bibinfo {author} {\bibfnamefont {L.~S.}\ \bibnamefont
  {Rothman}}\ and\ \bibinfo {author} {\bibfnamefont {A.}~\bibnamefont
  {Goldman}},\ }\href {\doibase 10.1364/AO.20.002182} {\bibfield  {journal}
  {\bibinfo  {journal} {Appl. Optics}\ }\textbf {\bibinfo {volume} {20}},\
  \bibinfo {pages} {2182} (\bibinfo {year} {1981})}\BibitemShut {NoStop}%
\bibitem [{\citenamefont {Gordon}\ \emph {et~al.}(2010)\citenamefont {Gordon},
  \citenamefont {Kassi}, \citenamefont {Campargue},\ and\ \citenamefont
  {Toon}}]{10GoKaCa.O2}%
  \BibitemOpen
  \bibfield  {author} {\bibinfo {author} {\bibfnamefont {I.~E.}\ \bibnamefont
  {Gordon}}, \bibinfo {author} {\bibfnamefont {S.}~\bibnamefont {Kassi}},
  \bibinfo {author} {\bibfnamefont {A.}~\bibnamefont {Campargue}}, \ and\
  \bibinfo {author} {\bibfnamefont {G.~C.}\ \bibnamefont {Toon}},\ }\href
  {\doibase 10.1016/j.jqsrt.2010.01.008} {\bibfield  {journal} {\bibinfo
  {journal} {J. Quant. Spectrosc. Radiat. Transf.}\ }\textbf {\bibinfo {volume}
  {111}},\ \bibinfo {pages} {1174} (\bibinfo {year} {2010})}\BibitemShut
  {NoStop}%
\bibitem [{\citenamefont {Leshchishina}\ \emph {et~al.}(2011)\citenamefont
  {Leshchishina}, \citenamefont {Kassi}, \citenamefont {Gordon}, \citenamefont
  {Yu},\ and\ \citenamefont {Campargue}}]{11LeKaGo.O2}%
  \BibitemOpen
  \bibfield  {author} {\bibinfo {author} {\bibfnamefont {O.}~\bibnamefont
  {Leshchishina}}, \bibinfo {author} {\bibfnamefont {S.}~\bibnamefont {Kassi}},
  \bibinfo {author} {\bibfnamefont {I.~E.}\ \bibnamefont {Gordon}}, \bibinfo
  {author} {\bibfnamefont {S.}~\bibnamefont {Yu}}, \ and\ \bibinfo {author}
  {\bibfnamefont {A.}~\bibnamefont {Campargue}},\ }\href {\doibase
  10.1016/j.jqsrt.2011.01.014} {\bibfield  {journal} {\bibinfo  {journal} {J.
  Quant. Spectrosc. Radiat. Transf.}\ }\textbf {\bibinfo {volume} {112}},\
  \bibinfo {pages} {1257} (\bibinfo {year} {2011})}\BibitemShut {NoStop}%
\bibitem [{\citenamefont {Reuter}, \citenamefont {Jennings},\ and\
  \citenamefont {Brault}(1986)}]{86ReJeBr.N2}%
  \BibitemOpen
  \bibfield  {author} {\bibinfo {author} {\bibfnamefont {D.}~\bibnamefont
  {Reuter}}, \bibinfo {author} {\bibfnamefont {D.~E.}\ \bibnamefont
  {Jennings}}, \ and\ \bibinfo {author} {\bibfnamefont {J.~W.}\ \bibnamefont
  {Brault}},\ }\href {\doibase 10.1016/0022-2852(86)90048-2} {\bibfield
  {journal} {\bibinfo  {journal} {J. Mol. Spectrosc.}\ }\textbf {\bibinfo
  {volume} {115}},\ \bibinfo {pages} {294} (\bibinfo {year}
  {1986})}\BibitemShut {NoStop}%
\bibitem [{\citenamefont {\v{C}erm\'{a}k}\ \emph {et~al.}(2017)\citenamefont
  {\v{C}erm\'{a}k}, \citenamefont {Vasilchenko}, \citenamefont {Mondelain},
  \citenamefont {Kassi},\ and\ \citenamefont {Campargue}}]{17CeVaMo.N2}%
  \BibitemOpen
  \bibfield  {author} {\bibinfo {author} {\bibfnamefont {P.}~\bibnamefont
  {\v{C}erm\'{a}k}}, \bibinfo {author} {\bibfnamefont {S.}~\bibnamefont
  {Vasilchenko}}, \bibinfo {author} {\bibfnamefont {D.}~\bibnamefont
  {Mondelain}}, \bibinfo {author} {\bibfnamefont {S.}~\bibnamefont {Kassi}}, \
  and\ \bibinfo {author} {\bibfnamefont {A.}~\bibnamefont {Campargue}},\ }\href
  {\doibase 10.1016/j.cplett.2016.11.002} {\bibfield  {journal} {\bibinfo
  {journal} {Chem. Phys. Lett.}\ }\textbf {\bibinfo {volume} {668}},\ \bibinfo
  {pages} {90} (\bibinfo {year} {2017})}\BibitemShut {NoStop}%
\bibitem [{\citenamefont {Herzberg}(1949)}]{49Herzberg.H2}%
  \BibitemOpen
  \bibfield  {author} {\bibinfo {author} {\bibfnamefont {G.}~\bibnamefont
  {Herzberg}},\ }\href {\doibase 10.1038/163170a0} {\bibfield  {journal}
  {\bibinfo  {journal} {Nature}\ }\textbf {\bibinfo {volume} {163}},\ \bibinfo
  {pages} {170} (\bibinfo {year} {1949})}\BibitemShut {NoStop}%
\bibitem [{\citenamefont {Wolniewicz}, \citenamefont {Simbotin},\ and\
  \citenamefont {Dalgarno}(1998)}]{98WoSiDa.H2}%
  \BibitemOpen
  \bibfield  {author} {\bibinfo {author} {\bibfnamefont {L.}~\bibnamefont
  {Wolniewicz}}, \bibinfo {author} {\bibfnamefont {I.}~\bibnamefont
  {Simbotin}}, \ and\ \bibinfo {author} {\bibfnamefont {A.}~\bibnamefont
  {Dalgarno}},\ }\href {\doibase 10.1086/313091} {\bibfield  {journal}
  {\bibinfo  {journal} {Astrophys. J. Suppl.}\ }\textbf {\bibinfo {volume}
  {115}},\ \bibinfo {pages} {293} (\bibinfo {year} {1998})}\BibitemShut
  {NoStop}%
\bibitem [{\citenamefont {Hu}\ \emph {et~al.}(2012)\citenamefont {Hu},
  \citenamefont {Pan}, \citenamefont {Cheng}, \citenamefont {Sun},
  \citenamefont {Li}, \citenamefont {Wang}, \citenamefont {Campargue},\ and\
  \citenamefont {Liu}}]{12HuPaCh.H2}%
  \BibitemOpen
  \bibfield  {author} {\bibinfo {author} {\bibfnamefont {S.-M.}\ \bibnamefont
  {Hu}}, \bibinfo {author} {\bibfnamefont {H.}~\bibnamefont {Pan}}, \bibinfo
  {author} {\bibfnamefont {C.-F.}\ \bibnamefont {Cheng}}, \bibinfo {author}
  {\bibfnamefont {Y.~R.}\ \bibnamefont {Sun}}, \bibinfo {author} {\bibfnamefont
  {X.-F.}\ \bibnamefont {Li}}, \bibinfo {author} {\bibfnamefont
  {J.}~\bibnamefont {Wang}}, \bibinfo {author} {\bibfnamefont {A.}~\bibnamefont
  {Campargue}}, \ and\ \bibinfo {author} {\bibfnamefont {A.-W.}\ \bibnamefont
  {Liu}},\ }\href {\doibase 10.1088/0004-637x/749/1/76} {\bibfield  {journal}
  {\bibinfo  {journal} {Astrophys. J.}\ }\textbf {\bibinfo {volume} {749}},\
  \bibinfo {pages} {76} (\bibinfo {year} {2012})}\BibitemShut {NoStop}%
\bibitem [{\citenamefont {Roueff}\ \emph {et~al.}(2019)\citenamefont {Roueff},
  \citenamefont {Abgrall}, \citenamefont {Czachorowski}, \citenamefont
  {Pachucki}, \citenamefont {Puchalski},\ and\ \citenamefont
  {Komasa}}]{19RoAbCz.H2}%
  \BibitemOpen
  \bibfield  {author} {\bibinfo {author} {\bibfnamefont {E.}~\bibnamefont
  {Roueff}}, \bibinfo {author} {\bibfnamefont {H.}~\bibnamefont {Abgrall}},
  \bibinfo {author} {\bibfnamefont {P.}~\bibnamefont {Czachorowski}}, \bibinfo
  {author} {\bibfnamefont {K.}~\bibnamefont {Pachucki}}, \bibinfo {author}
  {\bibfnamefont {M.}~\bibnamefont {Puchalski}}, \ and\ \bibinfo {author}
  {\bibfnamefont {J.}~\bibnamefont {Komasa}},\ }\href {\doibase
  10.1051/0004-6361/201936249} {\bibfield  {journal} {\bibinfo  {journal}
  {Astron. Astrophys.}\ }\textbf {\bibinfo {volume} {630}},\ \bibinfo {pages}
  {A58} (\bibinfo {year} {2019})}\BibitemShut {NoStop}%
\bibitem [{\citenamefont {Setzer}, \citenamefont {Kalb},\ and\ \citenamefont
  {Fink}(2003)}]{03SeKaFi.S2}%
  \BibitemOpen
  \bibfield  {author} {\bibinfo {author} {\bibfnamefont {K.~D.}\ \bibnamefont
  {Setzer}}, \bibinfo {author} {\bibfnamefont {M.}~\bibnamefont {Kalb}}, \ and\
  \bibinfo {author} {\bibfnamefont {E.~H.}\ \bibnamefont {Fink}},\ }\href
  {\doibase 10.1016/S0022-2852(03)00174-7} {\bibfield  {journal} {\bibinfo
  {journal} {J. Mol. Spectrosc.}\ }\textbf {\bibinfo {volume} {221}},\ \bibinfo
  {pages} {127} (\bibinfo {year} {2003})}\BibitemShut {NoStop}%
\bibitem [{\citenamefont {Augustovičová}\ \emph {et~al.}(2014)\citenamefont
  {Augustovičová}, \citenamefont {Kraemer}, \citenamefont {Špirko},\ and\
  \citenamefont {Soldán}}]{14AuKrSp.He2+}%
  \BibitemOpen
  \bibfield  {author} {\bibinfo {author} {\bibfnamefont {L.}~\bibnamefont
  {Augustovičová}}, \bibinfo {author} {\bibfnamefont {W.~P.}\ \bibnamefont
  {Kraemer}}, \bibinfo {author} {\bibfnamefont {V.}~\bibnamefont {Špirko}}, \
  and\ \bibinfo {author} {\bibfnamefont {P.}~\bibnamefont {Soldán}},\ }\href
  {\doibase 10.1093/mnras/stu2317} {\bibfield  {journal} {\bibinfo  {journal}
  {Monthly Notices of the Royal Astronomical Society}\ }\textbf {\bibinfo
  {volume} {446}},\ \bibinfo {pages} {2738} (\bibinfo {year}
  {2014})}\BibitemShut {NoStop}%
\bibitem [{\citenamefont {Liu}\ \emph {et~al.}(2014)\citenamefont {Liu},
  \citenamefont {Shi}, \citenamefont {Sun}, \citenamefont {Zhu},\ and\
  \citenamefont {Shulin}}]{14LiShSu.O2}%
  \BibitemOpen
  \bibfield  {author} {\bibinfo {author} {\bibfnamefont {H.}~\bibnamefont
  {Liu}}, \bibinfo {author} {\bibfnamefont {D.}~\bibnamefont {Shi}}, \bibinfo
  {author} {\bibfnamefont {J.}~\bibnamefont {Sun}}, \bibinfo {author}
  {\bibfnamefont {Z.}~\bibnamefont {Zhu}}, \ and\ \bibinfo {author}
  {\bibfnamefont {Z.}~\bibnamefont {Shulin}},\ }\href {\doibase
  10.1016/j.saa.2014.01.003} {\bibfield  {journal} {\bibinfo  {journal}
  {Spectra Chimica Acta A}\ }\textbf {\bibinfo {volume} {124}},\ \bibinfo
  {pages} {216} (\bibinfo {year} {2014})}\BibitemShut {NoStop}%
\bibitem [{\citenamefont {Yu}\ \emph {et~al.}(2012)\citenamefont {Yu},
  \citenamefont {Miller}, \citenamefont {Drouin},\ and\ \citenamefont
  {Müller}}]{12YuMiDr.O2}%
  \BibitemOpen
  \bibfield  {author} {\bibinfo {author} {\bibfnamefont {S.}~\bibnamefont
  {Yu}}, \bibinfo {author} {\bibfnamefont {C.~E.}\ \bibnamefont {Miller}},
  \bibinfo {author} {\bibfnamefont {B.~J.}\ \bibnamefont {Drouin}}, \ and\
  \bibinfo {author} {\bibfnamefont {H.~S.~P.}\ \bibnamefont {Müller}},\ }\href
  {\doibase 10.1063/1.4719170} {\bibfield  {journal} {\bibinfo  {journal} {J.
  Chem. Phys.}\ }\textbf {\bibinfo {volume} {137}},\ \bibinfo {pages} {024304}
  (\bibinfo {year} {2012})}\BibitemShut {NoStop}%
\bibitem [{\citenamefont {Tilford}\ and\ \citenamefont
  {Simmons}(1966)}]{66TiSixx.CO}%
  \BibitemOpen
  \bibfield  {author} {\bibinfo {author} {\bibfnamefont {S.~G.}\ \bibnamefont
  {Tilford}}\ and\ \bibinfo {author} {\bibfnamefont {J.~D.}\ \bibnamefont
  {Simmons}},\ }\href {\doibase 10.1063/1.1726596} {\bibfield  {journal}
  {\bibinfo  {journal} {J. Chem. Phys.}\ }\textbf {\bibinfo {volume} {44}},\
  \bibinfo {pages} {4145} (\bibinfo {year} {1966})}\BibitemShut {NoStop}%
\bibitem [{\citenamefont {Tilford}\ and\ \citenamefont
  {Simmons}(1972)}]{72TiSixx.CO}%
  \BibitemOpen
  \bibfield  {author} {\bibinfo {author} {\bibfnamefont {S.~G.}\ \bibnamefont
  {Tilford}}\ and\ \bibinfo {author} {\bibfnamefont {J.~D.}\ \bibnamefont
  {Simmons}},\ }\href {\doibase 10.1063/1.3253097} {\bibfield  {journal}
  {\bibinfo  {journal} {J. Phys. Chem. Ref. Data}\ }\textbf {\bibinfo {volume}
  {1}},\ \bibinfo {pages} {147} (\bibinfo {year} {1972})}\BibitemShut {NoStop}%
\bibitem [{\citenamefont {Campargue}\ \emph
  {et~al.}(2020{\natexlab{a}})\citenamefont {Campargue}, \citenamefont {Kassi},
  \citenamefont {Yachmenev}, \citenamefont {Kyuberis}, \citenamefont
  {K\"upper},\ and\ \citenamefont {Yurchenko}}]{20CaKaYa.H2O}%
  \BibitemOpen
  \bibfield  {author} {\bibinfo {author} {\bibfnamefont {A.}~\bibnamefont
  {Campargue}}, \bibinfo {author} {\bibfnamefont {S.}~\bibnamefont {Kassi}},
  \bibinfo {author} {\bibfnamefont {A.}~\bibnamefont {Yachmenev}}, \bibinfo
  {author} {\bibfnamefont {A.~A.}\ \bibnamefont {Kyuberis}}, \bibinfo {author}
  {\bibfnamefont {J.}~\bibnamefont {K\"upper}}, \ and\ \bibinfo {author}
  {\bibfnamefont {S.~N.}\ \bibnamefont {Yurchenko}},\ }\href {\doibase
  10.1103/PhysRevResearch.2.023091} {\bibfield  {journal} {\bibinfo  {journal}
  {Phys. Rev. Research}\ }\textbf {\bibinfo {volume} {2}},\ \bibinfo {pages}
  {023091} (\bibinfo {year} {2020}{\natexlab{a}})}\BibitemShut {NoStop}%
\bibitem [{\citenamefont {Campargue}\ \emph
  {et~al.}(2020{\natexlab{b}})\citenamefont {Campargue}, \citenamefont
  {Solodov}, \citenamefont {Solodov}, \citenamefont {Yachmenev},\ and\
  \citenamefont {Yurchenko}}]{20CaSoSo.H2O}%
  \BibitemOpen
  \bibfield  {author} {\bibinfo {author} {\bibfnamefont {A.}~\bibnamefont
  {Campargue}}, \bibinfo {author} {\bibfnamefont {A.~M.}\ \bibnamefont
  {Solodov}}, \bibinfo {author} {\bibfnamefont {A.~A.}\ \bibnamefont
  {Solodov}}, \bibinfo {author} {\bibfnamefont {A.}~\bibnamefont {Yachmenev}},
  \ and\ \bibinfo {author} {\bibfnamefont {S.~N.}\ \bibnamefont {Yurchenko}},\
  }\href {\doibase 10.1039/D0CP01667E} {\bibfield  {journal} {\bibinfo
  {journal} {Phys. Chem. Chem. Phys.}\ }\textbf {\bibinfo {volume} {22}},\
  \bibinfo {pages} {12476} (\bibinfo {year} {2020}{\natexlab{b}})}\BibitemShut
  {NoStop}%
\bibitem [{\citenamefont {Krupenie}(1966)}]{66Krupenie.CO}%
  \BibitemOpen
  \bibfield  {author} {\bibinfo {author} {\bibfnamefont {P.~H.}\ \bibnamefont
  {Krupenie}},\ }\href@noop {} {\emph {\bibinfo {title} {{The band spectrum of
  carbon monoxide}}}}\ (\bibinfo  {publisher} {U.S. Department of Commerce,
  National Bureau of Standards},\ \bibinfo {year} {1966})\BibitemShut {NoStop}%
\bibitem [{\citenamefont {Glinski}\ \emph {et~al.}(1996)\citenamefont
  {Glinski}, \citenamefont {Nuth}, \citenamefont {Reese},\ and\ \citenamefont
  {Sitko}}]{96GlNuRe.CO}%
  \BibitemOpen
  \bibfield  {author} {\bibinfo {author} {\bibfnamefont {R.~J.}\ \bibnamefont
  {Glinski}}, \bibinfo {author} {\bibfnamefont {J.~A.}\ \bibnamefont {Nuth}},
  \bibinfo {author} {\bibfnamefont {M.~D.}\ \bibnamefont {Reese}}, \ and\
  \bibinfo {author} {\bibfnamefont {M.~L.}\ \bibnamefont {Sitko}},\ }\href
  {\doibase 10.1086/310205} {\bibfield  {journal} {\bibinfo  {journal}
  {Astrophys. J.}\ }\textbf {\bibinfo {volume} {467}},\ \bibinfo {pages} {L109}
  (\bibinfo {year} {1996})}\BibitemShut {NoStop}%
\bibitem [{\citenamefont {Bunker}\ and\ \citenamefont
  {Jensen}(2006)}]{06BuJexx.method}%
  \BibitemOpen
  \bibfield  {author} {\bibinfo {author} {\bibfnamefont {P.~R.}\ \bibnamefont
  {Bunker}}\ and\ \bibinfo {author} {\bibfnamefont {P.}~\bibnamefont
  {Jensen}},\ }\href@noop {} {\emph {\bibinfo {title} {{Molecular Symmetry and
  Spectroscopy, Second Edition}}}}\ (\bibinfo  {publisher} {NRC Research
  Press},\ \bibinfo {address} {Ottawa, Canada},\ \bibinfo {year}
  {2006})\BibitemShut {NoStop}%
\bibitem [{\citenamefont {Goldman}\ \emph {et~al.}(1995)\citenamefont
  {Goldman}, \citenamefont {Rinsland}, \citenamefont {Canova}, \citenamefont
  {Zander},\ and\ \citenamefont {Dangnhu}}]{95GoRiCa.O2}%
  \BibitemOpen
  \bibfield  {author} {\bibinfo {author} {\bibfnamefont {A.}~\bibnamefont
  {Goldman}}, \bibinfo {author} {\bibfnamefont {C.~P.}\ \bibnamefont
  {Rinsland}}, \bibinfo {author} {\bibfnamefont {B.}~\bibnamefont {Canova}},
  \bibinfo {author} {\bibfnamefont {R.}~\bibnamefont {Zander}}, \ and\ \bibinfo
  {author} {\bibfnamefont {M.}~\bibnamefont {Dangnhu}},\ }\href {\doibase
  10.1016/0022-4073(95)00114-Z} {\bibfield  {journal} {\bibinfo  {journal} {J.
  Quant. Spectrosc. Radiat. Transf.}\ }\textbf {\bibinfo {volume} {54}},\
  \bibinfo {pages} {757} (\bibinfo {year} {1995})}\BibitemShut {NoStop}%
\bibitem [{\citenamefont {Domys{\l}awska}\ \emph {et~al.}(2016)\citenamefont
  {Domys{\l}awska}, \citenamefont {W{\'{o}}jtewicz}, \citenamefont
  {Mas{\l}owski}, \citenamefont {Cygan}, \citenamefont {Bielska}, \citenamefont
  {Trawi{\'{n}}ski}, \citenamefont {Ciury{\l}o},\ and\ \citenamefont
  {Lisak}}]{16DoWoMa.O2}%
  \BibitemOpen
  \bibfield  {author} {\bibinfo {author} {\bibfnamefont {J.}~\bibnamefont
  {Domys{\l}awska}}, \bibinfo {author} {\bibfnamefont {S.}~\bibnamefont
  {W{\'{o}}jtewicz}}, \bibinfo {author} {\bibfnamefont {P.}~\bibnamefont
  {Mas{\l}owski}}, \bibinfo {author} {\bibfnamefont {A.}~\bibnamefont {Cygan}},
  \bibinfo {author} {\bibfnamefont {K.}~\bibnamefont {Bielska}}, \bibinfo
  {author} {\bibfnamefont {R.~S.}\ \bibnamefont {Trawi{\'{n}}ski}}, \bibinfo
  {author} {\bibfnamefont {R.}~\bibnamefont {Ciury{\l}o}}, \ and\ \bibinfo
  {author} {\bibfnamefont {D.}~\bibnamefont {Lisak}},\ }\href {\doibase
  10.1016/j.jqsrt.2015.10.019} {\bibfield  {journal} {\bibinfo  {journal} {J.
  Quant. Spectrosc. Radiat. Transf.}\ }\textbf {\bibinfo {volume} {169}},\
  \bibinfo {pages} {111} (\bibinfo {year} {2016})}\BibitemShut {NoStop}%
\bibitem [{\citenamefont {Gordon}\ and\ \citenamefont {{et
  al.}}(2017)}]{HITRAN2016}%
  \BibitemOpen
  \bibfield  {author} {\bibinfo {author} {\bibfnamefont {I.~E.}\ \bibnamefont
  {Gordon}}\ and\ \bibinfo {author} {\bibnamefont {{et al.}}},\ }\href
  {\doibase 10.1016/j.jqsrt.2017.06.038} {\bibfield  {journal} {\bibinfo
  {journal} {J. Quant. Spectrosc. Radiat. Transf.}\ }\textbf {\bibinfo {volume}
  {203}},\ \bibinfo {pages} {3} (\bibinfo {year} {2017})}\BibitemShut {NoStop}%
\bibitem [{\citenamefont {Mishra}, \citenamefont {Balasubramanian},\ and\
  \citenamefont {Shetty}(2011)}]{11MiBaSh.O2}%
  \BibitemOpen
  \bibfield  {author} {\bibinfo {author} {\bibfnamefont {A.~P.}\ \bibnamefont
  {Mishra}}, \bibinfo {author} {\bibfnamefont {T.~K.}\ \bibnamefont
  {Balasubramanian}}, \ and\ \bibinfo {author} {\bibfnamefont {B.~J.}\
  \bibnamefont {Shetty}},\ }\href {\doibase 10.1016/j.jqsrt.2011.05.013}
  {\bibfield  {journal} {\bibinfo  {journal} {J. Quant. Spectrosc. Radiat.
  Transf.}\ }\textbf {\bibinfo {volume} {112}},\ \bibinfo {pages} {2303}
  (\bibinfo {year} {2011})}\BibitemShut {NoStop}%
\bibitem [{\citenamefont {Chiu}(1965)}]{65Chiu.quadrupole}%
  \BibitemOpen
  \bibfield  {author} {\bibinfo {author} {\bibfnamefont {Y.-N.}\ \bibnamefont
  {Chiu}},\ }\href {\doibase 10.1063/1.1703221} {\bibfield  {journal} {\bibinfo
   {journal} {J. Chem. Phys.}\ }\textbf {\bibinfo {volume} {42}},\ \bibinfo
  {pages} {2671} (\bibinfo {year} {1965})}\BibitemShut {NoStop}%
\bibitem [{\citenamefont {Balasubramanian}, \citenamefont {D'Cunha},\ and\
  \citenamefont {Rao}(1990)}]{90BaDCNa.O2}%
  \BibitemOpen
  \bibfield  {author} {\bibinfo {author} {\bibfnamefont {T.}~\bibnamefont
  {Balasubramanian}}, \bibinfo {author} {\bibfnamefont {R.}~\bibnamefont
  {D'Cunha}}, \ and\ \bibinfo {author} {\bibfnamefont {K.}~\bibnamefont
  {Rao}},\ }\href {\doibase 10.1016/0022-2852(90)90224-E} {\bibfield  {journal}
  {\bibinfo  {journal} {J. Mol. Spectrosc.}\ }\textbf {\bibinfo {volume}
  {144}},\ \bibinfo {pages} {374} (\bibinfo {year} {1990})}\BibitemShut
  {NoStop}%
\bibitem [{\citenamefont {Balasubramanian}\ and\ \citenamefont
  {Narayanan}(1994)}]{94BaNaxx.O2}%
  \BibitemOpen
  \bibfield  {author} {\bibinfo {author} {\bibfnamefont {T.~K.}\ \bibnamefont
  {Balasubramanian}}\ and\ \bibinfo {author} {\bibfnamefont {O.}~\bibnamefont
  {Narayanan}},\ }\href {\doibase 10.1007/BF03156406} {\bibfield  {journal}
  {\bibinfo  {journal} {Acta Physica Hungarica}\ }\textbf {\bibinfo {volume}
  {74}},\ \bibinfo {pages} {341} (\bibinfo {year} {1994})}\BibitemShut
  {NoStop}%
\bibitem [{\citenamefont {Tennyson}\ \emph {et~al.}(2020)\citenamefont
  {Tennyson}, \citenamefont {Yurchenko}, \citenamefont {Al-Refaie},
  \citenamefont {Clark}, \citenamefont {Chubb}, \citenamefont {Conway},
  \citenamefont {Dewan}, \citenamefont {Gorman}, \citenamefont {Hill},
  \citenamefont {Lynas-Gray}, \citenamefont {Mellor}, \citenamefont
  {McKemmish}, \citenamefont {Owens}, \citenamefont {Polyansky}, \citenamefont
  {Semenov}, \citenamefont {Somogyi}, \citenamefont {Tinetti}, \citenamefont
  {Upadhyay}, \citenamefont {Waldmann}, \citenamefont {Wang}, \citenamefont
  {Wright},\ and\ \citenamefont {Yurchenko}}]{20TeYuAl}%
  \BibitemOpen
  \bibfield  {author} {\bibinfo {author} {\bibfnamefont {J.}~\bibnamefont
  {Tennyson}}, \bibinfo {author} {\bibfnamefont {S.~N.}\ \bibnamefont
  {Yurchenko}}, \bibinfo {author} {\bibfnamefont {A.~F.}\ \bibnamefont
  {Al-Refaie}}, \bibinfo {author} {\bibfnamefont {V.~H.}\ \bibnamefont
  {Clark}}, \bibinfo {author} {\bibfnamefont {K.~L.}\ \bibnamefont {Chubb}},
  \bibinfo {author} {\bibfnamefont {E.~K.}\ \bibnamefont {Conway}}, \bibinfo
  {author} {\bibfnamefont {A.}~\bibnamefont {Dewan}}, \bibinfo {author}
  {\bibfnamefont {M.~N.}\ \bibnamefont {Gorman}}, \bibinfo {author}
  {\bibfnamefont {C.}~\bibnamefont {Hill}}, \bibinfo {author} {\bibfnamefont
  {A.}~\bibnamefont {Lynas-Gray}}, \bibinfo {author} {\bibfnamefont
  {T.}~\bibnamefont {Mellor}}, \bibinfo {author} {\bibfnamefont {L.~K.}\
  \bibnamefont {McKemmish}}, \bibinfo {author} {\bibfnamefont {A.}~\bibnamefont
  {Owens}}, \bibinfo {author} {\bibfnamefont {O.~L.}\ \bibnamefont
  {Polyansky}}, \bibinfo {author} {\bibfnamefont {M.}~\bibnamefont {Semenov}},
  \bibinfo {author} {\bibfnamefont {W.}~\bibnamefont {Somogyi}}, \bibinfo
  {author} {\bibfnamefont {G.}~\bibnamefont {Tinetti}}, \bibinfo {author}
  {\bibfnamefont {A.}~\bibnamefont {Upadhyay}}, \bibinfo {author}
  {\bibfnamefont {I.}~\bibnamefont {Waldmann}}, \bibinfo {author}
  {\bibfnamefont {Y.}~\bibnamefont {Wang}}, \bibinfo {author} {\bibfnamefont
  {S.}~\bibnamefont {Wright}}, \ and\ \bibinfo {author} {\bibfnamefont {O.~P.}\
  \bibnamefont {Yurchenko}},\ }\href {\doibase 10.1016/j.jqsrt.2020.107228}
  {\bibfield  {journal} {\bibinfo  {journal} {J. Quant. Spectrosc. Radiat.
  Transf.}\ }\textbf {\bibinfo {volume} {255}},\ \bibinfo {pages} {107228}
  (\bibinfo {year} {2020})}\BibitemShut {NoStop}%
\bibitem [{\citenamefont {Lammer}\ \emph {et~al.}(2019)\citenamefont {Lammer},
  \citenamefont {Sproß}, \citenamefont {Grenfell}, \citenamefont {Scherf},
  \citenamefont {Fossati}, \citenamefont {Lendl},\ and\ \citenamefont
  {Cubillos}}]{19LaSpGr.N2}%
  \BibitemOpen
  \bibfield  {author} {\bibinfo {author} {\bibfnamefont {H.}~\bibnamefont
  {Lammer}}, \bibinfo {author} {\bibfnamefont {L.}~\bibnamefont {Sproß}},
  \bibinfo {author} {\bibfnamefont {J.~L.}\ \bibnamefont {Grenfell}}, \bibinfo
  {author} {\bibfnamefont {M.}~\bibnamefont {Scherf}}, \bibinfo {author}
  {\bibfnamefont {L.}~\bibnamefont {Fossati}}, \bibinfo {author} {\bibfnamefont
  {M.}~\bibnamefont {Lendl}}, \ and\ \bibinfo {author} {\bibfnamefont {P.~E.}\
  \bibnamefont {Cubillos}},\ }\href {\doibase 10.1089/ast.2018.1914} {\bibfield
   {journal} {\bibinfo  {journal} {Astrobiology}\ }\textbf {\bibinfo {volume}
  {19}},\ \bibinfo {pages} {927} (\bibinfo {year} {2019})}\BibitemShut
  {NoStop}%
\bibitem [{\citenamefont {Meadows}\ \emph {et~al.}(2018)\citenamefont
  {Meadows}, \citenamefont {Reinhard}, \citenamefont {Arney}, \citenamefont
  {Parenteau}, \citenamefont {Schwieterman}, \citenamefont {Domagal-Goldman},
  \citenamefont {Lincowski}, \citenamefont {Stapelfeldt}, \citenamefont
  {Rauer}, \citenamefont {DasSarma}, \citenamefont {Hegde}, \citenamefont
  {Narita}, \citenamefont {Deitrick}, \citenamefont {Lustig-Yaeger},
  \citenamefont {Lyons}, \citenamefont {Siegler},\ and\ \citenamefont
  {Grenfell}}]{18MeReAr.O2}%
  \BibitemOpen
  \bibfield  {author} {\bibinfo {author} {\bibfnamefont {V.~S.}\ \bibnamefont
  {Meadows}}, \bibinfo {author} {\bibfnamefont {C.~T.}\ \bibnamefont
  {Reinhard}}, \bibinfo {author} {\bibfnamefont {G.~N.}\ \bibnamefont {Arney}},
  \bibinfo {author} {\bibfnamefont {M.~N.}\ \bibnamefont {Parenteau}}, \bibinfo
  {author} {\bibfnamefont {E.~W.}\ \bibnamefont {Schwieterman}}, \bibinfo
  {author} {\bibfnamefont {S.~D.}\ \bibnamefont {Domagal-Goldman}}, \bibinfo
  {author} {\bibfnamefont {A.~P.}\ \bibnamefont {Lincowski}}, \bibinfo {author}
  {\bibfnamefont {K.~R.}\ \bibnamefont {Stapelfeldt}}, \bibinfo {author}
  {\bibfnamefont {H.}~\bibnamefont {Rauer}}, \bibinfo {author} {\bibfnamefont
  {S.}~\bibnamefont {DasSarma}}, \bibinfo {author} {\bibfnamefont
  {S.}~\bibnamefont {Hegde}}, \bibinfo {author} {\bibfnamefont
  {N.}~\bibnamefont {Narita}}, \bibinfo {author} {\bibfnamefont
  {R.}~\bibnamefont {Deitrick}}, \bibinfo {author} {\bibfnamefont
  {J.}~\bibnamefont {Lustig-Yaeger}}, \bibinfo {author} {\bibfnamefont {T.~W.}\
  \bibnamefont {Lyons}}, \bibinfo {author} {\bibfnamefont {N.}~\bibnamefont
  {Siegler}}, \ and\ \bibinfo {author} {\bibfnamefont {J.~L.}\ \bibnamefont
  {Grenfell}},\ }\href {\doibase 10.1089/ast.2017.1727} {\bibfield  {journal}
  {\bibinfo  {journal} {Astrobiology}\ }\textbf {\bibinfo {volume} {18}},\
  \bibinfo {pages} {630} (\bibinfo {year} {2018})}\BibitemShut {NoStop}%
\bibitem [{\citenamefont {Doyle}\ \emph {et~al.}(2019)\citenamefont {Doyle},
  \citenamefont {Young}, \citenamefont {Klein}, \citenamefont {Zuckerman},\
  and\ \citenamefont {Schlichting}}]{19DoYoKl.O2}%
  \BibitemOpen
  \bibfield  {author} {\bibinfo {author} {\bibfnamefont {A.~E.}\ \bibnamefont
  {Doyle}}, \bibinfo {author} {\bibfnamefont {E.~D.}\ \bibnamefont {Young}},
  \bibinfo {author} {\bibfnamefont {B.}~\bibnamefont {Klein}}, \bibinfo
  {author} {\bibfnamefont {B.}~\bibnamefont {Zuckerman}}, \ and\ \bibinfo
  {author} {\bibfnamefont {H.~E.}\ \bibnamefont {Schlichting}},\ }\href
  {\doibase 10.1126/science.aax3901} {\bibfield  {journal} {\bibinfo  {journal}
  {Science}\ }\textbf {\bibinfo {volume} {366}},\ \bibinfo {pages} {356}
  (\bibinfo {year} {2019})}\BibitemShut {NoStop}%
\bibitem [{\citenamefont {Schaefer}\ \emph {et~al.}(2016)\citenamefont
  {Schaefer}, \citenamefont {Wordsworth}, \citenamefont {Berta-Thompson},\ and\
  \citenamefont {Sasselov}}]{16ScWoBe.O2}%
  \BibitemOpen
  \bibfield  {author} {\bibinfo {author} {\bibfnamefont {L.}~\bibnamefont
  {Schaefer}}, \bibinfo {author} {\bibfnamefont {R.~D.}\ \bibnamefont
  {Wordsworth}}, \bibinfo {author} {\bibfnamefont {Z.}~\bibnamefont
  {Berta-Thompson}}, \ and\ \bibinfo {author} {\bibfnamefont {D.}~\bibnamefont
  {Sasselov}},\ }\href {\doibase 10.3847/0004-637x/829/2/63} {\bibfield
  {journal} {\bibinfo  {journal} {Astrophys. J.}\ }\textbf {\bibinfo {volume}
  {829}},\ \bibinfo {pages} {63} (\bibinfo {year} {2016})}\BibitemShut
  {NoStop}%
\bibitem [{\citenamefont {Kite}\ and\ \citenamefont
  {Schaefer}(2021)}]{21KiScxx.O2}%
  \BibitemOpen
  \bibfield  {author} {\bibinfo {author} {\bibfnamefont {E.~S.}\ \bibnamefont
  {Kite}}\ and\ \bibinfo {author} {\bibfnamefont {L.}~\bibnamefont
  {Schaefer}},\ }\href {\doibase 10.3847/2041-8213/abe7dc} {\bibfield
  {journal} {\bibinfo  {journal} {Astrophys. J. Lett.}\ }\textbf {\bibinfo
  {volume} {909}},\ \bibinfo {pages} {L22} (\bibinfo {year}
  {2021})}\BibitemShut {NoStop}%
\bibitem [{\citenamefont {Lisse}\ \emph {et~al.}(2020)\citenamefont {Lisse},
  \citenamefont {Desch}, \citenamefont {Unterborn}, \citenamefont {Kane},
  \citenamefont {Young}, \citenamefont {Hartnett}, \citenamefont {Hinkel},
  \citenamefont {Shim}, \citenamefont {Mamajek},\ and\ \citenamefont
  {Izenberg}}]{20LiDeUn.O2}%
  \BibitemOpen
  \bibfield  {author} {\bibinfo {author} {\bibfnamefont {C.~M.}\ \bibnamefont
  {Lisse}}, \bibinfo {author} {\bibfnamefont {S.~J.}\ \bibnamefont {Desch}},
  \bibinfo {author} {\bibfnamefont {C.~T.}\ \bibnamefont {Unterborn}}, \bibinfo
  {author} {\bibfnamefont {S.~R.}\ \bibnamefont {Kane}}, \bibinfo {author}
  {\bibfnamefont {P.~R.}\ \bibnamefont {Young}}, \bibinfo {author}
  {\bibfnamefont {H.~E.}\ \bibnamefont {Hartnett}}, \bibinfo {author}
  {\bibfnamefont {N.~R.}\ \bibnamefont {Hinkel}}, \bibinfo {author}
  {\bibfnamefont {S.-H.}\ \bibnamefont {Shim}}, \bibinfo {author}
  {\bibfnamefont {E.~E.}\ \bibnamefont {Mamajek}}, \ and\ \bibinfo {author}
  {\bibfnamefont {N.~R.}\ \bibnamefont {Izenberg}},\ }\href {\doibase
  10.3847/2041-8213/ab9b91} {\bibfield  {journal} {\bibinfo  {journal}
  {Astrophys. J.}\ }\textbf {\bibinfo {volume} {898}},\ \bibinfo {pages} {L17}
  (\bibinfo {year} {2020})}\BibitemShut {NoStop}%
\bibitem [{\citenamefont {Yurchenko}\ \emph {et~al.}(2016)\citenamefont
  {Yurchenko}, \citenamefont {Lodi}, \citenamefont {Tennyson},\ and\
  \citenamefont {Stolyarov}}]{Duo}%
  \BibitemOpen
  \bibfield  {author} {\bibinfo {author} {\bibfnamefont {S.~N.}\ \bibnamefont
  {Yurchenko}}, \bibinfo {author} {\bibfnamefont {L.}~\bibnamefont {Lodi}},
  \bibinfo {author} {\bibfnamefont {J.}~\bibnamefont {Tennyson}}, \ and\
  \bibinfo {author} {\bibfnamefont {A.~V.}\ \bibnamefont {Stolyarov}},\ }\href
  {\doibase 10.1016/j.cpc.2015.12.021} {\bibfield  {journal} {\bibinfo
  {journal} {Comput. Phys. Commun.}\ }\textbf {\bibinfo {volume} {202}},\
  \bibinfo {pages} {262 } (\bibinfo {year} {2016})}\BibitemShut {NoStop}%
\bibitem [{\citenamefont {Kato}(1993)}]{93Kato.methods}%
  \BibitemOpen
  \bibfield  {author} {\bibinfo {author} {\bibfnamefont {H.}~\bibnamefont
  {Kato}},\ }\href {\doibase {10.1246/bcsj.66.3203}} {\bibfield  {journal}
  {\bibinfo  {journal} {Bull. Chem. Soc. Japan}\ }\textbf {\bibinfo {volume}
  {{66}}},\ \bibinfo {pages} {3203} (\bibinfo {year} {{1993}})}\BibitemShut
  {NoStop}%
\bibitem [{\citenamefont {Buckingham}(1959)}]{59Buckin.quadrupole}%
  \BibitemOpen
  \bibfield  {author} {\bibinfo {author} {\bibfnamefont {A.}~\bibnamefont
  {Buckingham}},\ }\href {\doibase 10.1039/qr9591300183} {\bibfield  {journal}
  {\bibinfo  {journal} {Quarterly review (London, England)}\ }\textbf {\bibinfo
  {volume} {13}},\ \bibinfo {pages} {183} (\bibinfo {year} {1959})}\BibitemShut
  {NoStop}%
\bibitem [{\citenamefont {Werner}\ \emph {et~al.}(2012)\citenamefont {Werner},
  \citenamefont {Knowles}, \citenamefont {Knizia}, \citenamefont {Manby},\ and\
  \citenamefont {Sch\"utz}}]{MOLPRO}%
  \BibitemOpen
  \bibfield  {author} {\bibinfo {author} {\bibfnamefont {H.-J.}\ \bibnamefont
  {Werner}}, \bibinfo {author} {\bibfnamefont {P.~J.}\ \bibnamefont {Knowles}},
  \bibinfo {author} {\bibfnamefont {G.}~\bibnamefont {Knizia}}, \bibinfo
  {author} {\bibfnamefont {F.~R.}\ \bibnamefont {Manby}}, \ and\ \bibinfo
  {author} {\bibfnamefont {M.}~\bibnamefont {Sch\"utz}},\ }\href {\doibase
  10.1002/wcms.82} {\bibfield  {journal} {\bibinfo  {journal} {WIREs Comput.
  Mol. Sci.}\ }\textbf {\bibinfo {volume} {2}},\ \bibinfo {pages} {242}
  (\bibinfo {year} {2012})}\BibitemShut {NoStop}%
\bibitem [{\citenamefont {Truhlar}(1972{\natexlab{a}})}]{72Truhlar.CO}%
  \BibitemOpen
  \bibfield  {author} {\bibinfo {author} {\bibfnamefont {D.}~\bibnamefont
  {Truhlar}},\ }\href {\doibase 10.1002/qua.560060515} {\bibfield  {journal}
  {\bibinfo  {journal} {Intern. J. Quantum Chem.}\ }\textbf {\bibinfo {volume}
  {6}},\ \bibinfo {pages} {975} (\bibinfo {year}
  {1972}{\natexlab{a}})}\BibitemShut {NoStop}%
\bibitem [{\citenamefont {Long}(2002)}]{02Longxx.quadrupole}%
  \BibitemOpen
  \bibfield  {author} {\bibinfo {author} {\bibfnamefont {D.~A.}\ \bibnamefont
  {Long}},\ }\href@noop {} {\emph {\bibinfo {title} {The {Raman} effect: {A}
  unified treatment of the theory of {Raman} scattering by molecules}}}\
  (\bibinfo  {publisher} {Wiley},\ \bibinfo {address} {Chichester ; New York},\
  \bibinfo {year} {2002})\BibitemShut {NoStop}%
\bibitem [{\citenamefont {Bunker}\ and\ \citenamefont
  {Jensen}(1998)}]{98BuJe.method}%
  \BibitemOpen
  \bibfield  {author} {\bibinfo {author} {\bibfnamefont {P.~R.}\ \bibnamefont
  {Bunker}}\ and\ \bibinfo {author} {\bibfnamefont {P.}~\bibnamefont
  {Jensen}},\ }\href@noop {} {\emph {\bibinfo {title} {Molecular Symmetry and
  Spectroscopy}}},\ \bibinfo {edition} {2nd}\ ed.\ (\bibinfo  {publisher} {NRC
  Research Press},\ \bibinfo {address} {Ottawa},\ \bibinfo {year}
  {1998})\BibitemShut {NoStop}%
\bibitem [{\citenamefont {Yachmenev}, \citenamefont {Thesing},\ and\
  \citenamefont {K\"{u}pper}(2019)}]{RichMol2}%
  \BibitemOpen
  \bibfield  {author} {\bibinfo {author} {\bibfnamefont {A.}~\bibnamefont
  {Yachmenev}}, \bibinfo {author} {\bibfnamefont {L.~V.}\ \bibnamefont
  {Thesing}}, \ and\ \bibinfo {author} {\bibfnamefont {J.}~\bibnamefont
  {K\"{u}pper}},\ }\href {\doibase 10.1063/1.5133837} {\bibfield  {journal}
  {\bibinfo  {journal} {J. Chem. Phys.}\ }\textbf {\bibinfo {volume} {151}},\
  \bibinfo {pages} {244118} (\bibinfo {year} {2019})}\BibitemShut {NoStop}%
\bibitem [{\citenamefont {Werner}\ \emph {et~al.}(2020)\citenamefont {Werner},
  \citenamefont {Knowles}, \citenamefont {Manby}, \citenamefont {Black},
  \citenamefont {Doll}, \citenamefont {Heßelmann}, \citenamefont {Kats},
  \citenamefont {Köhn}, \citenamefont {Korona}, \citenamefont {Kreplin},
  \citenamefont {Ma}, \citenamefont {Miller}, \citenamefont {Mitrushchenkov},
  \citenamefont {Peterson}, \citenamefont {Polyak}, \citenamefont {Rauhut},\
  and\ \citenamefont {Sibaev}}]{MOLPRO2020}%
  \BibitemOpen
  \bibfield  {author} {\bibinfo {author} {\bibfnamefont {H.-J.}\ \bibnamefont
  {Werner}}, \bibinfo {author} {\bibfnamefont {P.~J.}\ \bibnamefont {Knowles}},
  \bibinfo {author} {\bibfnamefont {F.~R.}\ \bibnamefont {Manby}}, \bibinfo
  {author} {\bibfnamefont {J.~A.}\ \bibnamefont {Black}}, \bibinfo {author}
  {\bibfnamefont {K.}~\bibnamefont {Doll}}, \bibinfo {author} {\bibfnamefont
  {A.}~\bibnamefont {Heßelmann}}, \bibinfo {author} {\bibfnamefont
  {D.}~\bibnamefont {Kats}}, \bibinfo {author} {\bibfnamefont {A.}~\bibnamefont
  {Köhn}}, \bibinfo {author} {\bibfnamefont {T.}~\bibnamefont {Korona}},
  \bibinfo {author} {\bibfnamefont {D.~A.}\ \bibnamefont {Kreplin}}, \bibinfo
  {author} {\bibfnamefont {Q.}~\bibnamefont {Ma}}, \bibinfo {author}
  {\bibfnamefont {T.~F.}\ \bibnamefont {Miller}}, \bibinfo {author}
  {\bibfnamefont {A.}~\bibnamefont {Mitrushchenkov}}, \bibinfo {author}
  {\bibfnamefont {K.~A.}\ \bibnamefont {Peterson}}, \bibinfo {author}
  {\bibfnamefont {I.}~\bibnamefont {Polyak}}, \bibinfo {author} {\bibfnamefont
  {G.}~\bibnamefont {Rauhut}}, \ and\ \bibinfo {author} {\bibfnamefont
  {M.}~\bibnamefont {Sibaev}},\ }\href {\doibase 10.1063/5.0005081} {\bibfield
  {journal} {\bibinfo  {journal} {The Journal of Chemical Physics}\ }\textbf
  {\bibinfo {volume} {152}},\ \bibinfo {pages} {144107} (\bibinfo {year}
  {2020})}\BibitemShut {NoStop}%
\bibitem [{\citenamefont {Stanton}\ \emph {et~al.}(2019)\citenamefont
  {Stanton}, \citenamefont {Gauss}, \citenamefont {Harding},\ and\
  \citenamefont {{et al.}}}]{CFOUR}%
  \BibitemOpen
  \bibfield  {author} {\bibinfo {author} {\bibfnamefont {J.}~\bibnamefont
  {Stanton}}, \bibinfo {author} {\bibfnamefont {J.}~\bibnamefont {Gauss}},
  \bibinfo {author} {\bibfnamefont {M.}~\bibnamefont {Harding}}, \ and\
  \bibinfo {author} {\bibnamefont {{et al.}}},\ }\href@noop {} {} (\bibinfo
  {year} {2019}),\ \bibinfo {note} {cFOUR, a quantum chemical program package
  written by J.F. Stanton, J. Gauss, M.E. Harding, P.G. Szalay with
  contributions from A.A. Auer, R.J. Bartlett, U. Benedikt, C. Berger, D.E.
  Bernholdt, Y.J. Bomble, L. Cheng, O. Christiansen, M. Heckert, O. Heun, C.
  Huber, T.-C. Jagau, D. Jonsson, J. Jus\'{e}lius, K. Klein, W.J. Lauderdale,
  D.A. Matthews, T. Metzroth, L.A. M\"{u}ck, D.P. O'Neill, D.R. Price, E.
  Prochnow, C. Puzzarini, K. Ruud, F. Schiffmann, W. Schwalbach, S. Stopkowicz,
  A. Tajti, J. V\'{a}zquez, F. Wang, J.D. Watts and the integral packages
  MOLECULE (J. Alml\"{o}f and P.R. Taylor), PROPS (P.R. Taylor), ABACUS (T.
  Helgaker, H.J. Aa. Jensen, P. J{\o}rgensen, and J. Olsen), and ECP routines
  by A. V. Mitin and C. van W\"{u}llen. For the current version, see
  http://www.cfour.de.}\BibitemShut {Stop}%
\bibitem [{\citenamefont {Johnson}\ \emph {et~al.}(1983)\citenamefont
  {Johnson}, \citenamefont {Goebel}, \citenamefont {Goorvitch},\ and\
  \citenamefont {Ridgway}}]{83JoGoGo.coolstars}%
  \BibitemOpen
  \bibfield  {author} {\bibinfo {author} {\bibfnamefont {H.~R.}\ \bibnamefont
  {Johnson}}, \bibinfo {author} {\bibfnamefont {J.~H.}\ \bibnamefont {Goebel}},
  \bibinfo {author} {\bibfnamefont {D.}~\bibnamefont {Goorvitch}}, \ and\
  \bibinfo {author} {\bibfnamefont {S.~T.}\ \bibnamefont {Ridgway}},\ }\href
  {\doibase 10.1086/184071} {\bibfield  {journal} {\bibinfo  {journal}
  {Astrophys. J. Lett.}\ }\textbf {\bibinfo {volume} {270}},\ \bibinfo {pages}
  {L63} (\bibinfo {year} {1983})}\BibitemShut {NoStop}%
\bibitem [{\citenamefont {Fraser}, \citenamefont {McCoustra},\ and\
  \citenamefont {Williams}(2002)}]{02FrMcWi.H2}%
  \BibitemOpen
  \bibfield  {author} {\bibinfo {author} {\bibfnamefont {H.~J.}\ \bibnamefont
  {Fraser}}, \bibinfo {author} {\bibfnamefont {M.~R.~S.}\ \bibnamefont
  {McCoustra}}, \ and\ \bibinfo {author} {\bibfnamefont {D.~A.}\ \bibnamefont
  {Williams}},\ }\href {\doibase 10.1046/j.1468-4004.2002.43210.x} {\bibfield
  {journal} {\bibinfo  {journal} {Astron. Geophys.}\ }\textbf {\bibinfo
  {volume} {43}},\ \bibinfo {pages} {2.10} (\bibinfo {year}
  {2002})}\BibitemShut {NoStop}%
\bibitem [{\citenamefont {Kreckel}\ \emph {et~al.}(2010)\citenamefont
  {Kreckel}, \citenamefont {Bruhns}, \citenamefont {Čížek}, \citenamefont
  {Glover}, \citenamefont {Miller}, \citenamefont {Urbain},\ and\ \citenamefont
  {Savin}}]{10KrBrCi.H2}%
  \BibitemOpen
  \bibfield  {author} {\bibinfo {author} {\bibfnamefont {H.}~\bibnamefont
  {Kreckel}}, \bibinfo {author} {\bibfnamefont {H.}~\bibnamefont {Bruhns}},
  \bibinfo {author} {\bibfnamefont {M.}~\bibnamefont {Čížek}}, \bibinfo
  {author} {\bibfnamefont {S.~C.~O.}\ \bibnamefont {Glover}}, \bibinfo {author}
  {\bibfnamefont {K.~A.}\ \bibnamefont {Miller}}, \bibinfo {author}
  {\bibfnamefont {X.}~\bibnamefont {Urbain}}, \ and\ \bibinfo {author}
  {\bibfnamefont {D.~W.}\ \bibnamefont {Savin}},\ }\href {\doibase
  10.1126/science.1187191} {\bibfield  {journal} {\bibinfo  {journal}
  {Science}\ }\textbf {\bibinfo {volume} {329}},\ \bibinfo {pages} {69}
  (\bibinfo {year} {2010})}\BibitemShut {NoStop}%
\bibitem [{\citenamefont {Nisini}\ \emph {et~al.}(2010)\citenamefont {Nisini},
  \citenamefont {Giannini}, \citenamefont {Neufeld}, \citenamefont {Yuan},
  \citenamefont {Antoniucci}, \citenamefont {Bergin},\ and\ \citenamefont
  {Melnick}}]{10NiGiNe.H2}%
  \BibitemOpen
  \bibfield  {author} {\bibinfo {author} {\bibfnamefont {B.}~\bibnamefont
  {Nisini}}, \bibinfo {author} {\bibfnamefont {T.}~\bibnamefont {Giannini}},
  \bibinfo {author} {\bibfnamefont {D.~A.}\ \bibnamefont {Neufeld}}, \bibinfo
  {author} {\bibfnamefont {Y.}~\bibnamefont {Yuan}}, \bibinfo {author}
  {\bibfnamefont {S.}~\bibnamefont {Antoniucci}}, \bibinfo {author}
  {\bibfnamefont {E.~A.}\ \bibnamefont {Bergin}}, \ and\ \bibinfo {author}
  {\bibfnamefont {G.~J.}\ \bibnamefont {Melnick}},\ }\href {\doibase
  10.1088/0004-637X/724/1/69} {\bibfield  {journal} {\bibinfo  {journal}
  {Astrophys. J.}\ }\textbf {\bibinfo {volume} {724}},\ \bibinfo {pages} {69}
  (\bibinfo {year} {2010})}\BibitemShut {NoStop}%
\bibitem [{\citenamefont {Islam}\ \emph {et~al.}(2010)\citenamefont {Islam},
  \citenamefont {Cecchi-Pestellini}, \citenamefont {Viti},\ and\ \citenamefont
  {Casu}}]{10IsCeVi.is}%
  \BibitemOpen
  \bibfield  {author} {\bibinfo {author} {\bibfnamefont {F.}~\bibnamefont
  {Islam}}, \bibinfo {author} {\bibfnamefont {C.}~\bibnamefont
  {Cecchi-Pestellini}}, \bibinfo {author} {\bibfnamefont {S.}~\bibnamefont
  {Viti}}, \ and\ \bibinfo {author} {\bibfnamefont {S.}~\bibnamefont {Casu}},\
  }\href {\doibase 10.1088/0004-637X/725/1/1111} {\bibfield  {journal}
  {\bibinfo  {journal} {Astrophys. J.}\ }\textbf {\bibinfo {volume} {725}},\
  \bibinfo {pages} {1111} (\bibinfo {year} {2010})}\BibitemShut {NoStop}%
\bibitem [{\citenamefont {Hollenbach}\ and\ \citenamefont
  {Tielens}(1999)}]{99HoTixx.is}%
  \BibitemOpen
  \bibfield  {author} {\bibinfo {author} {\bibfnamefont {D.}~\bibnamefont
  {Hollenbach}}\ and\ \bibinfo {author} {\bibfnamefont {A.}~\bibnamefont
  {Tielens}},\ }\href {\doibase 10.1103/RevModPhys.71.173} {\bibfield
  {journal} {\bibinfo  {journal} {Rev. Mod. Phys.}\ }\textbf {\bibinfo {volume}
  {71}},\ \bibinfo {pages} {173} (\bibinfo {year} {1999})}\BibitemShut
  {NoStop}%
\bibitem [{\citenamefont {Dalgarno}(2000)}]{00Dalgarno.H2}%
  \BibitemOpen
  \bibfield  {author} {\bibinfo {author} {\bibfnamefont {A.}~\bibnamefont
  {Dalgarno}},\ }in\ \href {\doibase 10.1017/CBO9780511564635.004} {\emph
  {\bibinfo {booktitle} {Molecular Hydrogen in Space}}},\ \bibinfo {series and
  number} {Cambridge Contemporary Astrophysics},\ \bibinfo {editor} {edited by\
  \bibinfo {editor} {\bibfnamefont {F.}~\bibnamefont {Combes}}\ and\ \bibinfo
  {editor} {\bibfnamefont {G.}~\bibnamefont {DesForets}}}\ (\bibinfo
  {organization} {{PCMI-CNRS; Collaborat Computat Project 7; Observ Paris;
  Minist Affaires Etrangeres; Univ Cergy Pontoise; Univ Paris XI; Inst
  Astrophys Paris}},\ \bibinfo {year} {2000})\ pp.\ \bibinfo {pages}
  {3--11}\BibitemShut {NoStop}%
\bibitem [{\citenamefont {Bowler}\ \emph {et~al.}(2010)\citenamefont {Bowler},
  \citenamefont {Liu}, \citenamefont {Dupuy},\ and\ \citenamefont
  {Cushing}}]{10BoLiDu.exo}%
  \BibitemOpen
  \bibfield  {author} {\bibinfo {author} {\bibfnamefont {B.~P.}\ \bibnamefont
  {Bowler}}, \bibinfo {author} {\bibfnamefont {M.~C.}\ \bibnamefont {Liu}},
  \bibinfo {author} {\bibfnamefont {T.~J.}\ \bibnamefont {Dupuy}}, \ and\
  \bibinfo {author} {\bibfnamefont {M.~C.}\ \bibnamefont {Cushing}},\ }\href
  {\doibase 10.1088/0004-637X/723/1/850} {\bibfield  {journal} {\bibinfo
  {journal} {Astrophys. J.}\ }\textbf {\bibinfo {volume} {723}},\ \bibinfo
  {pages} {850} (\bibinfo {year} {2010})}\BibitemShut {NoStop}%
\bibitem [{\citenamefont {Huitson}\ \emph {et~al.}(2012)\citenamefont
  {Huitson}, \citenamefont {Sing}, \citenamefont {Vidal-Madjar}, \citenamefont
  {Ballester}, \citenamefont {des Etangs}, \citenamefont {Désert},\ and\
  \citenamefont {Pont}}]{12HuSiVi.exo}%
  \BibitemOpen
  \bibfield  {author} {\bibinfo {author} {\bibfnamefont {C.~M.}\ \bibnamefont
  {Huitson}}, \bibinfo {author} {\bibfnamefont {D.~K.}\ \bibnamefont {Sing}},
  \bibinfo {author} {\bibfnamefont {A.}~\bibnamefont {Vidal-Madjar}}, \bibinfo
  {author} {\bibfnamefont {G.~E.}\ \bibnamefont {Ballester}}, \bibinfo {author}
  {\bibfnamefont {A.~L.}\ \bibnamefont {des Etangs}}, \bibinfo {author}
  {\bibfnamefont {J.-M.}\ \bibnamefont {Désert}}, \ and\ \bibinfo {author}
  {\bibfnamefont {F.}~\bibnamefont {Pont}},\ }\href {\doibase
  10.1111/j.1365-2966.2012.20805.x} {\bibfield  {journal} {\bibinfo  {journal}
  {Mon. Not. R. Astron. Soc.}\ }\textbf {\bibinfo {volume} {422}},\ \bibinfo
  {pages} {2477} (\bibinfo {year} {2012})}\BibitemShut {NoStop}%
\bibitem [{\citenamefont {Stevenson}\ \emph {et~al.}(2014)\citenamefont
  {Stevenson}, \citenamefont {Bean}, \citenamefont {Seifahrt}, \citenamefont
  {Desert}, \citenamefont {Madhusudhan}, \citenamefont {Bergmann},
  \citenamefont {Kreidberg},\ and\ \citenamefont {Homeier}}]{14StBeSe.exo}%
  \BibitemOpen
  \bibfield  {author} {\bibinfo {author} {\bibfnamefont {K.~B.}\ \bibnamefont
  {Stevenson}}, \bibinfo {author} {\bibfnamefont {J.~L.}\ \bibnamefont {Bean}},
  \bibinfo {author} {\bibfnamefont {A.}~\bibnamefont {Seifahrt}}, \bibinfo
  {author} {\bibfnamefont {J.-M.}\ \bibnamefont {Desert}}, \bibinfo {author}
  {\bibfnamefont {N.}~\bibnamefont {Madhusudhan}}, \bibinfo {author}
  {\bibfnamefont {M.}~\bibnamefont {Bergmann}}, \bibinfo {author}
  {\bibfnamefont {L.}~\bibnamefont {Kreidberg}}, \ and\ \bibinfo {author}
  {\bibfnamefont {D.}~\bibnamefont {Homeier}},\ }\href {\doibase
  10.1088/0004-6256/147/6/161} {\bibfield  {journal} {\bibinfo  {journal}
  {Astrophys. J.}\ }\textbf {\bibinfo {volume} {{147}}},\ \bibinfo {pages}
  {161} (\bibinfo {year} {{2014}})}\BibitemShut {NoStop}%
\bibitem [{\citenamefont {Ardaseva}\ \emph {et~al.}(2017)\citenamefont
  {Ardaseva}, \citenamefont {Rimmer}, \citenamefont {Waldmann}, \citenamefont
  {Rocchetto}, \citenamefont {Yurchenko}, \citenamefont {Helling},\ and\
  \citenamefont {Tennyson}}]{17ArRiWa.exo}%
  \BibitemOpen
  \bibfield  {author} {\bibinfo {author} {\bibfnamefont {A.}~\bibnamefont
  {Ardaseva}}, \bibinfo {author} {\bibfnamefont {P.~B.}\ \bibnamefont
  {Rimmer}}, \bibinfo {author} {\bibfnamefont {I.}~\bibnamefont {Waldmann}},
  \bibinfo {author} {\bibfnamefont {M.}~\bibnamefont {Rocchetto}}, \bibinfo
  {author} {\bibfnamefont {S.~N.}\ \bibnamefont {Yurchenko}}, \bibinfo {author}
  {\bibfnamefont {C.}~\bibnamefont {Helling}}, \ and\ \bibinfo {author}
  {\bibfnamefont {J.}~\bibnamefont {Tennyson}},\ }\href {\doibase
  10.1093/mnras/stx1012} {\bibfield  {journal} {\bibinfo  {journal} {Monthly
  Notices of The Royal Astronomical Society}\ }\textbf {\bibinfo {volume}
  {470}},\ \bibinfo {pages} {187} (\bibinfo {year} {2017})}\BibitemShut
  {NoStop}%
\bibitem [{\citenamefont {Ubachs}\ \emph
  {et~al.}(2016{\natexlab{a}})\citenamefont {Ubachs}, \citenamefont
  {Koelemeij}, \citenamefont {Eikema},\ and\ \citenamefont
  {Salumbides}}]{16UbKoEi.H2}%
  \BibitemOpen
  \bibfield  {author} {\bibinfo {author} {\bibfnamefont {W.}~\bibnamefont
  {Ubachs}}, \bibinfo {author} {\bibfnamefont {J.~C.~J.}\ \bibnamefont
  {Koelemeij}}, \bibinfo {author} {\bibfnamefont {K.~S.~E.}\ \bibnamefont
  {Eikema}}, \ and\ \bibinfo {author} {\bibfnamefont {E.~J.}\ \bibnamefont
  {Salumbides}},\ }\href {\doibase 10.1016/j.jms.2015.12.003} {\bibfield
  {journal} {\bibinfo  {journal} {J. Mol. Spectrosc.}\ }\textbf {\bibinfo
  {volume} {320}},\ \bibinfo {pages} {1} (\bibinfo {year}
  {2016}{\natexlab{a}})}\BibitemShut {NoStop}%
\bibitem [{\citenamefont {Ubachs}\ \emph
  {et~al.}(2016{\natexlab{b}})\citenamefont {Ubachs}, \citenamefont
  {Bagdonaite}, \citenamefont {Salumbides}, \citenamefont {Murphy},\ and\
  \citenamefont {Kaper}}]{16UbBaSa.H2}%
  \BibitemOpen
  \bibfield  {author} {\bibinfo {author} {\bibfnamefont {W.}~\bibnamefont
  {Ubachs}}, \bibinfo {author} {\bibfnamefont {J.}~\bibnamefont {Bagdonaite}},
  \bibinfo {author} {\bibfnamefont {E.~J.}\ \bibnamefont {Salumbides}},
  \bibinfo {author} {\bibfnamefont {M.~T.}\ \bibnamefont {Murphy}}, \ and\
  \bibinfo {author} {\bibfnamefont {L.}~\bibnamefont {Kaper}},\ }\href
  {\doibase 10.1103/RevModPhys.88.021003} {\bibfield  {journal} {\bibinfo
  {journal} {Rev. Mod. Phys.}\ }\textbf {\bibinfo {volume} {88}},\ \bibinfo
  {pages} {021003} (\bibinfo {year} {2016}{\natexlab{b}})}\BibitemShut
  {NoStop}%
\bibitem [{\citenamefont {Kolos}\ and\ \citenamefont
  {Wolniewicz}(1963)}]{63KoWoxx.H2}%
  \BibitemOpen
  \bibfield  {author} {\bibinfo {author} {\bibfnamefont {W.}~\bibnamefont
  {Kolos}}\ and\ \bibinfo {author} {\bibfnamefont {L.}~\bibnamefont
  {Wolniewicz}},\ }\href {\doibase 10.1103/RevModPhys.35.473} {\bibfield
  {journal} {\bibinfo  {journal} {Rev. Mod. Phys.}\ }\textbf {\bibinfo {volume}
  {35}},\ \bibinfo {pages} {473} (\bibinfo {year} {1963})}\BibitemShut
  {NoStop}%
\bibitem [{\citenamefont {Kolos}\ and\ \citenamefont
  {Wolniewicz}(1965)}]{65KoWoxx.H2}%
  \BibitemOpen
  \bibfield  {author} {\bibinfo {author} {\bibfnamefont {W.}~\bibnamefont
  {Kolos}}\ and\ \bibinfo {author} {\bibfnamefont {L.}~\bibnamefont
  {Wolniewicz}},\ }\href {\doibase 10.1063/1.1697142} {\bibfield  {journal}
  {\bibinfo  {journal} {J. Chem. Phys.}\ }\textbf {\bibinfo {volume} {43}},\
  \bibinfo {pages} {2429} (\bibinfo {year} {1965})}\BibitemShut {NoStop}%
\bibitem [{\citenamefont {LeRoy}\ and\ \citenamefont
  {Bernstein}(1968)}]{68LeBexx.H2}%
  \BibitemOpen
  \bibfield  {author} {\bibinfo {author} {\bibfnamefont {R.~J.}\ \bibnamefont
  {LeRoy}}\ and\ \bibinfo {author} {\bibfnamefont {R.~B.}\ \bibnamefont
  {Bernstein}},\ }\href {\doibase 10.1063/1.1669876} {\bibfield  {journal}
  {\bibinfo  {journal} {J. Chem. Phys.}\ }\textbf {\bibinfo {volume} {49}},\
  \bibinfo {pages} {4312} (\bibinfo {year} {1968})}\BibitemShut {NoStop}%
\bibitem [{\citenamefont {DalGarno}, \citenamefont {Allison},\ and\
  \citenamefont {Browne}(1969)}]{69DaAlBr.H2}%
  \BibitemOpen
  \bibfield  {author} {\bibinfo {author} {\bibfnamefont {A.}~\bibnamefont
  {DalGarno}}, \bibinfo {author} {\bibfnamefont {A.~C.}\ \bibnamefont
  {Allison}}, \ and\ \bibinfo {author} {\bibfnamefont {J.~C.}\ \bibnamefont
  {Browne}},\ }\href {\doibase 10.1175/1520-0469(1969)026<0946:RVQMEA>2.0.CO;2}
  {\bibfield  {journal} {\bibinfo  {journal} {J. Atmos. Sci.}\ }\textbf
  {\bibinfo {volume} {26}},\ \bibinfo {pages} {946} (\bibinfo {year}
  {1969})}\BibitemShut {NoStop}%
\bibitem [{\citenamefont {Truhlar}(1972{\natexlab{b}})}]{72Truhla.H2}%
  \BibitemOpen
  \bibfield  {author} {\bibinfo {author} {\bibfnamefont {D.}~\bibnamefont
  {Truhlar}},\ }\href {\doibase 10.1002/qua.560060515} {\bibfield  {journal}
  {\bibinfo  {journal} {Intern. J. Quantum Chem.}\ }\textbf {\bibinfo {volume}
  {6}},\ \bibinfo {pages} {975} (\bibinfo {year}
  {1972}{\natexlab{b}})}\BibitemShut {NoStop}%
\bibitem [{\citenamefont {Turner}, \citenamefont {Kirbydocken},\ and\
  \citenamefont {Dalgarno}(1977)}]{77TuKiDa.H2}%
  \BibitemOpen
  \bibfield  {author} {\bibinfo {author} {\bibfnamefont {J.}~\bibnamefont
  {Turner}}, \bibinfo {author} {\bibfnamefont {K.}~\bibnamefont {Kirbydocken}},
  \ and\ \bibinfo {author} {\bibfnamefont {A.}~\bibnamefont {Dalgarno}},\
  }\href {\doibase 10.1086/190481} {\bibfield  {journal} {\bibinfo  {journal}
  {Astrophys. J. Suppl.}\ }\textbf {\bibinfo {volume} {35}},\ \bibinfo {pages}
  {281} (\bibinfo {year} {1977})}\BibitemShut {NoStop}%
\bibitem [{\citenamefont {Komasa}\ \emph {et~al.}(2019)\citenamefont {Komasa},
  \citenamefont {Puchalski}, \citenamefont {Czachorowski}, \citenamefont
  {\L{}ach},\ and\ \citenamefont {Pachucki}}]{H2SPECTRE}%
  \BibitemOpen
  \bibfield  {author} {\bibinfo {author} {\bibfnamefont {J.}~\bibnamefont
  {Komasa}}, \bibinfo {author} {\bibfnamefont {M.}~\bibnamefont {Puchalski}},
  \bibinfo {author} {\bibfnamefont {P.}~\bibnamefont {Czachorowski}}, \bibinfo
  {author} {\bibfnamefont {G.}~\bibnamefont {\L{}ach}}, \ and\ \bibinfo
  {author} {\bibfnamefont {K.}~\bibnamefont {Pachucki}},\ }\href {\doibase
  10.1103/PhysRevA.100.032519} {\bibfield  {journal} {\bibinfo  {journal}
  {Phys. Rev. A}\ }\textbf {\bibinfo {volume} {100}},\ \bibinfo {pages}
  {032519} (\bibinfo {year} {2019})}\BibitemShut {NoStop}%
\bibitem [{\citenamefont {Pachucki}(2010)}]{10Pachucki.H2}%
  \BibitemOpen
  \bibfield  {author} {\bibinfo {author} {\bibfnamefont {K.}~\bibnamefont
  {Pachucki}},\ }\href {\doibase 10.1103/PhysRevA.82.032509} {\bibfield
  {journal} {\bibinfo  {journal} {Phys. Rev. A}\ }\textbf {\bibinfo {volume}
  {82}},\ \bibinfo {pages} {032509} (\bibinfo {year} {2010})}\BibitemShut
  {NoStop}%
\bibitem [{\citenamefont {Pachucki}\ and\ \citenamefont
  {Komasa}(2009)}]{09PaKoxx.H2}%
  \BibitemOpen
  \bibfield  {author} {\bibinfo {author} {\bibfnamefont {K.}~\bibnamefont
  {Pachucki}}\ and\ \bibinfo {author} {\bibfnamefont {J.}~\bibnamefont
  {Komasa}},\ }\href {\doibase 10.1063/1.3114680} {\bibfield  {journal}
  {\bibinfo  {journal} {J. Chem. Phys.}\ }\textbf {\bibinfo {volume} {130}},\
  \bibinfo {pages} {164113} (\bibinfo {year} {2009})}\BibitemShut {NoStop}%
\bibitem [{\citenamefont {Pachucki}\ and\ \citenamefont
  {Komasa}(2014)}]{14PaKoxx.H2}%
  \BibitemOpen
  \bibfield  {author} {\bibinfo {author} {\bibfnamefont {K.}~\bibnamefont
  {Pachucki}}\ and\ \bibinfo {author} {\bibfnamefont {J.}~\bibnamefont
  {Komasa}},\ }\href {\doibase 10.1063/1.4902981} {\bibfield  {journal}
  {\bibinfo  {journal} {J. Chem. Phys.}\ }\textbf {\bibinfo {volume} {141}},\
  \bibinfo {pages} {224103} (\bibinfo {year} {2014})}\BibitemShut {NoStop}%
\bibitem [{\citenamefont {Pachucki}\ and\ \citenamefont
  {Komasa}(2015)}]{15PaKoxx.H2}%
  \BibitemOpen
  \bibfield  {author} {\bibinfo {author} {\bibfnamefont {K.}~\bibnamefont
  {Pachucki}}\ and\ \bibinfo {author} {\bibfnamefont {J.}~\bibnamefont
  {Komasa}},\ }\href {\doibase 10.1063/1.4927079} {\bibfield  {journal}
  {\bibinfo  {journal} {J. Chem. Phys.}\ }\textbf {\bibinfo {volume} {143}},\
  \bibinfo {pages} {034111} (\bibinfo {year} {2015})}\BibitemShut {NoStop}%
\bibitem [{\citenamefont {Pachucki}(2012)}]{12Pachuc.H2}%
  \BibitemOpen
  \bibfield  {author} {\bibinfo {author} {\bibfnamefont {K.}~\bibnamefont
  {Pachucki}},\ }\href {\doibase 10.1103/PhysRevA.86.052514} {\bibfield
  {journal} {\bibinfo  {journal} {Phys. Rev. A}\ }\textbf {\bibinfo {volume}
  {86}},\ \bibinfo {pages} {052514} (\bibinfo {year} {2012})}\BibitemShut
  {NoStop}%
\bibitem [{\citenamefont {Puchalski}, \citenamefont {Komasa},\ and\
  \citenamefont {Pachucki}(2017)}]{17PuKoPu.H2}%
  \BibitemOpen
  \bibfield  {author} {\bibinfo {author} {\bibfnamefont {M.}~\bibnamefont
  {Puchalski}}, \bibinfo {author} {\bibfnamefont {J.}~\bibnamefont {Komasa}}, \
  and\ \bibinfo {author} {\bibfnamefont {K.}~\bibnamefont {Pachucki}},\ }\href
  {\doibase 10.1103/PhysRevA.95.052506} {\bibfield  {journal} {\bibinfo
  {journal} {Phys. Rev. A}\ }\textbf {\bibinfo {volume} {95}},\ \bibinfo
  {pages} {052506} (\bibinfo {year} {2017})}\BibitemShut {NoStop}%
\bibitem [{\citenamefont {Bragg}, \citenamefont {Brault},\ and\ \citenamefont
  {Smith}(1982)}]{82BrBrSm.H2}%
  \BibitemOpen
  \bibfield  {author} {\bibinfo {author} {\bibfnamefont {S.~L.}\ \bibnamefont
  {Bragg}}, \bibinfo {author} {\bibfnamefont {J.~W.}\ \bibnamefont {Brault}}, \
  and\ \bibinfo {author} {\bibfnamefont {W.~H.}\ \bibnamefont {Smith}},\ }\href
  {\doibase 10.1086/160568} {\bibfield  {journal} {\bibinfo  {journal}
  {Astrophys. J.}\ }\textbf {\bibinfo {volume} {263}},\ \bibinfo {pages} {999}
  (\bibinfo {year} {1982})}\BibitemShut {NoStop}%
\bibitem [{\citenamefont {Campargue}\ \emph {et~al.}(2012)\citenamefont
  {Campargue}, \citenamefont {Kassi}, \citenamefont {Pachucki},\ and\
  \citenamefont {Komasa}}]{12CaKaPa.H2}%
  \BibitemOpen
  \bibfield  {author} {\bibinfo {author} {\bibfnamefont {A.}~\bibnamefont
  {Campargue}}, \bibinfo {author} {\bibfnamefont {S.}~\bibnamefont {Kassi}},
  \bibinfo {author} {\bibfnamefont {K.}~\bibnamefont {Pachucki}}, \ and\
  \bibinfo {author} {\bibfnamefont {J.}~\bibnamefont {Komasa}},\ }\href
  {\doibase 10.1039/c1cp22912e} {\bibfield  {journal} {\bibinfo  {journal}
  {Phys. Chem. Chem. Phys.}\ }\textbf {\bibinfo {volume} {14}},\ \bibinfo
  {pages} {802} (\bibinfo {year} {2012})}\BibitemShut {NoStop}%
\bibitem [{\citenamefont {Komasa}\ \emph {et~al.}(2011)\citenamefont {Komasa},
  \citenamefont {Piszczatowski}, \citenamefont {Lach}, \citenamefont
  {Przybytek}, \citenamefont {Jeziorski},\ and\ \citenamefont
  {Pachucki}}]{11KoPiLa.H2}%
  \BibitemOpen
  \bibfield  {author} {\bibinfo {author} {\bibfnamefont {J.}~\bibnamefont
  {Komasa}}, \bibinfo {author} {\bibfnamefont {K.}~\bibnamefont
  {Piszczatowski}}, \bibinfo {author} {\bibfnamefont {G.}~\bibnamefont {Lach}},
  \bibinfo {author} {\bibfnamefont {M.}~\bibnamefont {Przybytek}}, \bibinfo
  {author} {\bibfnamefont {B.}~\bibnamefont {Jeziorski}}, \ and\ \bibinfo
  {author} {\bibfnamefont {K.}~\bibnamefont {Pachucki}},\ }\href {\doibase
  10.1021/ct200438t} {\bibfield  {journal} {\bibinfo  {journal} {J. Chem.
  Theory Comput.}\ }\textbf {\bibinfo {volume} {7}},\ \bibinfo {pages} {3105}
  (\bibinfo {year} {2011})}\BibitemShut {NoStop}%
\bibitem [{\citenamefont {Chetty}\ and\ \citenamefont
  {Couling}(2011)}]{11ChCoxx.CO}%
  \BibitemOpen
  \bibfield  {author} {\bibinfo {author} {\bibfnamefont {N.}~\bibnamefont
  {Chetty}}\ and\ \bibinfo {author} {\bibfnamefont {V.~W.}\ \bibnamefont
  {Couling}},\ }\href {\doibase 10.1063/1.3585605} {\bibfield  {journal}
  {\bibinfo  {journal} {J. Chem. Phys.}\ }\textbf {\bibinfo {volume} {134}},\
  \bibinfo {pages} {164307} (\bibinfo {year} {2011})}\BibitemShut {NoStop}%
\bibitem [{\citenamefont {Li}\ \emph {et~al.}(2015)\citenamefont {Li},
  \citenamefont {Gordon}, \citenamefont {Rothman}, \citenamefont {Tan},
  \citenamefont {Hu}, \citenamefont {Kassi}, \citenamefont {Campargue},\ and\
  \citenamefont {Medvedev}}]{15LiGoRo.CO}%
  \BibitemOpen
  \bibfield  {author} {\bibinfo {author} {\bibfnamefont {G.}~\bibnamefont
  {Li}}, \bibinfo {author} {\bibfnamefont {I.~E.}\ \bibnamefont {Gordon}},
  \bibinfo {author} {\bibfnamefont {L.~S.}\ \bibnamefont {Rothman}}, \bibinfo
  {author} {\bibfnamefont {Y.}~\bibnamefont {Tan}}, \bibinfo {author}
  {\bibfnamefont {S.-M.}\ \bibnamefont {Hu}}, \bibinfo {author} {\bibfnamefont
  {S.}~\bibnamefont {Kassi}}, \bibinfo {author} {\bibfnamefont
  {A.}~\bibnamefont {Campargue}}, \ and\ \bibinfo {author} {\bibfnamefont
  {E.~S.}\ \bibnamefont {Medvedev}},\ }\href {\doibase
  10.1088/0067-0049/216/1/15} {\bibfield  {journal} {\bibinfo  {journal}
  {Astrophys. J. Suppl.}\ }\textbf {\bibinfo {volume} {216}},\ \bibinfo {pages}
  {15} (\bibinfo {year} {2015})}\BibitemShut {NoStop}%
\bibitem [{\citenamefont {Goorvitch}(1994)}]{94Goorvich.CO}%
  \BibitemOpen
  \bibfield  {author} {\bibinfo {author} {\bibfnamefont {D.}~\bibnamefont
  {Goorvitch}},\ }\href {\doibase 10.1086/192110} {\bibfield  {journal}
  {\bibinfo  {journal} {Astrophys. J. Suppl.}\ }\textbf {\bibinfo {volume}
  {95}},\ \bibinfo {pages} {535} (\bibinfo {year} {1994})}\BibitemShut
  {NoStop}%
\bibitem [{\citenamefont {Hur\'e}\ and\ \citenamefont
  {Roueff}(1996)}]{96HuRoxx.CO}%
  \BibitemOpen
  \bibfield  {author} {\bibinfo {author} {\bibfnamefont {J.~M.}\ \bibnamefont
  {Hur\'e}}\ and\ \bibinfo {author} {\bibfnamefont {E.}~\bibnamefont
  {Roueff}},\ }\href {\doibase 10.1051/aas:1996174} {\bibfield  {journal}
  {\bibinfo  {journal} {Astron. Astrophys. Suppl.}\ }\textbf {\bibinfo {volume}
  {117}},\ \bibinfo {pages} {561} (\bibinfo {year} {1996})}\BibitemShut
  {NoStop}%
\bibitem [{\citenamefont {Chackerian~Jr.}\ \emph {et~al.}(1984)\citenamefont
  {Chackerian~Jr.}, \citenamefont {Farrenq}, \citenamefont {Guelachvili},
  \citenamefont {Rossetti},\ and\ \citenamefont {Urban}}]{84ChFaGu.CO}%
  \BibitemOpen
  \bibfield  {author} {\bibinfo {author} {\bibfnamefont {C.}~\bibnamefont
  {Chackerian~Jr.}}, \bibinfo {author} {\bibfnamefont {R.}~\bibnamefont
  {Farrenq}}, \bibinfo {author} {\bibfnamefont {G.}~\bibnamefont
  {Guelachvili}}, \bibinfo {author} {\bibfnamefont {C.}~\bibnamefont
  {Rossetti}}, \ and\ \bibinfo {author} {\bibfnamefont {W.}~\bibnamefont
  {Urban}},\ }\href {\doibase 10.1139/p84-202} {\bibfield  {journal} {\bibinfo
  {journal} {Can. J. Phys.}\ }\textbf {\bibinfo {volume} {62}},\ \bibinfo
  {pages} {1579} (\bibinfo {year} {1984})}\BibitemShut {NoStop}%
\bibitem [{\citenamefont {Langhoff}\ and\ \citenamefont
  {Bauschlicher}(1995)}]{95LaBaxx.CO}%
  \BibitemOpen
  \bibfield  {author} {\bibinfo {author} {\bibfnamefont {S.~R.}\ \bibnamefont
  {Langhoff}}\ and\ \bibinfo {author} {\bibfnamefont {C.~W.}\ \bibnamefont
  {Bauschlicher}},\ }\href {\doibase 10.1063/1.469247} {\bibfield  {journal}
  {\bibinfo  {journal} {J. Chem. Phys.}\ }\textbf {\bibinfo {volume} {102}},\
  \bibinfo {pages} {5220} (\bibinfo {year} {1995})}\BibitemShut {NoStop}%
\bibitem [{\citenamefont {Coxon}\ and\ \citenamefont
  {Hajigeorgiou}(2004)}]{04CoHaxx.CO}%
  \BibitemOpen
  \bibfield  {author} {\bibinfo {author} {\bibfnamefont {J.~A.}\ \bibnamefont
  {Coxon}}\ and\ \bibinfo {author} {\bibfnamefont {P.~G.}\ \bibnamefont
  {Hajigeorgiou}},\ }\href {\doibase 10.1063/1.1768167} {\bibfield  {journal}
  {\bibinfo  {journal} {J. Chem. Phys.}\ }\textbf {\bibinfo {volume} {121}},\
  \bibinfo {pages} {2992} (\bibinfo {year} {2004})}\BibitemShut {NoStop}%
\bibitem [{\citenamefont {Coriani}\ \emph {et~al.}(2003)\citenamefont
  {Coriani}, \citenamefont {Halkier}, \citenamefont {Jonsson}, \citenamefont
  {Gauss}, \citenamefont {Rizzo},\ and\ \citenamefont
  {Christiansen}}]{03CoHaJo.CO}%
  \BibitemOpen
  \bibfield  {author} {\bibinfo {author} {\bibfnamefont {S.}~\bibnamefont
  {Coriani}}, \bibinfo {author} {\bibfnamefont {A.}~\bibnamefont {Halkier}},
  \bibinfo {author} {\bibfnamefont {D.}~\bibnamefont {Jonsson}}, \bibinfo
  {author} {\bibfnamefont {J.}~\bibnamefont {Gauss}}, \bibinfo {author}
  {\bibfnamefont {A.}~\bibnamefont {Rizzo}}, \ and\ \bibinfo {author}
  {\bibfnamefont {O.}~\bibnamefont {Christiansen}},\ }\href {\doibase
  10.1063/1.1562198} {\bibfield  {journal} {\bibinfo  {journal} {J. Chem.
  Phys.}\ }\textbf {\bibinfo {volume} {118}},\ \bibinfo {pages} {7329}
  (\bibinfo {year} {2003})}\BibitemShut {NoStop}%
\bibitem [{\citenamefont {Meerts}, \citenamefont {Leeuw},\ and\ \citenamefont
  {Dymanus}(1977)}]{77MeLeDy.CO}%
  \BibitemOpen
  \bibfield  {author} {\bibinfo {author} {\bibfnamefont {W.~L.}\ \bibnamefont
  {Meerts}}, \bibinfo {author} {\bibfnamefont {F.~H.~D.}\ \bibnamefont
  {Leeuw}}, \ and\ \bibinfo {author} {\bibfnamefont {A.}~\bibnamefont
  {Dymanus}},\ }\href {\doibase 10.1016/0301-0104(77)87016-X} {\bibfield
  {journal} {\bibinfo  {journal} {Chem. Phys.}\ }\textbf {\bibinfo {volume}
  {22}},\ \bibinfo {pages} {319} (\bibinfo {year} {1977})}\BibitemShut
  {NoStop}%
\bibitem [{\citenamefont {Mohr}, \citenamefont {Newell},\ and\ \citenamefont
  {Taylor}(2016)}]{CODATA}%
  \BibitemOpen
  \bibfield  {author} {\bibinfo {author} {\bibfnamefont {P.~J.}\ \bibnamefont
  {Mohr}}, \bibinfo {author} {\bibfnamefont {D.~B.}\ \bibnamefont {Newell}}, \
  and\ \bibinfo {author} {\bibfnamefont {B.~N.}\ \bibnamefont {Taylor}},\
  }\href {http://link.aps.org/doi/10.1103/RevModPhys.88.035009} {\bibfield
  {journal} {\bibinfo  {journal} {Rev. Mod. Phys.}\ }\textbf {\bibinfo {volume}
  {88}},\ \bibinfo {pages} {035009} (\bibinfo {year} {2016})}\BibitemShut
  {NoStop}%
\bibitem [{\citenamefont {Graham}, \citenamefont {Imrie},\ and\ \citenamefont
  {Raab}(1998)}]{98GrImRa.CO}%
  \BibitemOpen
  \bibfield  {author} {\bibinfo {author} {\bibfnamefont {C.}~\bibnamefont
  {Graham}}, \bibinfo {author} {\bibfnamefont {D.~A.}\ \bibnamefont {Imrie}}, \
  and\ \bibinfo {author} {\bibfnamefont {R.~E.}\ \bibnamefont {Raab}},\ }\href
  {\doibase 10.1080/002689798169429} {\bibfield  {journal} {\bibinfo  {journal}
  {Mol. Phys.}\ }\textbf {\bibinfo {volume} {93}},\ \bibinfo {pages} {49}
  (\bibinfo {year} {1998})}\BibitemShut {NoStop}%
\bibitem [{\citenamefont {Buckingham}, \citenamefont {Disch},\ and\
  \citenamefont {Dunmur}(1968)}]{68BuDiDu.CO}%
  \BibitemOpen
  \bibfield  {author} {\bibinfo {author} {\bibfnamefont {A.~D.}\ \bibnamefont
  {Buckingham}}, \bibinfo {author} {\bibfnamefont {R.~L.}\ \bibnamefont
  {Disch}}, \ and\ \bibinfo {author} {\bibfnamefont {D.~A.}\ \bibnamefont
  {Dunmur}},\ }\href {\doibase 10.1021/ja01014a023} {\bibfield  {journal}
  {\bibinfo  {journal} {J. Am. Chem. Soc.}\ }\textbf {\bibinfo {volume} {90}},\
  \bibinfo {pages} {3104} (\bibinfo {year} {1968})}\BibitemShut {NoStop}%
\bibitem [{\citenamefont {Meshkov}\ \emph {et~al.}(2018)\citenamefont
  {Meshkov}, \citenamefont {Stolyarov}, \citenamefont {Ermilov}, \citenamefont
  {Medvedev}, \citenamefont {Ushakov},\ and\ \citenamefont
  {Gordon}}]{18MeStEr.CO}%
  \BibitemOpen
  \bibfield  {author} {\bibinfo {author} {\bibfnamefont {V.~V.}\ \bibnamefont
  {Meshkov}}, \bibinfo {author} {\bibfnamefont {A.~V.}\ \bibnamefont
  {Stolyarov}}, \bibinfo {author} {\bibfnamefont {A.~Y.}\ \bibnamefont
  {Ermilov}}, \bibinfo {author} {\bibfnamefont {E.~S.}\ \bibnamefont
  {Medvedev}}, \bibinfo {author} {\bibfnamefont {V.~G.}\ \bibnamefont
  {Ushakov}}, \ and\ \bibinfo {author} {\bibfnamefont {I.~E.}\ \bibnamefont
  {Gordon}},\ }\href {\doibase 10.1016/j.jqsrt.2018.06.001} {\bibfield
  {journal} {\bibinfo  {journal} {J. Quant. Spectrosc. Radiat. Transf.}\
  }\textbf {\bibinfo {volume} {217}},\ \bibinfo {pages} {262} (\bibinfo {year}
  {2018})}\BibitemShut {NoStop}%
\bibitem [{\citenamefont {Medvedev}\ \emph {et~al.}(2015)\citenamefont
  {Medvedev}, \citenamefont {Meshkov}, \citenamefont {Stolyarov},\ and\
  \citenamefont {Gordon}}]{15MeMeSt.CO}%
  \BibitemOpen
  \bibfield  {author} {\bibinfo {author} {\bibfnamefont {E.~S.}\ \bibnamefont
  {Medvedev}}, \bibinfo {author} {\bibfnamefont {V.~V.}\ \bibnamefont
  {Meshkov}}, \bibinfo {author} {\bibfnamefont {A.~V.}\ \bibnamefont
  {Stolyarov}}, \ and\ \bibinfo {author} {\bibfnamefont {I.~E.}\ \bibnamefont
  {Gordon}},\ }\href {\doibase 10.1063/1.4933136} {\bibfield  {journal}
  {\bibinfo  {journal} {J. Chem. Phys.}\ }\textbf {\bibinfo {volume} {143}},\
  \bibinfo {pages} {154301} (\bibinfo {year} {2015})}\BibitemShut {NoStop}%
\bibitem [{\citenamefont {{Yurchenko}}, \citenamefont {{Al-Refaie}},\ and\
  \citenamefont {{Tennyson}}(2018)}]{ExoCross}%
  \BibitemOpen
  \bibfield  {author} {\bibinfo {author} {\bibfnamefont {S.~N.}\ \bibnamefont
  {{Yurchenko}}}, \bibinfo {author} {\bibfnamefont {A.~F.}\ \bibnamefont
  {{Al-Refaie}}}, \ and\ \bibinfo {author} {\bibfnamefont {J.}~\bibnamefont
  {{Tennyson}}},\ }\href {\doibase 10.1051/0004-6361/201732531} {\bibfield
  {journal} {\bibinfo  {journal} {Astron. Astrophys.}\ }\textbf {\bibinfo
  {volume} {614}},\ \bibinfo {pages} {A131} (\bibinfo {year}
  {2018})}\BibitemShut {NoStop}%
\bibitem [{\citenamefont {Brown}\ \emph {et~al.}(1975)\citenamefont {Brown},
  \citenamefont {Hougen}, \citenamefont {Huber}, \citenamefont {Johns},
  \citenamefont {Kopp}, \citenamefont {Lefebvre-Brion}, \citenamefont {Merer},
  \citenamefont {Ramsay}, \citenamefont {Rostas},\ and\ \citenamefont
  {Zare}}]{75BrHoHu.diatom}%
  \BibitemOpen
  \bibfield  {author} {\bibinfo {author} {\bibfnamefont {J.~M.}\ \bibnamefont
  {Brown}}, \bibinfo {author} {\bibfnamefont {J.~T.}\ \bibnamefont {Hougen}},
  \bibinfo {author} {\bibfnamefont {K.~P.}\ \bibnamefont {Huber}}, \bibinfo
  {author} {\bibfnamefont {J.~W.~C.}\ \bibnamefont {Johns}}, \bibinfo {author}
  {\bibfnamefont {I.}~\bibnamefont {Kopp}}, \bibinfo {author} {\bibfnamefont
  {H.}~\bibnamefont {Lefebvre-Brion}}, \bibinfo {author} {\bibfnamefont
  {A.~J.}\ \bibnamefont {Merer}}, \bibinfo {author} {\bibfnamefont {D.~A.}\
  \bibnamefont {Ramsay}}, \bibinfo {author} {\bibfnamefont {J.}~\bibnamefont
  {Rostas}}, \ and\ \bibinfo {author} {\bibfnamefont {R.~N.}\ \bibnamefont
  {Zare}},\ }\href {\doibase 10.1016/0022-2852(75)90291-X} {\bibfield
  {journal} {\bibinfo  {journal} {J. Mol. Spectrosc.}\ }\textbf {\bibinfo
  {volume} {55}},\ \bibinfo {pages} {500} (\bibinfo {year} {1975})}\BibitemShut
  {NoStop}%
\bibitem [{\citenamefont {Weiss}(1963)}]{63Weissx.HF}%
  \BibitemOpen
  \bibfield  {author} {\bibinfo {author} {\bibfnamefont {R.}~\bibnamefont
  {Weiss}},\ }\href {\doibase 10.1103/PhysRev.131.659} {\bibfield  {journal}
  {\bibinfo  {journal} {Phys. Rev.}\ }\textbf {\bibinfo {volume} {131}},\
  \bibinfo {pages} {659} (\bibinfo {year} {1963})}\BibitemShut {NoStop}%
\bibitem [{\citenamefont {de~Leeuw}\ and\ \citenamefont
  {Dymanus}(1973)}]{73deDyxx.HF}%
  \BibitemOpen
  \bibfield  {author} {\bibinfo {author} {\bibfnamefont {F.~H.}\ \bibnamefont
  {de~Leeuw}}\ and\ \bibinfo {author} {\bibfnamefont {A.}~\bibnamefont
  {Dymanus}},\ }\href {\doibase 10.1016/0022-2852(73)90107-0} {\bibfield
  {journal} {\bibinfo  {journal} {J. Mol. Spectrosc.}\ }\textbf {\bibinfo
  {volume} {48}},\ \bibinfo {pages} {427} (\bibinfo {year} {1973})}\BibitemShut
  {NoStop}%
\bibitem [{\citenamefont {Piecuch}\ \emph {et~al.}(1996)\citenamefont
  {Piecuch}, \citenamefont {Kondo}, \citenamefont {Špirko},\ and\
  \citenamefont {Paldus}}]{96PiKoSp.HF}%
  \BibitemOpen
  \bibfield  {author} {\bibinfo {author} {\bibfnamefont {P.}~\bibnamefont
  {Piecuch}}, \bibinfo {author} {\bibfnamefont {A.~E.}\ \bibnamefont {Kondo}},
  \bibinfo {author} {\bibfnamefont {V.}~\bibnamefont {Špirko}}, \ and\
  \bibinfo {author} {\bibfnamefont {J.}~\bibnamefont {Paldus}},\ }\href
  {\doibase 10.1063/1.471164} {\bibfield  {journal} {\bibinfo  {journal} {J.
  Chem. Phys.}\ }\textbf {\bibinfo {volume} {104}},\ \bibinfo {pages} {4699}
  (\bibinfo {year} {1996})}\BibitemShut {NoStop}%
\bibitem [{\citenamefont {Maroulis}(2003)}]{03Maroul.HF}%
  \BibitemOpen
  \bibfield  {author} {\bibinfo {author} {\bibfnamefont {G.}~\bibnamefont
  {Maroulis}},\ }\href {\doibase 10.1016/S0166-1280(03)00273-2} {\bibfield
  {journal} {\bibinfo  {journal} {J. Molec. Struct. (THEOCHEM)}\ }\textbf
  {\bibinfo {volume} {633}},\ \bibinfo {pages} {177} (\bibinfo {year}
  {2003})}\BibitemShut {NoStop}%
\bibitem [{\citenamefont {Harrison}(2008)}]{08Harris.HF}%
  \BibitemOpen
  \bibfield  {author} {\bibinfo {author} {\bibfnamefont {J.~F.}\ \bibnamefont
  {Harrison}},\ }\href {\doibase 10.1063/1.2897445} {\bibfield  {journal}
  {\bibinfo  {journal} {J. Chem. Phys.}\ }\textbf {\bibinfo {volume} {128}},\
  \bibinfo {pages} {114320} (\bibinfo {year} {2008})}\BibitemShut {NoStop}%
\bibitem [{\citenamefont {Sadlej}(1988)}]{88Sadlej.ai}%
  \BibitemOpen
  \bibfield  {author} {\bibinfo {author} {\bibfnamefont {A.~J.}\ \bibnamefont
  {Sadlej}},\ }\href {\doibase 10.1135/cccc19881995} {\bibfield  {journal}
  {\bibinfo  {journal} {Collection of Czechoslovak Chemical Communications}\
  }\textbf {\bibinfo {volume} {53}},\ \bibinfo {pages} {1995} (\bibinfo {year}
  {1988})}\BibitemShut {NoStop}%
\bibitem [{\citenamefont {Coxon}\ and\ \citenamefont
  {Hajigeorgiou}(2015)}]{15CoHaxx.HF}%
  \BibitemOpen
  \bibfield  {author} {\bibinfo {author} {\bibfnamefont {J.~A.}\ \bibnamefont
  {Coxon}}\ and\ \bibinfo {author} {\bibfnamefont {P.~G.}\ \bibnamefont
  {Hajigeorgiou}},\ }\href {\doibase 10.1016/j.jqsrt.2014.08.028} {\bibfield
  {journal} {\bibinfo  {journal} {J. Quant. Spectrosc. Radiat. Transf.}\
  }\textbf {\bibinfo {volume} {151}},\ \bibinfo {pages} {133} (\bibinfo {year}
  {2015})}\BibitemShut {NoStop}%
\bibitem [{\citenamefont {Noxon}(1961)}]{61Noxonx.O2}%
  \BibitemOpen
  \bibfield  {author} {\bibinfo {author} {\bibfnamefont {J.~F.}\ \bibnamefont
  {Noxon}},\ }\href {\doibase 10.1139/p61-126} {\bibfield  {journal} {\bibinfo
  {journal} {Can. J. Phys.}\ }\textbf {\bibinfo {volume} {39}},\ \bibinfo
  {pages} {1110} (\bibinfo {year} {1961})}\BibitemShut {NoStop}%
\bibitem [{\citenamefont {F\"{o}ldes}\ \emph {et~al.}(2009)\citenamefont
  {F\"{o}ldes}, \citenamefont {{\v{C}}erm{\'{a}}k}, \citenamefont {Macko},
  \citenamefont {Veis},\ and\ \citenamefont {Macko}}]{09FoCeMa.O2}%
  \BibitemOpen
  \bibfield  {author} {\bibinfo {author} {\bibfnamefont {T.}~\bibnamefont
  {F\"{o}ldes}}, \bibinfo {author} {\bibfnamefont {P.}~\bibnamefont
  {{\v{C}}erm{\'{a}}k}}, \bibinfo {author} {\bibfnamefont {M.}~\bibnamefont
  {Macko}}, \bibinfo {author} {\bibfnamefont {P.}~\bibnamefont {Veis}}, \ and\
  \bibinfo {author} {\bibfnamefont {P.}~\bibnamefont {Macko}},\ }\href
  {\doibase 10.1016/j.cplett.2008.11.040} {\bibfield  {journal} {\bibinfo
  {journal} {Chem. Phys. Lett.}\ }\textbf {\bibinfo {volume} {467}},\ \bibinfo
  {pages} {233} (\bibinfo {year} {2009})}\BibitemShut {NoStop}%
\bibitem [{\citenamefont {Fink}\ \emph {et~al.}(1986)\citenamefont {Fink},
  \citenamefont {Kruse}, \citenamefont {Ramsay},\ and\ \citenamefont
  {Vervloet}}]{86FiKrRa.O2}%
  \BibitemOpen
  \bibfield  {author} {\bibinfo {author} {\bibfnamefont {E.~H.}\ \bibnamefont
  {Fink}}, \bibinfo {author} {\bibfnamefont {H.}~\bibnamefont {Kruse}},
  \bibinfo {author} {\bibfnamefont {D.~A.}\ \bibnamefont {Ramsay}}, \ and\
  \bibinfo {author} {\bibfnamefont {M.}~\bibnamefont {Vervloet}},\ }\href
  {\doibase 10.1139/p86-044} {\bibfield  {journal} {\bibinfo  {journal}
  {Canadian Journal of Physics}\ }\textbf {\bibinfo {volume} {64}},\ \bibinfo
  {pages} {242} (\bibinfo {year} {1986})}\BibitemShut {NoStop}%
\bibitem [{\citenamefont {Werner}\ \emph {et~al.}(2015)\citenamefont {Werner},
  \citenamefont {Knowles}, \citenamefont {Knizia}, \citenamefont {Manby},
  \citenamefont {Sch{\"{u}}tz}, \citenamefont {Celani}, \citenamefont
  {Gy{\"{o}}rffy}, \citenamefont {Kats}, \citenamefont {Korona}, \citenamefont
  {Lindh}, \citenamefont {Mitrushenkov}, \citenamefont {Rauhut}, \citenamefont
  {Shamasundar}, \citenamefont {Adler}, \citenamefont {Amos}, \citenamefont
  {Bernhardsson}, \citenamefont {Berning}, \citenamefont {Cooper},
  \citenamefont {Deegan}, \citenamefont {Dobbyn}, \citenamefont {Eckert},
  \citenamefont {Goll}, \citenamefont {Hampel}, \citenamefont {Hesselmann},
  \citenamefont {Hetzer}, \citenamefont {Hrenar}, \citenamefont {Jansen},
  \citenamefont {K{\"{o}}ppl}, \citenamefont {Liu}, \citenamefont {Lloyd},
  \citenamefont {Mata}, \citenamefont {May}, \citenamefont {McNicholas},
  \citenamefont {Meyer}, \citenamefont {Mura}, \citenamefont {Nicklass},
  \citenamefont {O'Neill}, \citenamefont {Palmieri}, \citenamefont {Peng},
  \citenamefont {Pfl{\"{u}}ger}, \citenamefont {Pitzer}, \citenamefont
  {Reiher}, \citenamefont {Shiozaki}, \citenamefont {Stoll}, \citenamefont
  {Stone}, \citenamefont {Tarroni}, \citenamefont {Thorsteinsson},\ and\
  \citenamefont {Wang}}]{MOLPRO2015}%
  \BibitemOpen
  \bibfield  {author} {\bibinfo {author} {\bibfnamefont {H.~J.}\ \bibnamefont
  {Werner}}, \bibinfo {author} {\bibfnamefont {P.~J.}\ \bibnamefont {Knowles}},
  \bibinfo {author} {\bibfnamefont {G.}~\bibnamefont {Knizia}}, \bibinfo
  {author} {\bibfnamefont {F.~R.}\ \bibnamefont {Manby}}, \bibinfo {author}
  {\bibfnamefont {M.}~\bibnamefont {Sch{\"{u}}tz}}, \bibinfo {author}
  {\bibfnamefont {P.}~\bibnamefont {Celani}}, \bibinfo {author} {\bibfnamefont
  {W.}~\bibnamefont {Gy{\"{o}}rffy}}, \bibinfo {author} {\bibfnamefont
  {D.}~\bibnamefont {Kats}}, \bibinfo {author} {\bibfnamefont {T.}~\bibnamefont
  {Korona}}, \bibinfo {author} {\bibfnamefont {R.}~\bibnamefont {Lindh}},
  \bibinfo {author} {\bibfnamefont {A.}~\bibnamefont {Mitrushenkov}}, \bibinfo
  {author} {\bibfnamefont {G.}~\bibnamefont {Rauhut}}, \bibinfo {author}
  {\bibfnamefont {K.~R.}\ \bibnamefont {Shamasundar}}, \bibinfo {author}
  {\bibfnamefont {T.~B.}\ \bibnamefont {Adler}}, \bibinfo {author}
  {\bibfnamefont {R.~D.}\ \bibnamefont {Amos}}, \bibinfo {author}
  {\bibfnamefont {A.}~\bibnamefont {Bernhardsson}}, \bibinfo {author}
  {\bibfnamefont {A.}~\bibnamefont {Berning}}, \bibinfo {author} {\bibfnamefont
  {D.~L.}\ \bibnamefont {Cooper}}, \bibinfo {author} {\bibfnamefont {M.~J.~O.}\
  \bibnamefont {Deegan}}, \bibinfo {author} {\bibfnamefont {A.~J.}\
  \bibnamefont {Dobbyn}}, \bibinfo {author} {\bibfnamefont {F.}~\bibnamefont
  {Eckert}}, \bibinfo {author} {\bibfnamefont {E.}~\bibnamefont {Goll}},
  \bibinfo {author} {\bibfnamefont {C.}~\bibnamefont {Hampel}}, \bibinfo
  {author} {\bibfnamefont {A.}~\bibnamefont {Hesselmann}}, \bibinfo {author}
  {\bibfnamefont {G.}~\bibnamefont {Hetzer}}, \bibinfo {author} {\bibfnamefont
  {T.}~\bibnamefont {Hrenar}}, \bibinfo {author} {\bibfnamefont
  {G.}~\bibnamefont {Jansen}}, \bibinfo {author} {\bibfnamefont
  {C.}~\bibnamefont {K{\"{o}}ppl}}, \bibinfo {author} {\bibfnamefont
  {Y.}~\bibnamefont {Liu}}, \bibinfo {author} {\bibfnamefont {A.~W.}\
  \bibnamefont {Lloyd}}, \bibinfo {author} {\bibfnamefont {R.~A.}\ \bibnamefont
  {Mata}}, \bibinfo {author} {\bibfnamefont {A.~J.}\ \bibnamefont {May}},
  \bibinfo {author} {\bibfnamefont {S.~J.}\ \bibnamefont {McNicholas}},
  \bibinfo {author} {\bibfnamefont {W.}~\bibnamefont {Meyer}}, \bibinfo
  {author} {\bibfnamefont {M.~E.}\ \bibnamefont {Mura}}, \bibinfo {author}
  {\bibfnamefont {A.}~\bibnamefont {Nicklass}}, \bibinfo {author}
  {\bibfnamefont {D.~P.}\ \bibnamefont {O'Neill}}, \bibinfo {author}
  {\bibfnamefont {P.}~\bibnamefont {Palmieri}}, \bibinfo {author}
  {\bibfnamefont {D.}~\bibnamefont {Peng}}, \bibinfo {author} {\bibfnamefont
  {K.}~\bibnamefont {Pfl{\"{u}}ger}}, \bibinfo {author} {\bibfnamefont
  {R.}~\bibnamefont {Pitzer}}, \bibinfo {author} {\bibfnamefont
  {M.}~\bibnamefont {Reiher}}, \bibinfo {author} {\bibfnamefont
  {T.}~\bibnamefont {Shiozaki}}, \bibinfo {author} {\bibfnamefont
  {H.}~\bibnamefont {Stoll}}, \bibinfo {author} {\bibfnamefont {A.~J.}\
  \bibnamefont {Stone}}, \bibinfo {author} {\bibfnamefont {R.}~\bibnamefont
  {Tarroni}}, \bibinfo {author} {\bibfnamefont {T.}~\bibnamefont
  {Thorsteinsson}}, \ and\ \bibinfo {author} {\bibfnamefont {M.}~\bibnamefont
  {Wang}},\ }\href@noop {} {\enquote {\bibinfo {title} {Molpro, version 2015.1,
  a package of ab initio programs},}\ }\bibinfo {howpublished}
  {http://www.molpro.net} (\bibinfo {year} {2015})\BibitemShut {NoStop}%
\end{thebibliography}%

\end{document}